\documentclass[a4paper,11pt]{article}

\pdfoutput=1
\usepackage{color}
\usepackage{amsmath,graphicx}
\usepackage{simplewick}
\usepackage{epsf}
\usepackage{graphicx,epsfig}
\usepackage{grffile}
\usepackage{amsfonts}
\usepackage{amssymb}
\usepackage[nosort]{cite}
\usepackage{setspace}

\usepackage[export]{adjustbox}
\usepackage{jcappub}
\usepackage[T1]{fontenc}

\def\CO{{\cal O}}

\def\high{\vphantom{\Biggl(}\displaystyle}

\newcommand{\be}{\begin{equation}}
\newcommand{\ee}{\end{equation}}
\newcommand{\bea}{\begin{eqnarray}}
\newcommand{\eea}{\end{eqnarray}}
\newcommand{\barr}{\begin{array}}
\newcommand{\earr}{\end{array}}
\def\bal#1\eal{\begin{align}#1\end{align}}

\title{\boldmath The Future of Primordial Features with 21 cm Tomography}

\author[a,b]{Xingang Chen,}
\author[c]{P. Daniel Meerburg,}
\author[d,e]{and Moritz  M{\"u}nchmeyer}

\affiliation[a]{Institute for Theory and Computation, Harvard-Smithsonian Center for Astrophysics, 60 Garden Street, Cambridge, MA 02138, USA}
\affiliation[b]{Department of Physics, The University of Texas at Dallas, Richardson, TX 75083, USA}
\affiliation[c]{CITA, University of Toronto, 60 St. George Street, Toronto, Canada}
\affiliation[d]{Sorbonne Universit\'{e}s, UPMC Univ Paris 06, UMR7095}
\affiliation[e]{CNRS, UMR7095, Institut d'Astrophysique de Paris, F-75014, Paris, France}

\emailAdd{xingang.chen@cfa.harvard.edu}
\emailAdd{meerburg@cita.utoronto.ca}
\emailAdd{munchmey@iap.fr}

\abstract{Detecting a deviation from a featureless primordial power spectrum of fluctuations would give profound insight into the physics of the primordial Universe. Depending on their nature, primordial features can either provide direct evidence for the inflation scenario or pin down details of the inflation model. Thus far, using the cosmic microwave background (CMB) we have only been able to put stringent constraints on the amplitude of features, but no significant evidence has been found for such signals. Here we explore the limit of the experimental reach in constraining such features using 21 cm tomography at high redshift. A measurement of the 21 cm power spectrum from the Dark Ages is generally considered as the ideal experiment for early Universe physics, with potentially access to a large number of modes. We consider three different categories of theoretically motivated models: the sharp feature models, resonance models, and standard clock models. We study the improvements on bounds on features as a function of the total number of observed modes and identify parameter degeneracies. The detectability depends critically on the amplitude, frequency and scale-location of the features, as well as the angular and redshift resolution of the experiment. We quantify these effects by considering different fiducial models. Our forecast shows that a cosmic variance limited 21 cm experiment measuring fluctuations in the redshift range  $30\leq z \leq 100$ with a 0.01-MHz bandwidth and sub-arcminute angular resolution could potentially improve bounds by several orders of magnitude for most features compared to current Planck bounds. At the same time, 21 cm tomography also opens up a unique window into features that are located on very small scales.}

\begin{document}
\maketitle
\flushbottom

\section{Introduction}\label{sec:intro}
\setcounter{equation}{0}

In the post-CMB era one of the most exciting observational frontiers is the measurement of the high-redshift 21 cm
spin-flip transition of neutral hydrogen. Measuring small fluctuations of the brightness temperature of the 21 cm field will be challenging
but could potentially provide the most detailed picture of the primordial universe \cite{Fuetal2006,Fur2009proc,Pritchard:2011xb} on scales that are non-perturbative today (e.g.~low-$z$ large scale structure) or below the CMB damping scale. The 21 cm absorption field direct probe of large-scale structure during the cosmic ``Dark Ages'' \cite{Loeb_2004}. This era follows the last scattering of photons of the CMB and precedes the formation of the
first luminous objects, sometimes referred to as the ``Cosmic Dawn".

Since most scales are not yet non-linear and the signal is not limited by a narrow visibility function (i.e. is effectively a 3D field), the 21 cm absorption field  in principle contains much more information than CMB
anisotropies. Fluctuations in the 21 cm absorption field are limited only by the baryonic Jeans scale, $k_{\rm J} \sim 300$ Mpc$^{-1}$, and since overdensities remain small during the Dark Ages their growth is very well described by linear perturbation theory down to redshift 30 \cite{Naoz_2005}. Non-linear corrections can become important at later times \cite{Lewis_2007}; however, contrary to the present-day density
field which reaches order unity fluctuations on scales $k \gtrsim k_{\rm NL} \sim0.1$ Mpc$^{-1}$, for $z \gtrsim 30$ non-linear corrections remain perturbative on all scales of interest and the Dark Ages 21 cm power spectrum can in principle be computed analytically.

The 21 cm power spectrum as a cosmological probe was first considered in Ref.~\cite{Loeb_2004}.  This initial computation did not include fluctuations of the local velocity gradient and the gas temperature, shown to be important in Ref.~\cite{Bharadwaj_2004}. The authors of Ref.~\cite{Lewis_2007} have provided the most detailed calculation, including relativistic and velocity corrections, as well as approximate non-linear corrections. Relative velocities can lead to correction of order $10\%$, both on very large scales $k \ll 0.01$ Mpc$^{-1}$ \cite{Yacine2013} as well as small scales \cite{Tsel_2010,Fialkov_2012,Fialkov_2013b,Bittner_2011}. The potential of low $z$ reionization signal for cosmology has been considered in various papers \cite{Fuetal2006,Fur2009proc,Mao:2008ug} and recently in the context of low redshift intensity mapping \cite{ChimeShaw}.

In this paper we will explore the potential of the highly redshifted 21 cm signal to constrain models of the early Universe. In particular, we are interested in an important class of models that predict primordial scale-dependent oscillatory features. The precise characteristics of these oscillations could provide valuable information on the precise mechanism that source the seeds of structure formation. They may be used to distinguish inflation from alternative scenarios, or provide specific details of the inflation scenario. Many different feature models have been considered and constrained using CMB data, both at the level of the power spectrum \cite{PSOscillationsMartin2004,PSOscillationsHamann2007,PSoscillationsPahud2009,PSOscillationsBenetti2011,PSfeaturesDvorkin2011,PSfeaturesHazra2013,PSOscillationsAich2013,PlanckInflation2013,Meerburg2014a,Meerburg2014b,FeaturesBenitti,BICEPoscillations2014,MeerburgOscillations2014,PlanckInflation2014} and the bispectrum \cite{PlanckNGs2013,PlanckNGs2015}. Our main goal is to estimate how much additional information can be claimed, when mapping out the 21 cm signal from the era between the last scattering surface and the formation of the first stars \cite{Mao:2008ug}. Since the signal we are searching for is small and there is no theoretical lower-limit from model building, our initial analysis will focus on the most optimistic scenario; one in which we are cosmic variance limited out to very small scales, and where we have been able to remove foregrounds entirely. The reason to consider such an optimistic scenario is that it should give us a benchmark on how much information can maximally be obtained. Since no experiment that can map out this highly redshifted signal is planned in the immediate future, it is well motivated to first consider such an ideal case. A more detailed analysis should be performed once initial low$-z$ analysis has proven to be able to correctly characterize and subsequently remove foregrounds (i.e. practical limitations are better understood). Techniques to effectively clean the 21 cm observation from foregrounds have been studied in detail both in the theory \cite{Foregrounds1,Foregrounds4,Foregrounds5,Foregrounds7} and to some extend in practice \cite{Foregrounds2,Foregrounds3,Foregrounds6,Foregrounds8} but it remains to be seen how well these (tested) methods work at high $z > 20$. At too low frequencies, the ionosphere becomes opaque and a measurement of 21 cm can only be done from space, e.g. the moon \cite{Carilli_2007}. Experimental noise can be estimated by considering different baseline, antenna configuration, collecting area, scintillation effects and integration time (see e.g. \cite{Loeb:2008hg,Mao:2008ug,2016MNRAS.458.3099V,2012ApJ...745..176V,2013ApJ...779..124M} and Ch.12 of the book \cite{LoebBook} for review).

This paper is organized as follows. In Sec.~\ref{Sec:Models} we review different classes of oscillatory signatures, which can be broadly classified as being generated by sharp features, resonance features and primordial standard clocks. In Sec.~\ref{Sec:21cmforecast} we first estimate the cosmic variance limits on the available signal in the matter perturbations, and then perform a full Fisher analysis for the 21 cm signal, taking into account all model parameters. We identify possible degeneracies and limitations of 21 cm tomography as a probe of primordial features. We conclude in Sec.~\ref{Sec:conclusions}.

\section{Oscillatory features in primordial density perturbations}
\label{Sec:Models}
\begin{figure}[tbp] 
   \centering
   \includegraphics[width=6in]{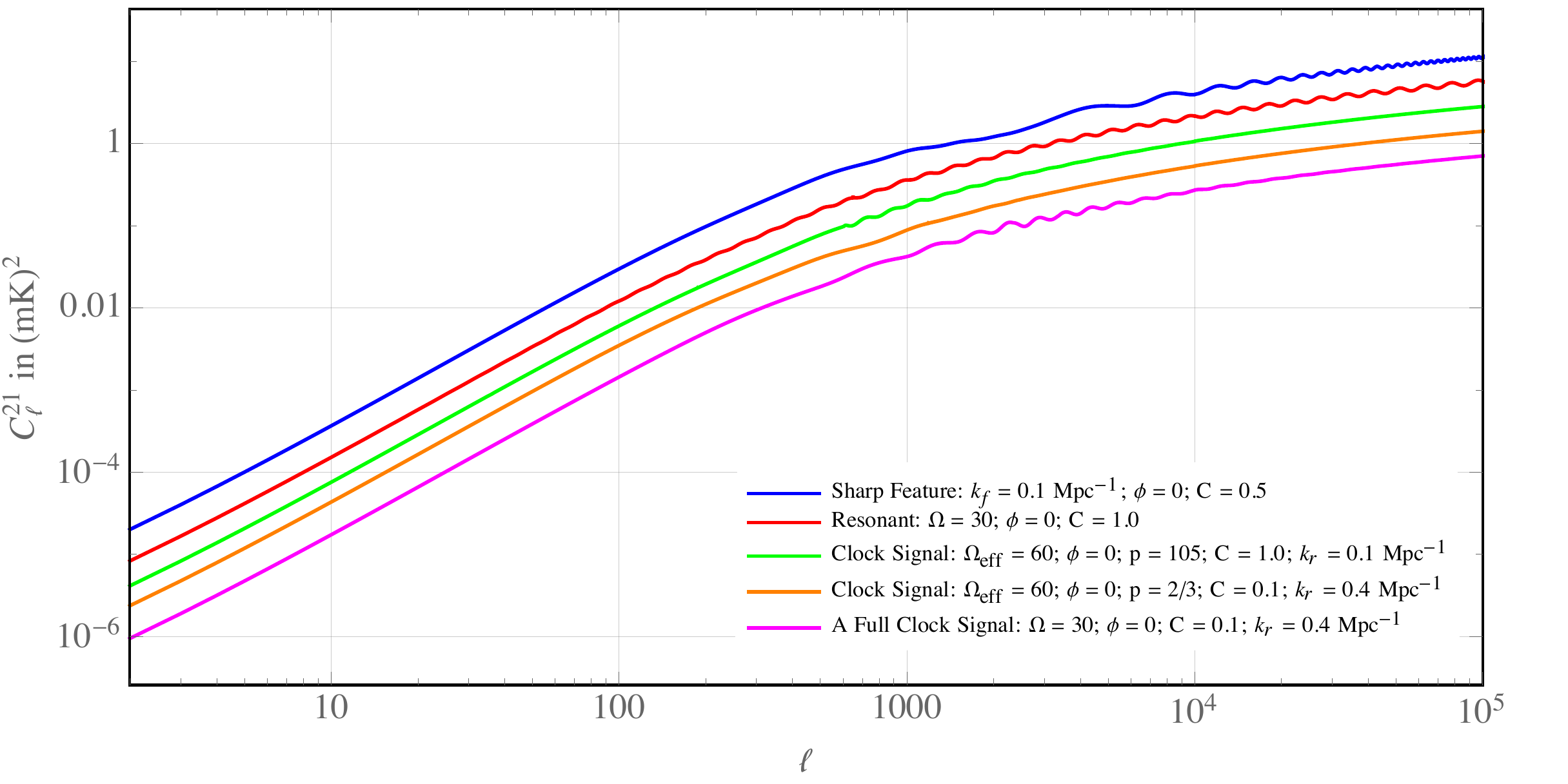}
   \caption{5 examples of the 3 different spectra considered in this paper at $z = 50$. The amplitudes have been made large such that the features can easily be distinguished by eye from the default featureless spectrum. The spectra are also artificially offset not to overlap.  }
   \label{fig:ModelOverview}
\end{figure}

\begin{figure}[tbp] 
   \centering
   \includegraphics[width=6in]{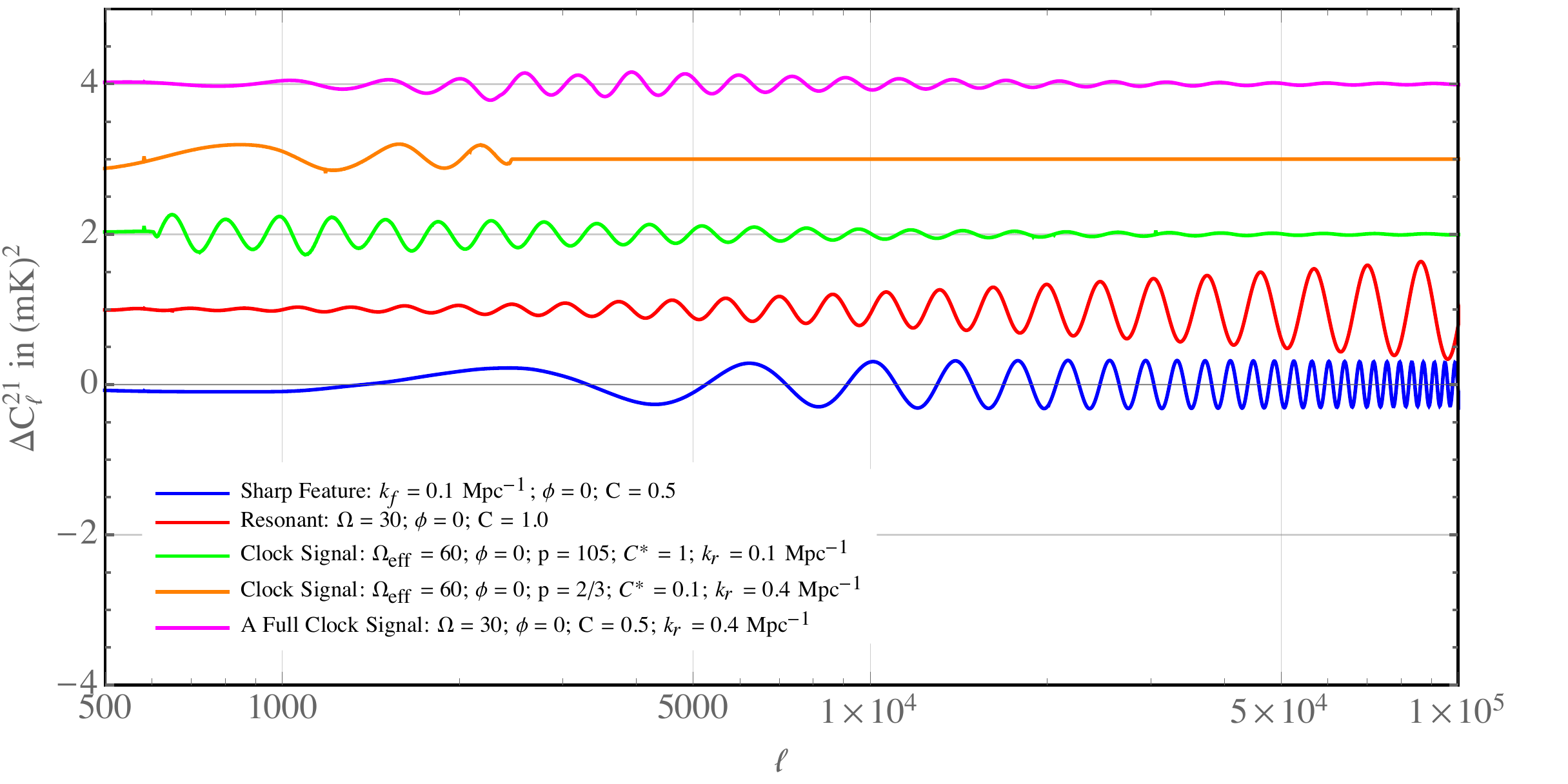}
   \caption{Same spectra as Fig.~\ref{fig:ModelOverview} explicitly showing the difference with the default featureless spectrum at $z = 50$. Linear, resonance, clock and full clock signals can now be distinguished more clearly. Note that the clock signal in a matter contraction Universe (orange) has features up to some scale $k_r$.  As before, for illustration purpose, in this figure the amplitudes of the features have been chosen such that we can easily see the effect; in reality, these large amplitudes are completely excluded by CMB data analysis. }
   \label{fig:ModelOverviewDiffs}
\end{figure}

In this section we give a short overview of models of primordial features with scale-dependent oscillatory signatures. We divide them into several classes, and in each class define a simple template that captures the essential physics. These templates will be used for our Fisher forecast.
The details of these feature models are reviewed in \cite{Chen:2010xka,Chluba:2015bqa,Chen:2016cbe}.
For each of these models we plot examples of corresponding 21 cm power spectra in Fig. \ref{fig:ModelOverview} and Fig. \ref{fig:ModelOverviewDiffs}.

\subsection{Sharp feature signal}
\label{Sec:Sharp}

Sharp features refer to localized features in the potential or internal field space of inflation models. The sharp feature temporarily breaks the slow-roll conditions and excites the quantum fluctuations of the curvature mode near and inside the horizon, generating the ``sharp feature" signal in density perturbations with special oscillatory scale-dependence \cite{Starobinsky:1992ts}.

There can be a large variety of different types of sharp features in model building. For example, the sharp feature can be a kink, step or bump in the single field inflationary potential \cite{Starobinsky:1992ts,Adams:2001vc,Chen:2006xjb,Adshead:2011jq,Hazra:2014goa,Romano:2014kla}, or the same type of features in the internal field space such as the sound speed of the inflaton \cite{Bean:2008na,Miranda:2012rm,Bartolo:2013exa}; the step/kink/bump can also occur in multi-field space \cite{Chen:2014joa,Chen:2014cwa}; a sharp bending of the inflaton trajectory in multi-field space introduces another type of sharp feature \cite{Achucarro:2010da,Gao:2012uq,CorrelatedFeaturesCs}; and many more.
In addition, the profiles of signals are also affected by the sharpness of the features in each type of model.
Despite this model sensitivity, the sharp feature signals share the common characteristic that their running behavior in momentum space is sinusoidal, and this running behavior appears in both the power spectrum and non-Gaussianities and are highly correlated \cite{Chen:2006xjb,Chen:2008wn,CorrelatedFeaturesCs4,CorrelatedFeaturesCs3,Appleby:2015bpw}.
The presence of such a character can be explained as follows. Sharp feature generates a localized feature in the evolution of the various background parameters and
the sinusoidal running behavior is essentially a consequence of a Fourier transform of this evolution.
On top of this common character, the sinusoidal running has a highly model-dependent envelop.

We refer to this type of signals as the ``sharp feature signal'', and we use the following template for its power spectrum \cite{Chen:2008wn,Chen:2011zf,Fergusson:2014hya,Fergusson:2014tza}:
\begin{align}
  \frac{\Delta P_\zeta}{P_{\zeta0}} =
    C \sin \left( \frac{2 k}{k_f} + \phi \right)
    ~,
    \label{Template_Sharp}
\end{align}
where $P_{\zeta 0}$ is the featureless power spectrum including the usual non-oscillatory running, and $\Delta P_\zeta$ is the correction due to the feature. This template has 3 parameters: the relative amplitude $C$, the parameter $k_f$ specifying the starting location, as well as frequency, of the feature, and the phase $\phi$. We emphasize that, for simplicity, this template only captures the leading property of the sharp feature signal, namely the sinusoidal running, while neglecting the envelop behavior. We make a few comments on the above approximation:

\begin{itemize}
\item
The sharpness of the feature determines how local the feature signal is in  $k$-space. The above template is a better approximation for those features that are sufficiently sharp such that the signals are very broad in $k$ space. In the opposite limit, in which the feature is not very sharp such that it causes very few oscillations in the $k$-space, the envelop becomes important. Such features have been considered to explain the mildly significant CMB glitch between $\ell=20-30$ \cite{Ade:2015lrj}. The observed power deficit is very localized and on very large scales.  We do not expect the  21 cm experiment to significantly improve these constraints. For this reason, we will not consider this type of models in this paper.

\item
For very sharp features, the scale dependence of this envelop is milder comparing to that of the sinusoidal running. More importantly, as the envelop behavior is highly model-dependent, for a model-independent data analysis, it is often more effective to ignore it and instead only analyze the common sinusoidal behavior. If a candidate signal is identified, the envelop behavior may be added and put important additional constraints on subsequent model selection, in which case we expect to be able to distinguish the sharpness of the feature better than the nature of the feature.

\item
For the purpose of forecast, the fully extended template represents the best scenario case. In cases where the effect of the envelop is important and the sinusoidal running is cut off at a certain scale, the error bars are expected to be larger. We will plot the error as a function of scales for this template, so that even for these cutoff cases, the forecasted error bars can be directly read off.

\end{itemize}


There are couple of interesting candidates in the CMB data for the sharp feature signal. There is a candidate at around $\ell \sim 20-40$ with marginal statistical significance \cite{Ade:2015lrj}, which was first found by WMAP. There is another candidate around $\ell \sim 700-800$ \cite{Chen:2014cwa}.  Depending on the frequency,the relative amplitude of the sharp feature signals $C$ has been constrained to be below a few percent.

\subsection{Resonance feature signal}
\label{Sec:Resonance}

Resonance features refer to periodic or semi-periodic features in inflation models. The most important property of this type of features is not the sharpness but its periodicity.
These periodic features induce time-dependent oscillatory components in various background parameters. If its frequency is much larger than the Hubble parameter $H$, the background oscillation will resonate with quantum fluctuations of fields that are deep inside the horizon, generating a ``resonance feature" signal in density perturbations with another kind of special oscillatory scale-dependence \cite{Chen:2008wn}.

In model-building, resonance models may be realized in terms of large field string theory models such as the axion monodromy inflation \cite{Flauger:2009ab}, small field string theory models such as brane inflation \cite{Bean:2008na}, or in particle physics in terms of axion inflation \cite{Wang:2002hf}. In these models the periodic features appear in the inflationary potential or the inflaton sound speed. Special types of resonance signals can arise due to other (semi-)periodic oscillations such as the oscillation of massive fields \cite{Chen:2011zf}, which we shall study in Sec.~\ref{Sec:SC}. Like the sharp feature signals, the resonance features also have highly correlated signals in non-Gaussianities \cite{Chen:2008wn,Flauger:2010ja,Chen:2010bka}.

The template for the resonance feature power spectrum is given by \cite{Chen:2008wn}
\begin{align}
\frac{\Delta P_\zeta}{P_{\zeta0}} = C \sin \left[ \Omega \log \left( 2k \right) + \phi \right] .
\label{Template_Resonance}
\end{align}
This template has 3 parameters: the relative amplitude $C$, the frequency $\Omega$ and the phase $\phi$. The discrete symmetry of the periodic features manifests as the discrete rescaling symmetry of the momentum in this template.

The Planck 2015 analysis presented a statistically insignificant best-fit for the resonance feature signal with $\Omega \sim 30$ \cite{Ade:2015lrj}. \\

The above two types of feature templates, namely the sharp feature and resonance feature templates, can also arise in terms of models of non-Bunch-Davies (non-BD) vacua. See \cite{Ade:2015lrj} for a summary. In these models, a new physics scale is introduced hypothetically; the quantum mode coming out from this scale takes a specific non-BD vacuum form and then follows the equation of motion in the usual low-energy field theory. If the new physics scale
is introduced at a specific time, the signal generated in the power spectrum due to the non-BD vacuum takes the form of the sharp feature signal in Sec.~\ref{Sec:Sharp}.
If the new physics scale is introduce at a specific energy scale for each mode, the signal takes the form of the resonance feature signal in Sec.~\ref{Sec:Resonance}.
This correspondence can be readily understood, because the sharp and resonance features can be viewed as the concrete realizations of the new physics scale hypothesized in the above models. Again, correlations between the power spectrum and higher order spectra are predicted \cite{NonBDBispectrum2009,NonBDBispectrum2010,NonBDBispectrum2010b,nonBDbispectrum2015}.

\subsection{Primordial standard clock signals}
\label{Sec:SC}

In the previous two subsections, we have assumed inflation as the driving mechanism for producing the background expansion and used small primordial features to learn about the details of the model. In this subsection, we do not assume inflation and we shall study a different class of primordial feature models whose properties can be used to distinguish between inflation and possible alternative scenarios.
These features are generated by classical or quantum oscillations of massive fields. These oscillations can be used as ``standard clocks" to directly measure the scale factor evolution $a(t)$ of the primordial universe \cite{Chen:2011zf,Chen:2011tu,Chen:2012ja,Chen:2014joa,Chen:2014cwa,Chen:2015lza,Chen:2016cbe}. For this reason the massive fields are called the primordial standard clocks. There are two general classes of the primordial standard clock models. The first one is the classical standard clocks \cite{Chen:2011zf,Chen:2011tu,Chen:2012ja,Chen:2014joa,Chen:2014cwa}, in which case the massive fields oscillate classically due to some kind of kicks from sharp features. The second one is the quantum primordial standard clocks \cite{Chen:2015lza,Chen:2016cbe} where the massive field oscillate automatically due to quantum fluctuations. The latter is a more general phenomenon, but to observe it requires a measurement of primordial non-Gaussianities \cite{Arkani-Hamed:2015bza,Chen:2009we,Chen:2009zp,Baumann:2011nk,Assassi:2012zq,Noumi:2012vr,Dimastrogiovanni:2015pla}. In this work, we concentrate on the classical one, which induces scale-dependent oscillatory features in the power spectrum.

Besides measuring $a(t)$, discovering such clock signals in the power spectrum also means the discovery of new massive particles, which are likely going to be the heaviest particles ever to be found.

The classical standard clock models are a class of feature models that involve special mixture of two types of features smoothly connected to each other.
The first is the signal generated by whatever sharp feature that excites the massive field; it is of the sharp feature type and it oscillates on large scales within the entire signal. The second part is generated by the subsequent oscillation of the massive field; this signal is of the resonance type and it rings on small scales.
The latter part is referred to as the clock signal and is the most important part of the full signal.

The oscillation of massive fields in any time-dependent background is standard and can be regarded as a clock that generates standard ticks. These ticks get imprinted in the density perturbations, and directly record the scale factor of the primordial universe $a$ as a function of time $t$. The function $a(t)$ is the defining property of the primordial universe scenario, and this is the important information that the clock signal carries. See Ref.~\cite{Chen:2014cwa} for a review on classical primordial standard clocks.

The standard clock signal has the potential to provide a unified explanation to the two anomalous glitches in the CMB data, namely the glitches at $\ell\sim 20-40$ and $\ell\sim 700-800$, with the former being the sharp feature signal and the latter the clock signal \cite{Chen:2011zf,Chen:2014cwa}.

\subsubsection{Clock signals}

We first study the most important part of the standard clock signals, namely the clock signal, in different primordial universe scenarios. Notice that, in the clock signal, we have artificially cut off the sharp feature signal part which in reality should be smoothly connected to the clock signal. A profile of the full standard clock signal has not yet been computed in the most general case, but in special examples we can use a closed-fitting approximation that matches all the properties. Such a full clock signal will be studied in Sec.~\ref{Sec:fullSC}.

The approximate clock signal can be written as follows \cite{Chen:2011zf,Chen:2014cwa}
\begin{align}
  \frac{\Delta P_\zeta}{P_{\zeta0}} = \begin{cases}
    0, & \text{$k <  k_r/2$ (expanding)},  \\
    0, & \text{$k >  k_r/2$ and $ k < k_r/\Omega_{\rm eff}$ (contracting)}, \\
    \high{
    C \left( \frac{2k}{k_r}~  \right)^{-\frac{3}{2}+\frac{1}{2p}}
    \sin \left[ p \frac{\Omega_{\rm eff}}{2} \left( \frac{2k}{k_r}  \right)^{\frac{1}{p} } + \phi \right]},  & \text{otherwise}.
  \end{cases}
  \label{clock_template}
\end{align}
This template has 5 parameters. The parameter $p$ specifies different primordial universe scenarios:
$|p|>1$ corresponds to inflation; $0<p\sim \CO(1)<1$ corresponds to the fast contraction scenario, such as matter contraction scenario; $0<p\ll 1$ corresponds to the slow-contraction scenario, such as the ekpyrotic scenario; and $-1\ll p<0$ corresponds to the slow-expansion scenario.

For inflation models, $p\gg 1$, and the last line reduces to
\bea
\frac{\Delta P_\zeta}{P_{\zeta0}} \xrightarrow{p\gg1} C \left( \frac{2k}{k_r}  \right)^{-\frac{3}{2}}
\sin \left( \frac{\Omega_{\rm eff}}{2} \ln \frac{2k}{k_r} + \phi \right) ~.
\label{eq:clock_template_large_p}
\eea
We then effectively have 4 parameters left instead of 5.

The templates of the clock signal are compared with Planck 2013 data in \cite{Chen:2014cwa}, and a best-fit (yet statistically insignificant) model for the inflationary clock is found. The scale location of this candidate will be used as the fiducial value in our later analysis.

\subsubsection{An example of full standard clock signal}
\label{Sec:fullSC}

As mentioned, general full standard clock signal that includes both the sharp feature and clock signal is currently unavailable.
The following is a special example template derived for a full standard clock signal in an inflation model, denoted as T2 in \cite{Chen:2014cwa},
\bea
\frac{\Delta P_\zeta}{P_{\zeta0}} =
\left\{
\begin{array}{ll}
\high{
C \left[ 7\times 10^{-4} \left( \frac{2k_1}{k_0} \right)^2 + 0.5 \right]
\cos \left[ \frac{2k_1}{k_0} + 0.55\pi \right] ~,
}
&
k_1 < k_a ~,
\\
\high{
\frac{14}{13} C \left( \frac{2k_1}{k_r} \right)^{-3/2}
\sin \left[ \Omega \ln \frac{2k_1}{k_r} + 0.75\pi \right] ~,
}
&
k_b >k_1 \ge k_a ~,
\\
\high{
\frac{19}{13} C \left( \frac{2k_1}{k_r} \right)^{-3/2}
\sin \left[ \Omega \ln \frac{2k_1}{k_r} + 0.75\pi \right] ~,
}
&
k_1 \ge k_b ~,
\end{array}
\right.
\label{Template_SC_full}
\eea
where
\bea
k_0 = \frac{k_r}{1.05\Omega} ~, ~~~
k_a = \frac{67}{140}k_r ~, ~~~
k_b = \frac{24}{35}k_r ~, ~~~
\Omega=30 ~.
\eea

In \cite{Chen:2014cwa} the following procedure is used to obtain this template. The best-fit for the inflationary clock signal \eqref{eq:clock_template_large_p} is first obtained and used to determine all the parameters in a standard clock model. Then, numerical simulation is used to work out the full prediction of this model on the power spectrum. Eq.~\eqref{Template_SC_full} is an analytical fit to this numerical result. In this process, some of the free parameters have been fixed. There are only two parameters left in this special example. Since the full clock signal should have the same number of parameters as in the clock signal, it is a drawback of this procedure that not all non-degenerate parameters can be made variables.

\section{21-cm forecast}
\label{Sec:21cmforecast}

In this section we perform a Fisher forecast for the models discussed above. We first give a brief description of the 21 cm signal as observed from Dark Ages, a period of 21 cm absorption against the CMB, and describe our forecasting methodology. We then study the three model classes one by one, forecasting the experimental sensitivity and commenting on the peculiarities of each class.

\subsection{The 21 cm signal}
We start by reviewing the 21 cm signal from the Dark Ages. A neutral hydrogen atom has two states with slightly different energy; one in which the proton and electron spin are aligned (singlet, $n_0$), and one in which they are opposite (triplet, $n_1$). Through the Boltzmann factor we associate a spin temperature $T_s$ relating the abundance of these two states, i.e. \cite{Wouthuysen_1952,Field_1958},
\be
\frac{n_1}{n_0} \equiv 3 \exp\left(- \frac{E_{10}}{T_s}\right) \approx 3
\left(1 - \frac{E_{10}}{T_s}\right).
\ee
The energy difference $E \sim 0.068$K allows the transition to produce or absorb a photon with a wavelength $\lambda = 21$ cm. The spin temperature can not be measured directly. Instead, we measure the brightness of 21 cm fluctuations against the background of CMB photons. At the same time, the brightness has to cross the inter-galactic medium, which has a finite optical depth for the 21 cm photon, so that
\be
T_b^{21} = \tau_{21} \frac{T_s-T_{\rm cmb}}{1+z},
\ee
gives the brightness temperature measured today by an observer on earth. We have used that the optical depth is small which is true from the observer to the Dark Ages. The optical depth is given by
\be
\tau_{21} = \frac{3 E_{10}}{32 \pi T_s} x_{\rm HI} n_{\rm H} \lambda_{10}^3
\frac{A_{10}}{H + \partial_{\parallel} v_{\parallel}}.
\label{eq:tau}
\ee
Here $\lambda_{10}=21$ cm, $x_{\rm HI}$ is the fraction of neutral hydrogen and $\partial_{\parallel} v_{\parallel}$ is the line-of-sight gradient of the component of the peculiar velocity along the line of sight.

Fluctuations in the brightness are therefore sensitive to fluctuations in the radiation field, the neutral hydrogen field (baryons), the velocity field and the spin temperature. The spin temperature is determined from a balance between collisional transitions, which drive $T_s \rightarrow T_{\rm gas}$, and radiative
transitions mediated by CMB photons, which drive $T_s \rightarrow T_{\rm cmb}$ (see e.g. \cite{Pritchard:2011xb} for a detailed discussion). It is given by
\be
T_{s} = T_{\rm cmb} + (T_{\rm gas} - T_{\rm cmb}) \frac{C_{10}}{C_{10}
  + A_{10}\frac{T_{\rm gas}}{E_{10}}}, \label{eq:Ts}
\ee
with $C_{10}$ the collisional transition rate and $A_{10} \approx 2.85 \times 10^{-15}$ s$^{-1}$ is the spontaneous decay rate. $C_{10}$ itself is a function of $T_{\rm gas}$ \cite{Kuhlen_2006}.  The gas temperature is a function of the baryon density and the free electron fraction \cite{Lewis_2007}. The radiative transitions are captured by $A_{10}$. To lowest order in fluctuations, we then have
\be
\delta T_b^{21} = \mathcal{T}_H\delta_H + \mathcal{T}_{T_{\rm gas}} \delta_{T_{\rm gas}}+\bar{T}_b^{21} (1-\delta_v) + \mathcal{O}(\delta^2).
\label{eq:21cmfluctuations}
\ee
$\bar{T}_b$, $\mathcal{T}_H$ and $\mathcal{T}_{T_{\rm gas}}$ are functions of redshift only. This  first  term  was initially computed in Ref.~\cite{Loeb_2004} and later extended to include the last two terms in Ref.~\cite{Bharadwaj_2004}. The full computation to linear order, including fluctuations in the electron fraction, post-newtonian effects and non-linear effects due to gravitational lensing, was performed in Ref.~\cite{Lewis_2007} (for higher order effects see e.g. \cite{Pillepich_2007,Yacine2013,21cmFNL}).

We will compute the 21 power spectra in the flat sky limit as in Ref. \cite{21cmFNL}. Here the full spectrum is given by
\be
P_{21} (k,z) \simeq \left(\alpha(z) + \beta(z) \left(\frac{P_{\rm gas}}{P_{\rm b}}\right)^2 + \bar{T}_b^{21} \left(\frac{k_{\parallel}}{k}\right)^2\right)^2 P_{\rm b}(k,z)
\ee
with $\alpha(z)$ the derivative of the 21 cm brightness temperature w.r.t. to the hydrogen fluctuations and $\beta(z)$ the derivative w.r.t. the fluctuations in the gas. The third term comes from the velocity fluctuations and $P_{\rm b}$ the baryon power spectrum. 

\subsection{Fisher matrix and forecast parameters}

We want to estimate the sensitivity in an ideal 21 cm experiment limited only by cosmic variance. We therefore consider a hypothetical tomographic 21 cm experiment providing a 3-dimensional reconstruction. The information about parameters $p_i$ contained in a comoving 3-dimensional box of size $V$ is given by the Fisher matrix
\begin{equation}
F_{ij} = V \int_{k_{\rm min}}^{k_{\rm max}} \frac{k^2 dk}{(2\pi)^2}
\frac{\partial P(k)}{\partial p_i} \, \frac{1}{P(k)^2} \, \frac{\partial P(k)}{\partial p_j},
\label{eq:LSSFisher}
\end{equation}
where we assumed the covariance matrix is given by cosmic variance without additional noise. Let us first estimate the ultimate constraint on a small feature amplitude $C$ such that $P(k) = P_0(k) + C \delta P(k)$ (where for example $\delta P(k) = P_0 \sin\left(\omega k\right)$) for which the Fisher matrix reduces to
\begin{equation}
F_{CC} = V \int_{k_{\rm min}}^{k_{\rm max}} \frac{k^2 dk}{(2\pi)^2} \frac{\delta P(k)^2}{P(k)^2}.
\label{eq:LSSFisher2}
\end{equation}
For a power spectrum that is approximately scale invariant (where we also allow for oscillations, but not for an envelope factor), one finds from eq. \eqref{eq:LSSFisher2} and $\sigma_C=(F)^{-1/2}$ that one can detect amplitudes of order $\sigma_C \sim \sqrt{\left( \frac{k_{\rm min}}{k_{\rm max}} \right)^3}$. The smallest mode we can ever hope to use (practical limitations aside) is the Jeans scale, which is $k_J \sim 300$ Mpc$^{-1}$ \cite{Yacine2013} for the redshift regime of interest. The largest
mode is limited by the survey volume, and we take it to be $k_{\rm min} = 2\pi(3V(z_{\rm min},z_{\rm max})/4\pi)^{-1/3}$ with
\be
V(z_{\rm min},z_{\rm max}) = \frac{4\pi}{3}(d(z_{\rm max})^3-d(z_{\rm min})^3)
\ee
and
\be
d(z) = c \int dz'/H(z')
\ee
Inserting those numbers, and assuming a redshift range $30 \leq z \leq 100$ we obtain $\sigma_C \simeq 10^{-9}$. For the suggested volume, this value provides an upper limit (or an ultimate constraint) on non-localized features.

For a more realistic calculation, including correlations between different feature parameters and cosmological parameters, we need to take into account the transfer functions to include the redshift evolution of the probe. The 21 cm field during the Dark Ages is a biased tracer of the linear matter power spectrum and we can relate $P_{21}(k,z) = T_{21}^2(k,z) P^{\mathrm{initial}}(k)$ with 21 cm transfer function $T_{21}(k,z)$. We will divide the resolution into radial modes (along the line of sight) and angular modes (perpendicular along the line of sight). The redshift dependence can be taken into account most easily by dividing the redshift range in several large bins. Our idealized experimental setup is a cosmic variance limited experiment with a frequency band observing 21 cm fluctuations from  $30 \leq z \leq 100$ with equal bins of $\Delta z = 5$. We thus have to sum over 14 bins, a number small enough to treat them independently without losing substantial cross-correlation information. Furthermore, the angular and radial resolution are set independently, where the former is set by the baseline ($b$) in km as
\be
k_{\rm max}^{\perp} \simeq 2 \pi \nu_0 b \frac{1}{d(z) (1+z)} \frac{1}{c}
\ee with $\nu_0$ the rest frequency of the 21 cm line in Hz and $c$ the speed of light in km/s and $d(z)$ in Mpc.  and the latter by the frequency window of the experiment, i.e. \cite{21cmFNL},
\be
k_{\rm max}^{\parallel} \simeq \sqrt{\frac{17}{3}} \left(20\, \delta \nu \sqrt{1+z)}\right)^{-1}.
\ee
with $\delta \nu$ in MHz. The Fisher matrix is then given by
\begin{equation}
F_{ij} = \sum_z V(z) \int_{k_{\rm min}}^{k_{\rm max}^{\parallel,\perp}} \frac{d^3 \vec{k}}{(2\pi)^3}
\frac{\partial P_{21}(k,z)}{\partial p_i} \, \frac{1}{P_{21}(k,z)^2} \, \frac{\partial P_{21}(k,z)}{\partial p_j},
\label{eq:LSSFisher4}
\end{equation}
This is the equation we will use for our Fisher forecast below with the minimal and maximal observed $k$ are set by the volume and the resolution in the angular and line-of-sight direction as explained above. The integral in the flat-sky becomes a 2-dimensional integral over parallel and perpendicular modes.

In the following analysis we vary the width of the window $\delta \nu$, which sets the radial resolution and the baseline $b$, which sets the angular resolution, of a future array (without specifying further details of the experiment). When estimating the correlations between parameters we consider $\delta \nu = 0.01$ MHz and a baseline of 1 km (all contour plots are based on those settings). For the marginalized estimates of the errors of feature parameters we will vary the baseline from $1$ to $100$ km (roughly corresponding to $\ell \sim 10^3-10^5$ at $z = 30$), and $0.01 \leq \delta \nu \leq 1$ and will show marginalized errors on primordial parameters in various panels. We consider 6 baselines and 5 windows, resulting in a total of 30 different experimental configurations. In principle, it is more realistic to further narrow the window function, but in practice the noise is a function of the width of the window function (see e.g. Ref. \cite{2013ApJ...779..124M} Eq. (56)) and one has to optimize this width based on the specific goals of the experiment. The 2-dimensional plots in the next section will visualize how the baseline and window affect the forecasted constraints; in combination with noise estimates, one could in principle determine the optimal experimental setup (including costs and realizability).   

In the following we will consider the 3 different models: sharp features, resonance features and clock signals and in addition a full clock example. For the sake of simplicity we only vary parameters that are known to correlate, which are the primordial parameters that describe the feature and $H_0$. Although correlations with cosmological parameters are not expected to be strong, at very low frequencies correlations do appear. In particular, for linear features, since the Baryonic Acoustic Oscillations (BAO's) are linear, it is expected that for a primordial frequency similar to the BAO frequency, correlations with late time cosmological parameters will appear (see e.g. \cite{Fergusson:2014hya} and \cite{Meerburg2015b} for discussions). We will use a fiducial Planck cosmology fixing the remaining five standard cosmological parameters: $\Omega_b$, $\Omega_c$, $\tau$, $n_s$ and $A_s$. We have 2-4 feature parameters depending on the model we consider. Our Fisher matrix will thus have a minimal of $3\times3$ components (clock example) and a maximum of $5\times5$ components (clock signal). 

We will also make plots that show explicit degeneracies (which will be present for some frequencies) through contours, i.e. $C(p_n,p_m) = (p_n - p_n^0) Q_{nm} (p_n-p_m^0)$ where $Q_{nm}$ is a matrix that that can be derived from the full Fisher matrix $F_{ij}$. Inverting $F_{ij}$ and considering only the elements associated with parameter $p_n$ and $p_m$ and inverting this again gives the matrix $Q_{nm}$. The constructed contours present the one and two sigma bounds of the parameters $p_n$ and $p_m$ marginalized over all other parameters when equated to $C(p_n,p_m)  = 1.5137^2$ ($\sigma$) and $C(p_n,p_m)  = 2.448^2$ ($2\sigma$).



\subsection{Influence of the spherical survey geometry}

The redshift dependence and spherical geometry of a realistic 21 cm survey makes a treatment in terms of spherical harmonics attractive, and we considered this setup in an earlier version of the paper. However for a tomographic survey, it is computationally very challenging to implement a $C_{\ell}$ based forecast. For a very low radial resolution (wide frequency bands), one can approximate the signal as a sum of several independent redshift shells (see e.g. \cite{21cmFNL}) so that the signal scales as
\be
\label{eq:modes2}
\sigma_{C'} \sim \sqrt{\frac{1}{N_z} \left(\frac{1}{l_{\rm max}} \right)^2},
\ee
where $N_z$ is the number of redshift shells. However this calculation is no longer justified once the redshift resolution is comparable or larger than the angular resolution. In that case, the fact that modes correlate redshift shells along the line-of-sight cannot be neglected. A full Fisher analysis in terms of $C_{\ell}$ and $z$ taking into account cross-correlation between shells is technically very difficult and should not lead to qualitatively different results than the forecast in terms of $P_k$ we present here. The radial resolution of a future experiment mapping the Dark Ages is unknown, but naively it seems that radial (frequency) resolution is easier to obtain than angular resolution, the latter of which requires a very large observatory. In this paper we therefore present results as a function of both the angular and frequency resolutions.

We point out a further important difference between a $C_{\ell}$ based forecast and the full 3-dimensional $P(k,z)$ result. The detectability of features will predominantly be a function of the amplitude and the frequency of the feature. Similar analysis of the CMB has shown that as the frequency increases, the effective amplitude after projection decreases. This effect is caused by the geometric properties of the transfer functions, and will also be true for 21 cm astronomy; the primordial amplitude of the feature is suppressed (on large scales) due to the convolution of the oscillating signal with rapidly oscillating transfer functions. In our flat-sky $k$-space analysis the transfer function is much smoother in both $k$ and $z$. As a result we do not expect that the ability to constrain the amplitude will be suppressed as a function of primordial frequency. Of course, once cross correlation between shells is taken into account and the radial resolution is large, the full 3-dimensional information can in principle also be recovered in the $C_{\ell}$ treatment.

\subsection{Sharp feature signal (linear feature)}

\begin{figure}[h] 
   \centering
   \includegraphics[width=3in,valign=t]{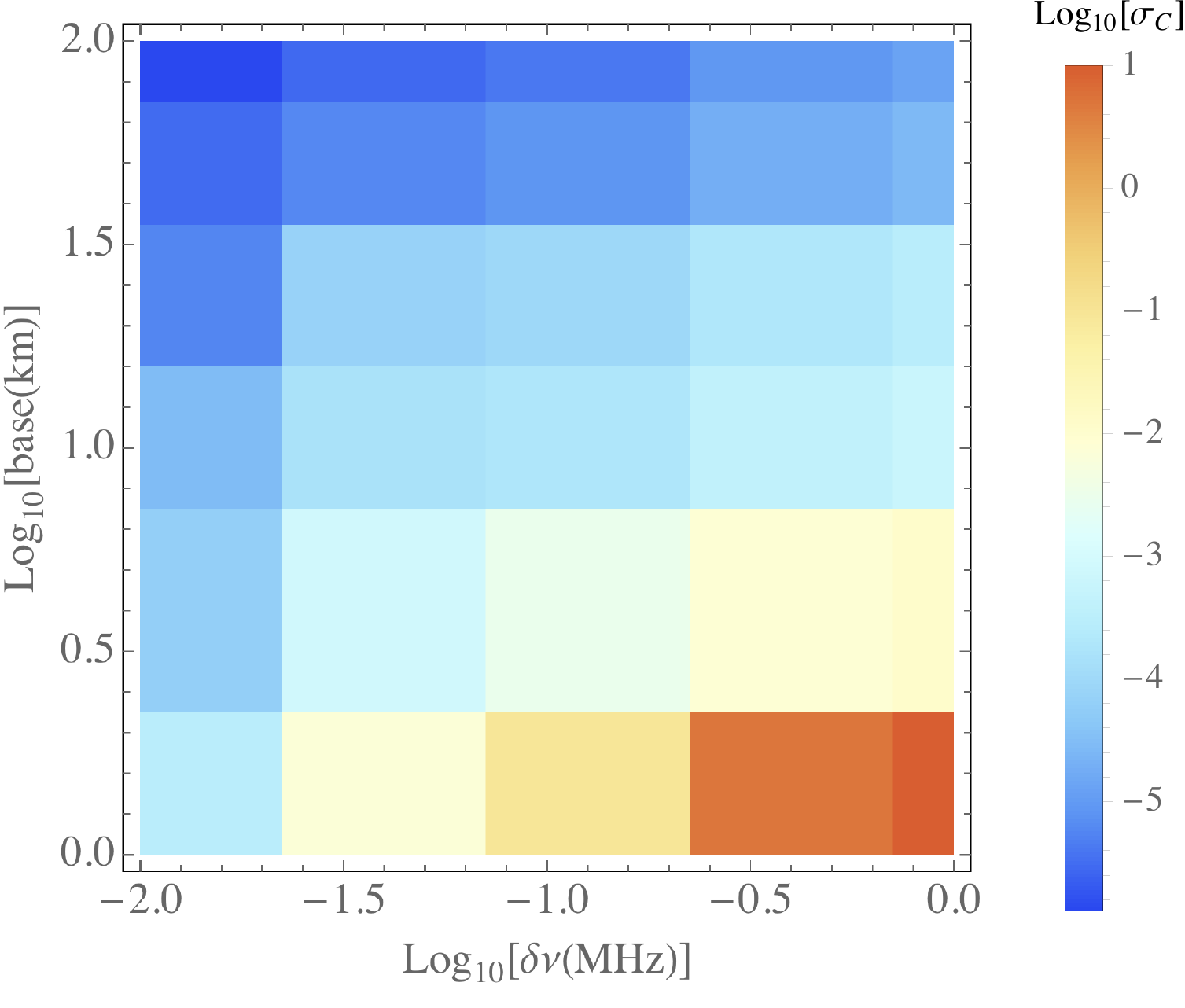}
   \includegraphics[width=3in,valign=t]{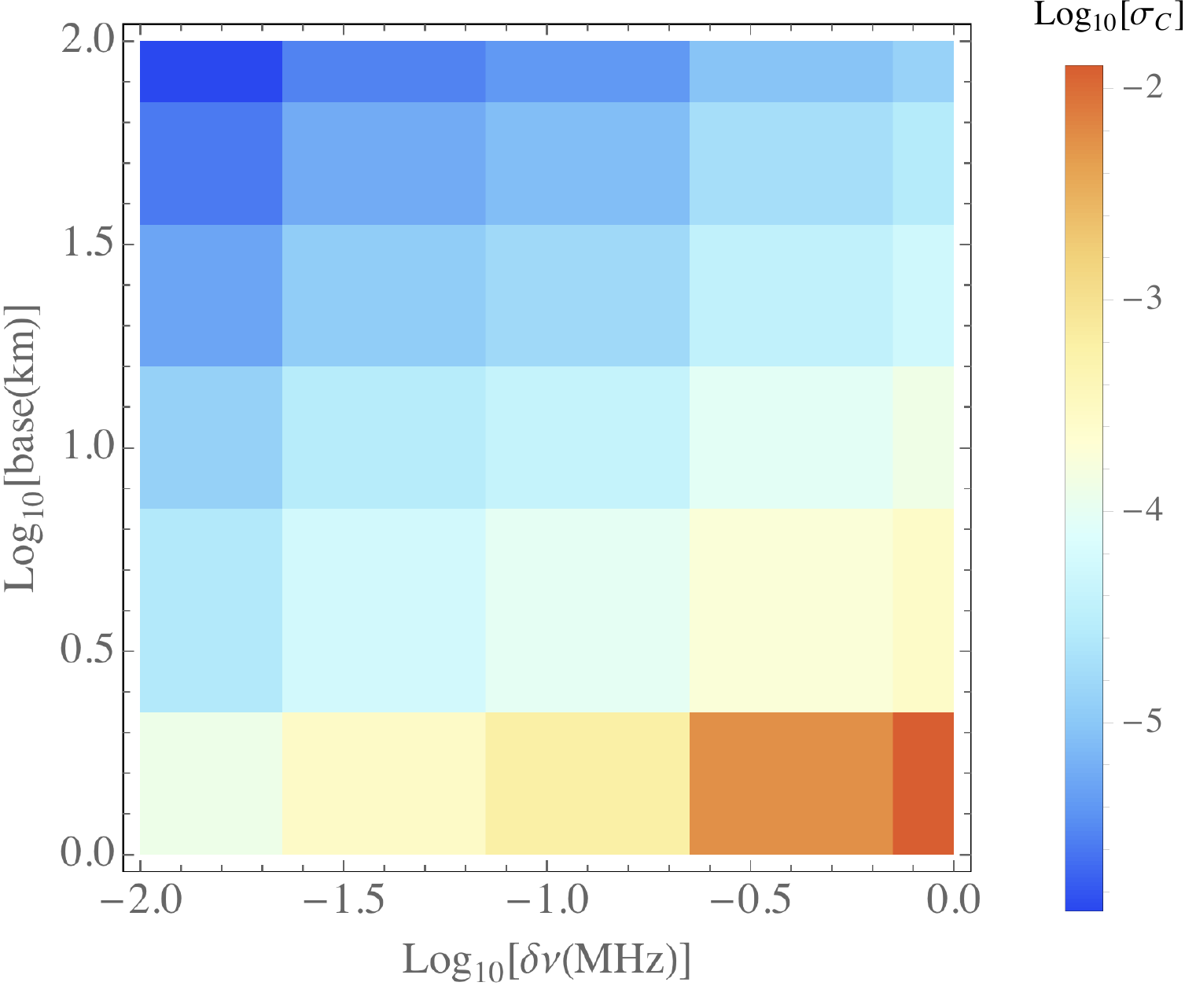}
   \caption{{\bf Sharp feature:} The marginalized absolute error in the amplitude $\sigma_C$ versus the baseline which sets the angular resolution and the frequency window of an experiment $\delta \nu$ for $k_f = 0.9$ Mpc$^{-1}$ (left) and $k_f = 0.1$ Mpc$^{-1}$ (right).  Degeneracies lead to a jump depending on the frequency of the feature in $\sigma_C$.  Because of the absence of correlations between $C$ and other parameters and because the effective amplitude is not affected by the frequency, the plot for $k_f = 0.9$ Mpc$^{-1}$ is practically the same as that for $k_f = 0.1$ Mpc$^{-1}$, with differences only for the shortest baseline and widest window function. This difference, as we will see also for $\sigma_{k_f}$ is the result of not resolving enough oscillations. For a futuristic experiment with a baseline of 100 km and a window of $\delta \nu = 0.01$ MHz, we should be able to constrain features down to $10^{-6}$ of the total primordial amplitude, which is 4-5 orders of magnitude below current constraints. }
   \label{fig:sigmac_feature}
\end{figure}

\begin{figure}[htbp] 
   \centering
   \includegraphics[width=3in,valign=t]{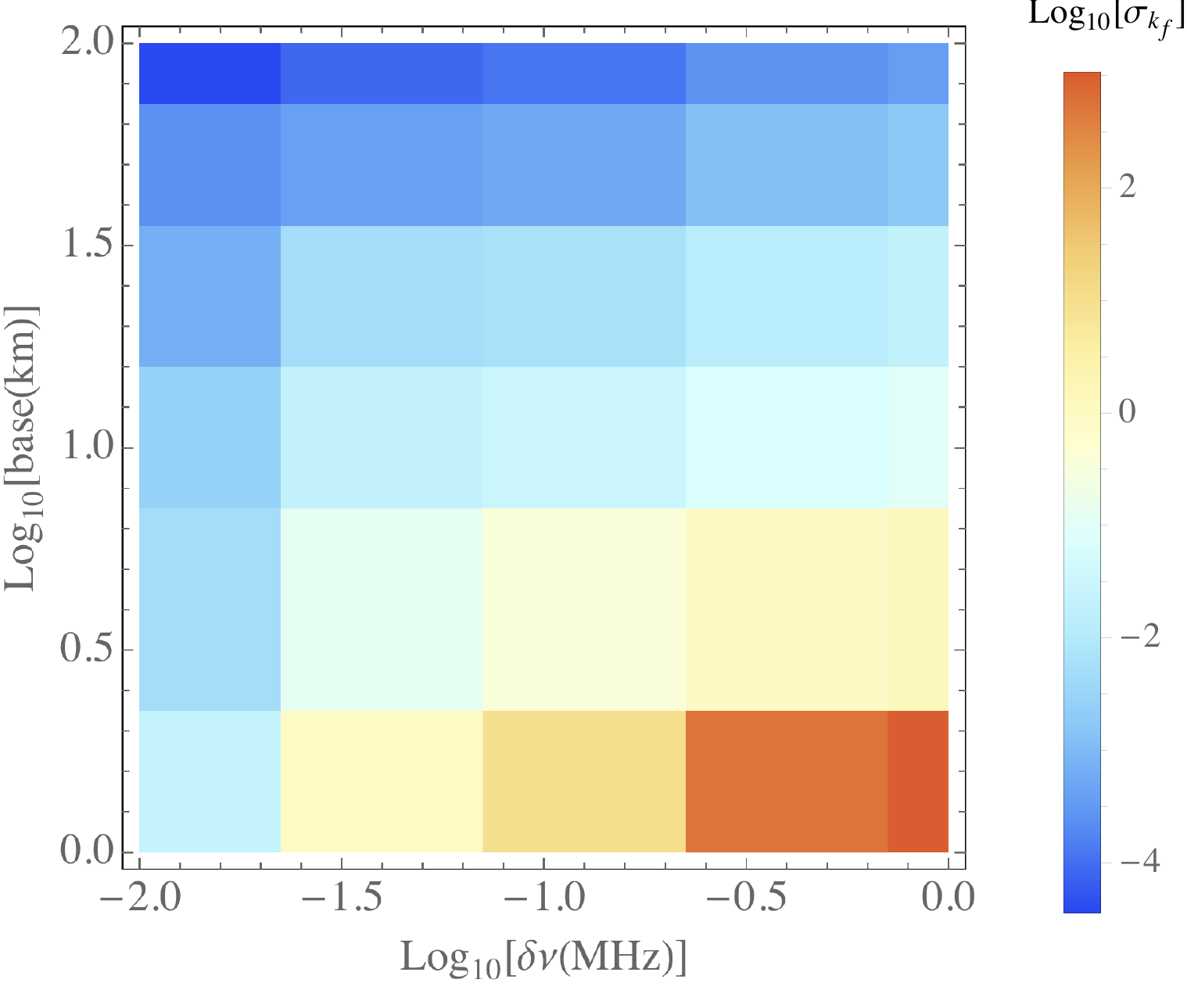}
   \includegraphics[width=3in,valign=t]{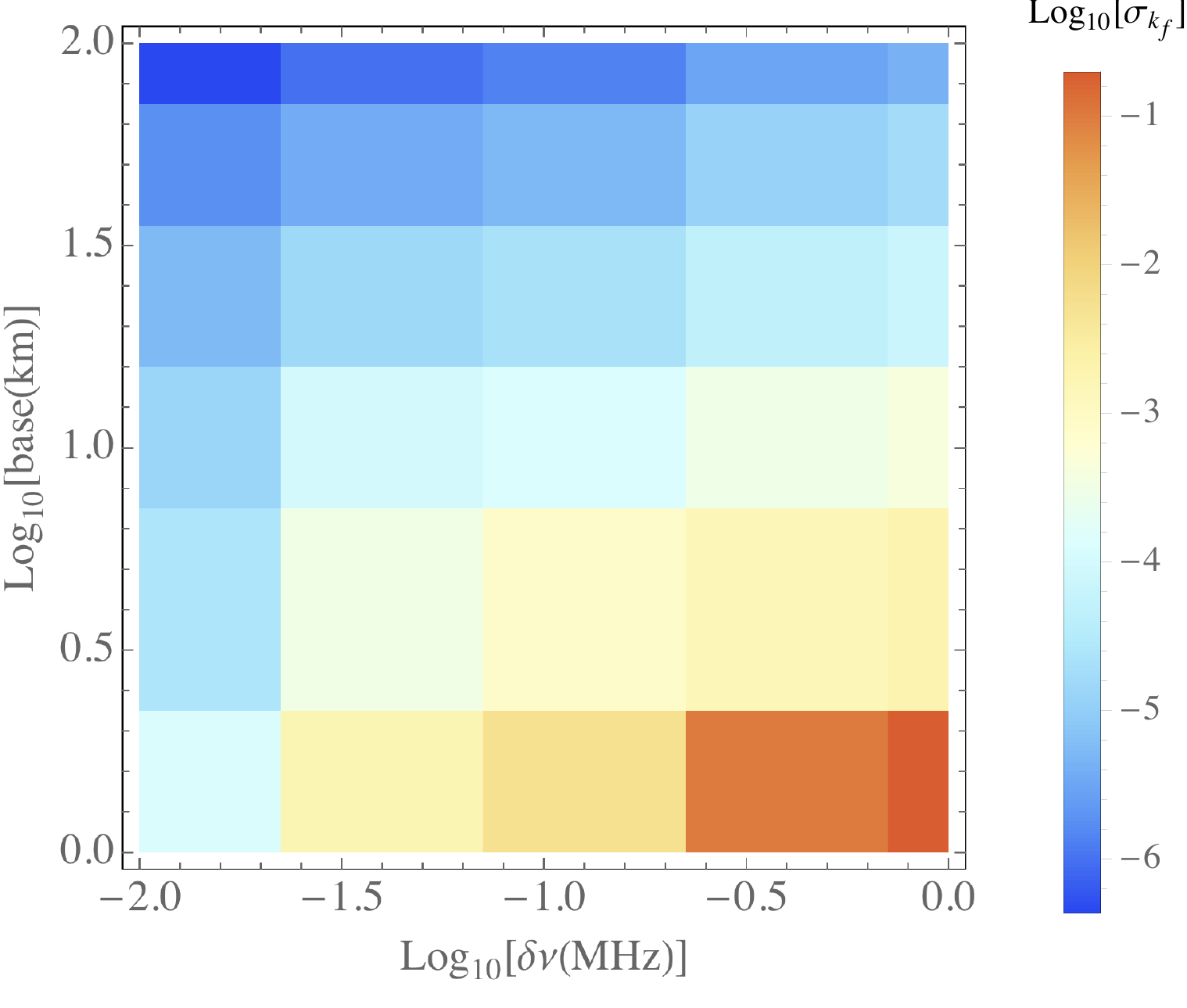}
   \caption{{\bf Sharp feature:} The marginalized absolute error in the frequency $\sigma_{k_f}$ versus the baseline which sets the angular resolution and the frequency window of an experiment $\delta \nu$ for $k_f = 0.9$ Mpc$^{-1}$ (left) and $k_f = 0.1$ Mpc$^{-1}$ (right).  Degeneracies lead to a jump depending on the frequency of the feature in $\sigma_{k_f}$. The higher the frequency (the smaller $k_f$) the more oscillations you can resolve when increasing the baseline or decreasing the width of the window function. By resolving more oscillations, the constraints improve. By computing the ratio of the $\sigma_{\rm k_f}$ for each frequency, we find the improvement is close to constant for $\sigma_{k_f = 0.1}/\sigma_{k_f = 0.03} \simeq 11$,  except for very short baselines and wide windows. Because of degeneracies, this is not the case for the lowest frequency, i.e. $k_f = 0.9$ Mpc$^{-1}$. }
   \label{fig:sigmakf_feature}
\end{figure}

The key quantity of interest is the sensitivity $\sigma_C$ on the amplitude parameter. The Fisher analysis presented here assumes a primordial amplitude of $C=0.01$ at which we compute the derivatives. A reason to consider this amplitude is that the current CMB data puts constraints on the amplitude of certain oscillatory features to be lower than a few percent. We do not expect a strong dependence on this assumption as longs as $C\ll1$. For the linear feature we consider 3  different frequencies, $k_f = 0.9$ Mpc$^{-1}$, $k_f = 0.1$ Mpc$^{-1}$ and $k_f = 0.03$ Mpc$^{-1}$. The phase does not significantly change the results and we choose $\phi = 0$. The cosmological parameters are set to Planck 2015 best-fit values, except $H_0 = 100 h$ which is varied. We show the marginalized error of the amplitude of the feature, $\sigma_C \equiv (F^{-1})_{CC}$ as a function of the baseline in km and $\delta \nu$ in MHz in Fig.~\ref{fig:sigmac_feature}. As expected, a longer baseline and a narrow frequency width improve the constraints. For a baseline of 100 km and a frequency resolution of $\delta \nu = 0.01$ MHz we find $\sigma_C \simeq 10^{-6}$, which is almost independent of the frequency as expected. In Fig.~\ref{fig:sigmakf_feature} we show the same grid for the error on $k_f$. The finer the resolution in both directions, the more oscillations can be resolved and thus the more accurate the frequency can be determined. A take-away point is that one would prefer $\delta \nu < 1$ MHz to obtain good resolution for features.


\begin{figure}[h] 
   \centering
   \includegraphics[width=3in,valign=t]{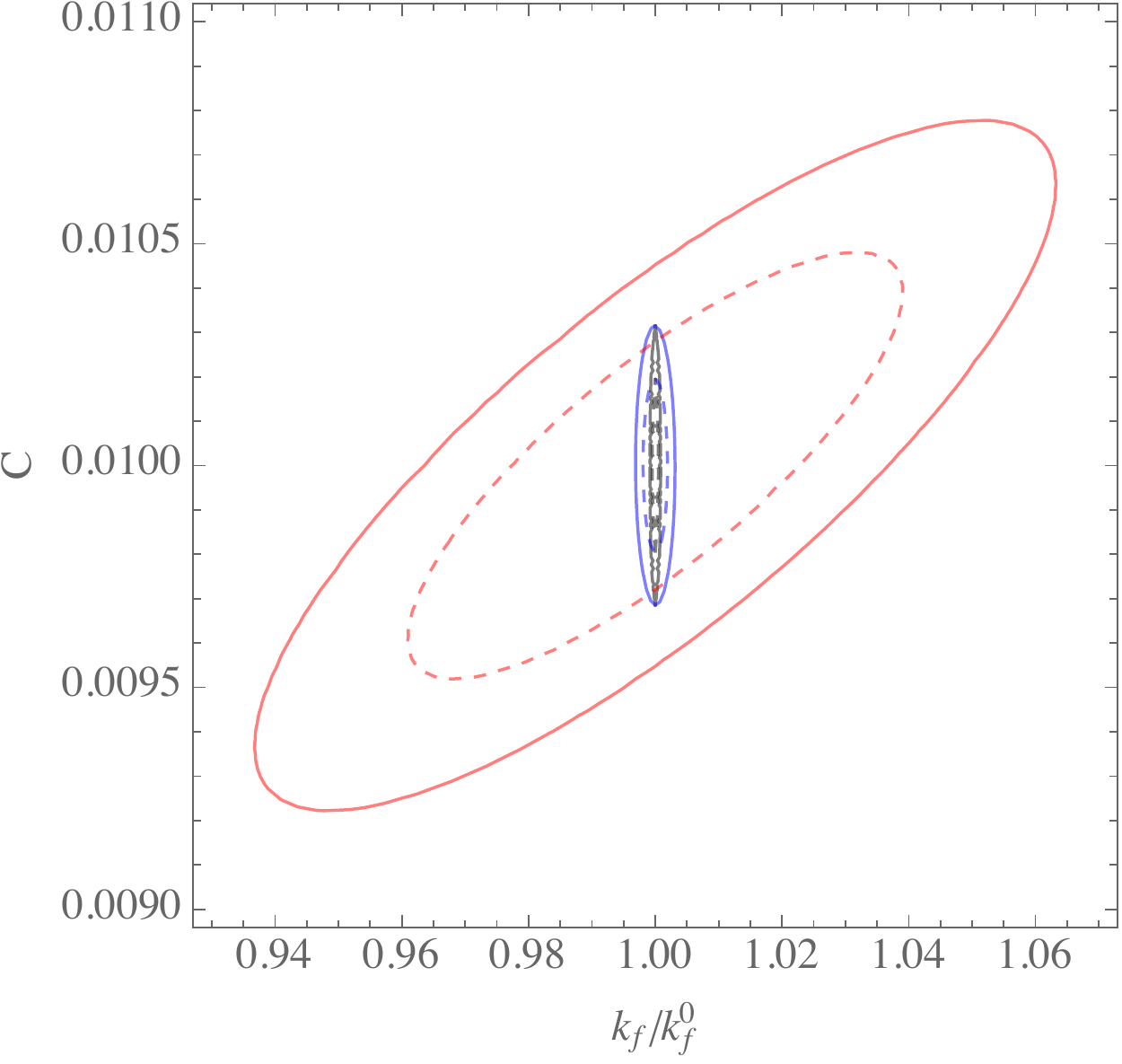}
   \includegraphics[width=3in,valign=t]{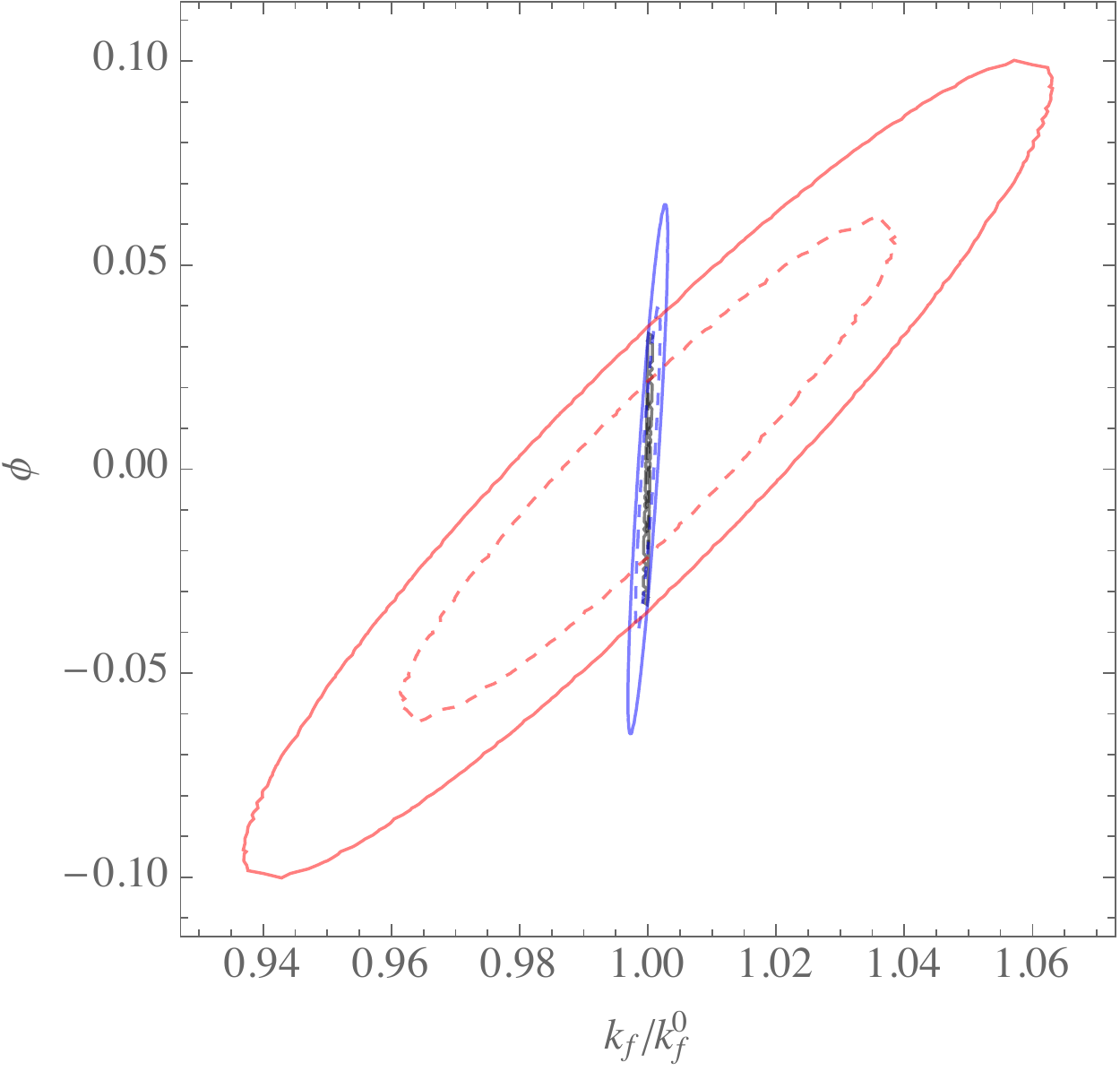}
   \includegraphics[width=3in,valign=t]{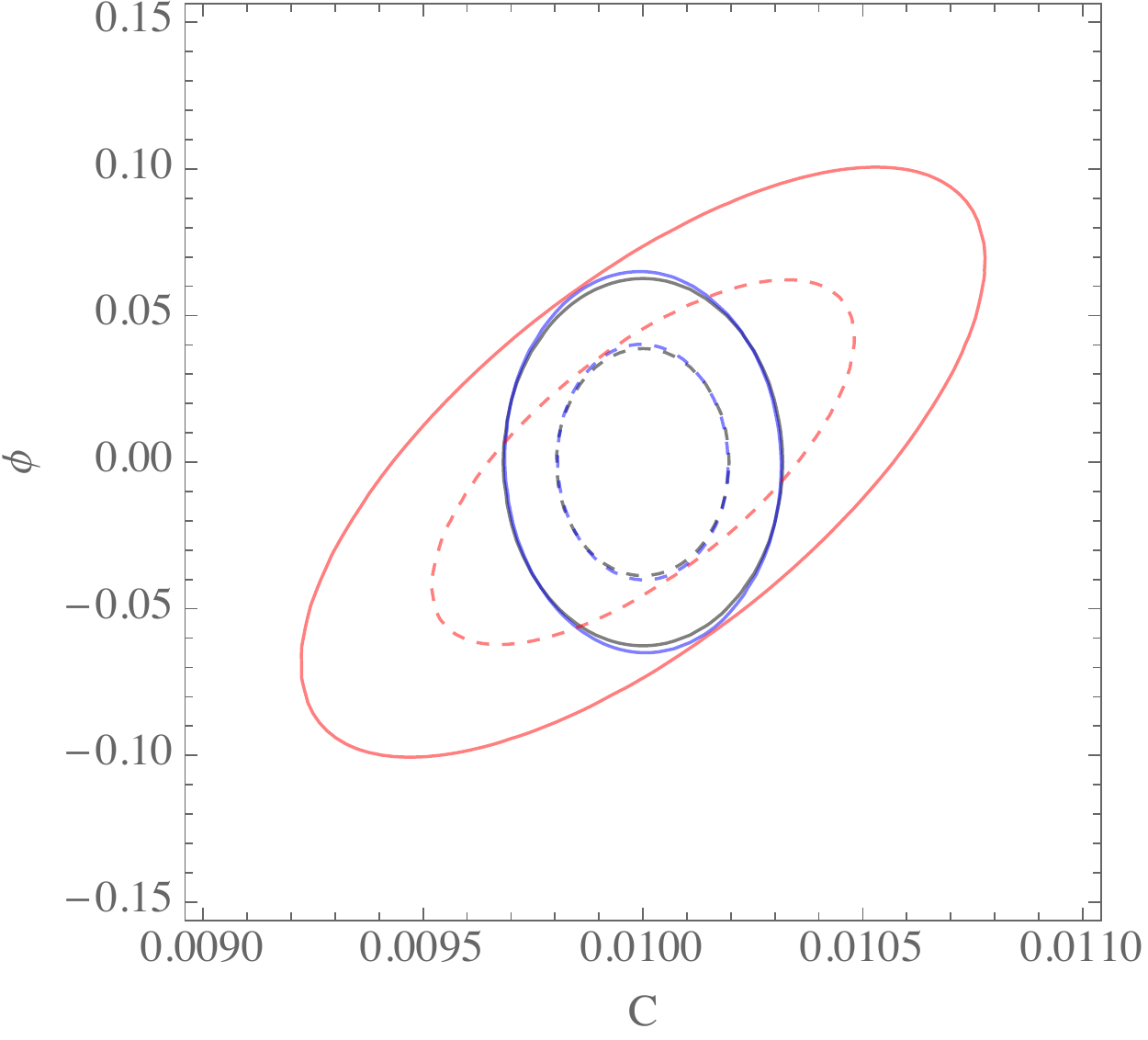}
   \includegraphics[width=3in,valign=t]{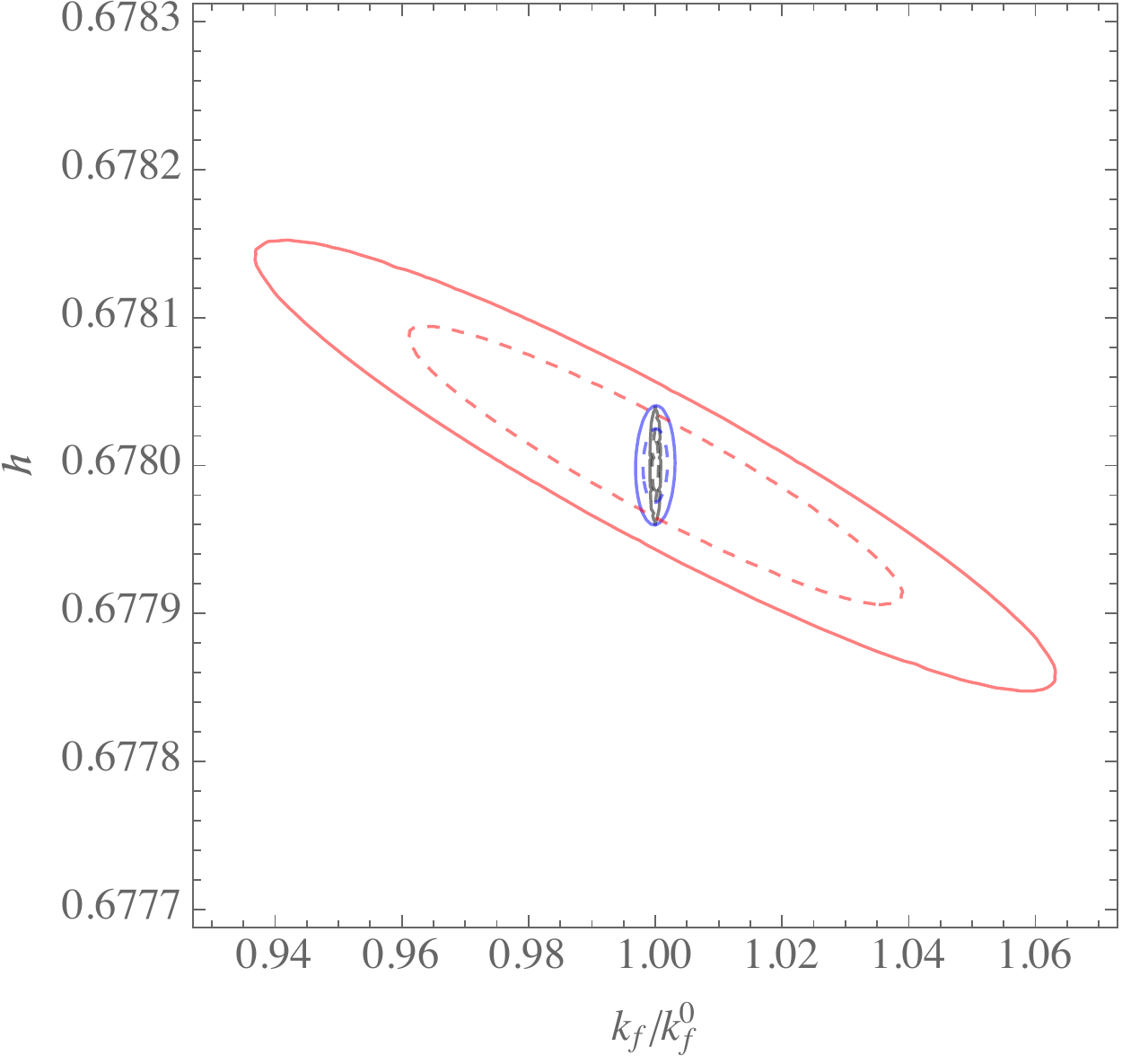}
   \caption{ {\bf Sharp feature:} Figure showing the one and two sigma contours for the amplitude $C$ vs $k_f$(top left),  $\phi$ vs $k_f$ (top right), $\phi$ vs $C$ (bottom left) and $k_f$ vs $h_0$ (bottom right) for the sharp feature model.  The various contours are in red ($k_f = 0.9$  Mpc$^{-1}$), blue ($k_f = 0.1$  Mpc$^{-1}$) and in black ($k_f = 0.03$  Mpc$^{-1}$). Degeneracies between primordial parameters as well as degeneracies between $h_0$ and primordial parameters lead to inflated error bars for the lowest frequencies ($k_f = 0.9$  Mpc$^{-1}$). We set $\delta \nu = 0.01$ MHz and a baseline of 1 km. }
   \label{fig:Contours}
\end{figure}


To visualize the effect of degeneracies we show a set of marginalized contours in Fig.~\ref{fig:Contours}. For the lowest frequency, we find degeneracies between the primordial parameter and the Hubble rate. This results in a inflated error bar on all feature parameters. The frequency and the phase always show a small correlation, which is expected given that the derivatives of an oscillating function with respect to the phase and frequency are proportional.  The higher the frequency, the more this correlation will fade, since information that distinguishes the two derivatives will increase.

The degeneracy of the Hubble rate and the primordial parameters disappear for higher frequencies, as the BAO should decouple from high frequency oscillations. Increasing the resolution will also break degeneracies, which can be concluded from Fig.~\ref{fig:sigmac_feature} (left).



\subsection{Resonance feature signal (log feature)}

\begin{figure}[htbp] 
   \centering
  \includegraphics[width=3in,valign=t]{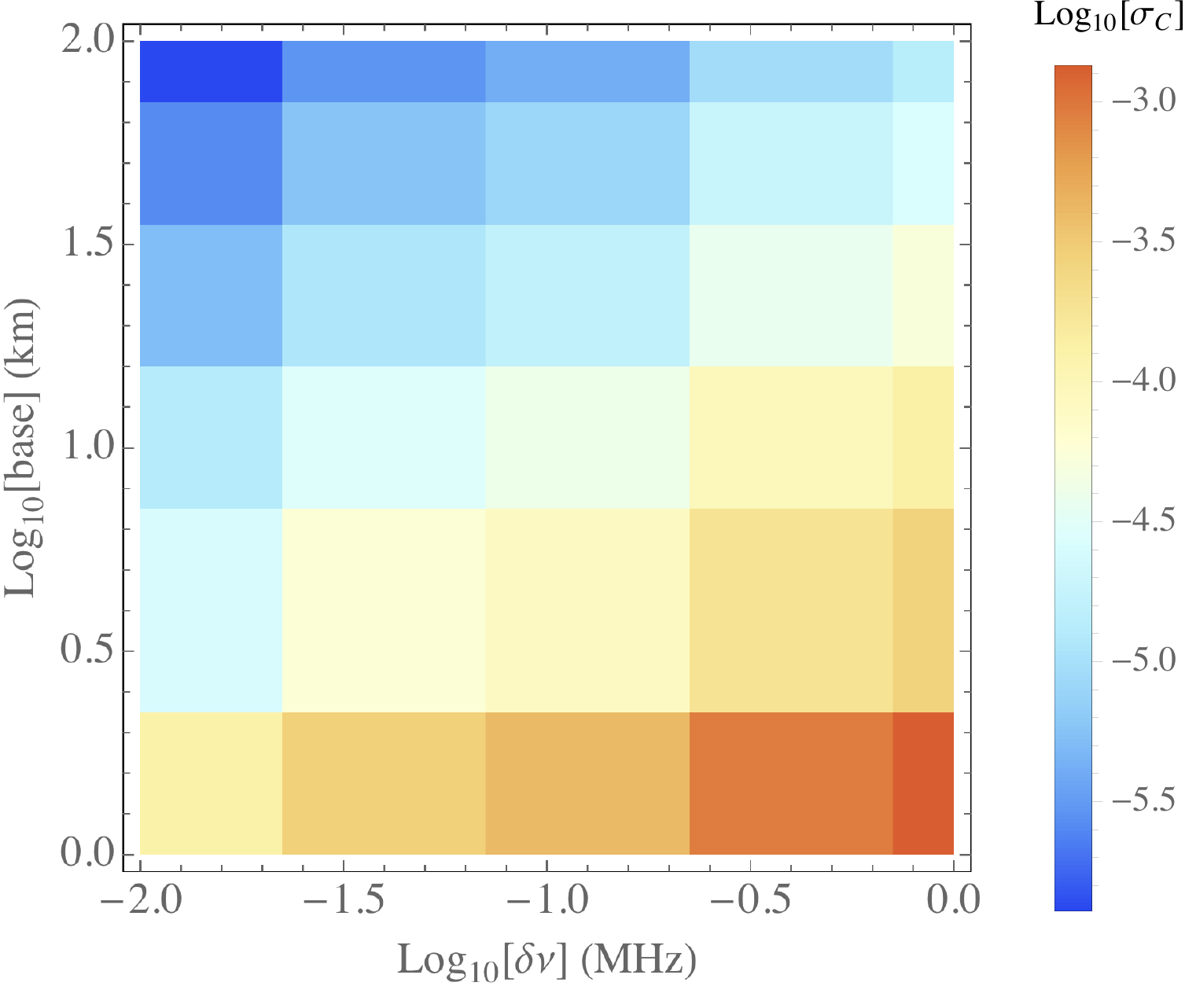}
   \includegraphics[width=3in,valign=t]{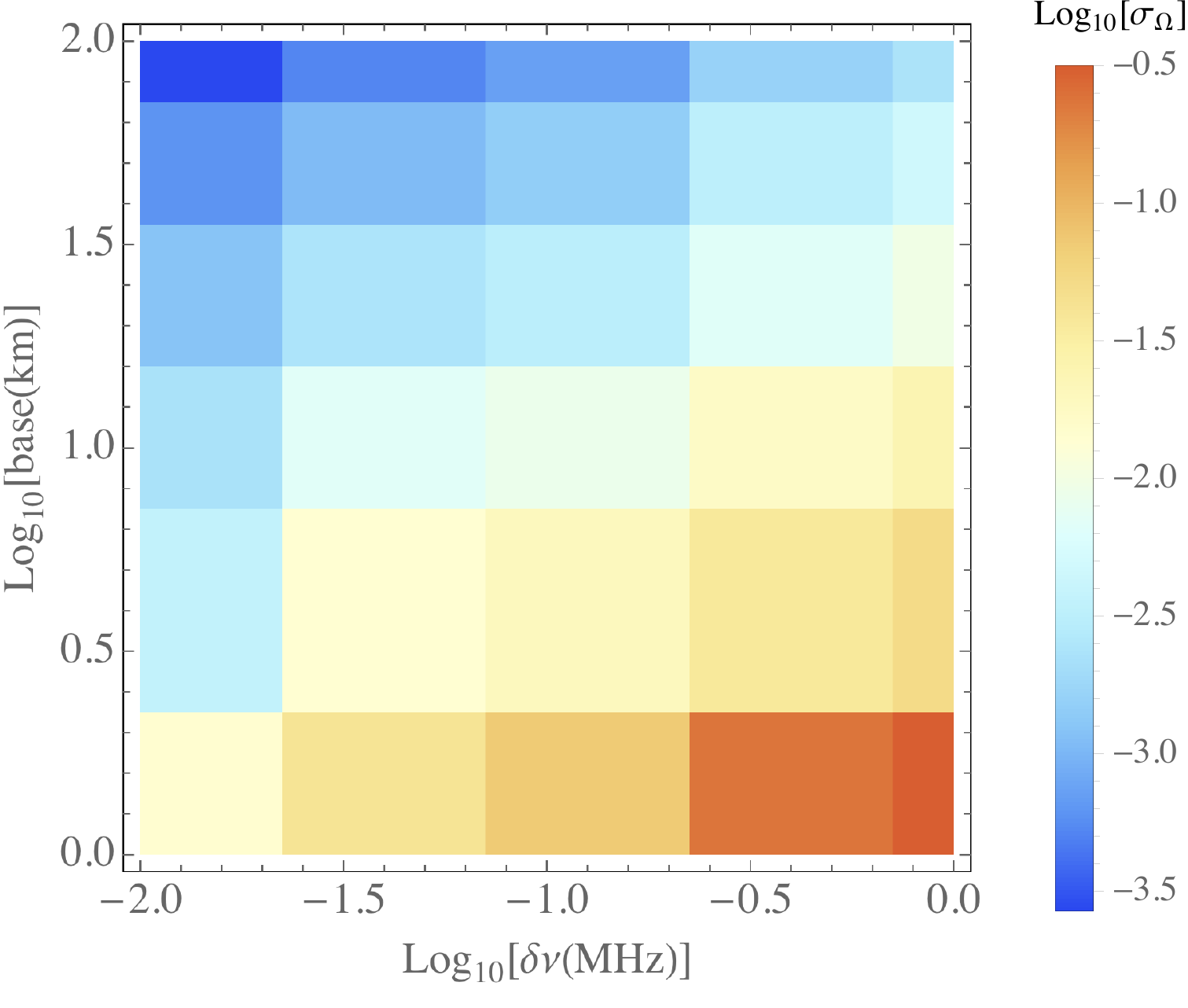}
   \caption{{\bf Resonance feature:} The marginalized absolute error in the amplitude $\sigma_C$ (left) and frequency $\sigma_{\Omega}$ versus the baseline which sets the angular resolution and the frequency window of an experiment for $\Omega_{\rm eff} = 100$. Because of the absence of degeneracies between $C$ and other parameters there is no dependence of frequency.  Similarly, due to both the absence of strong degeneracies and because the logarithmic nature hardly improves the number of resolves oscillation comparatively between different frequencies, the error on $\Omega$ is almost independent of frequency. }
   \label{fig:sigmac_resonant}
\end{figure}

\begin{figure}[htbp] 
   \centering
   \includegraphics[width=3in,valign=t]{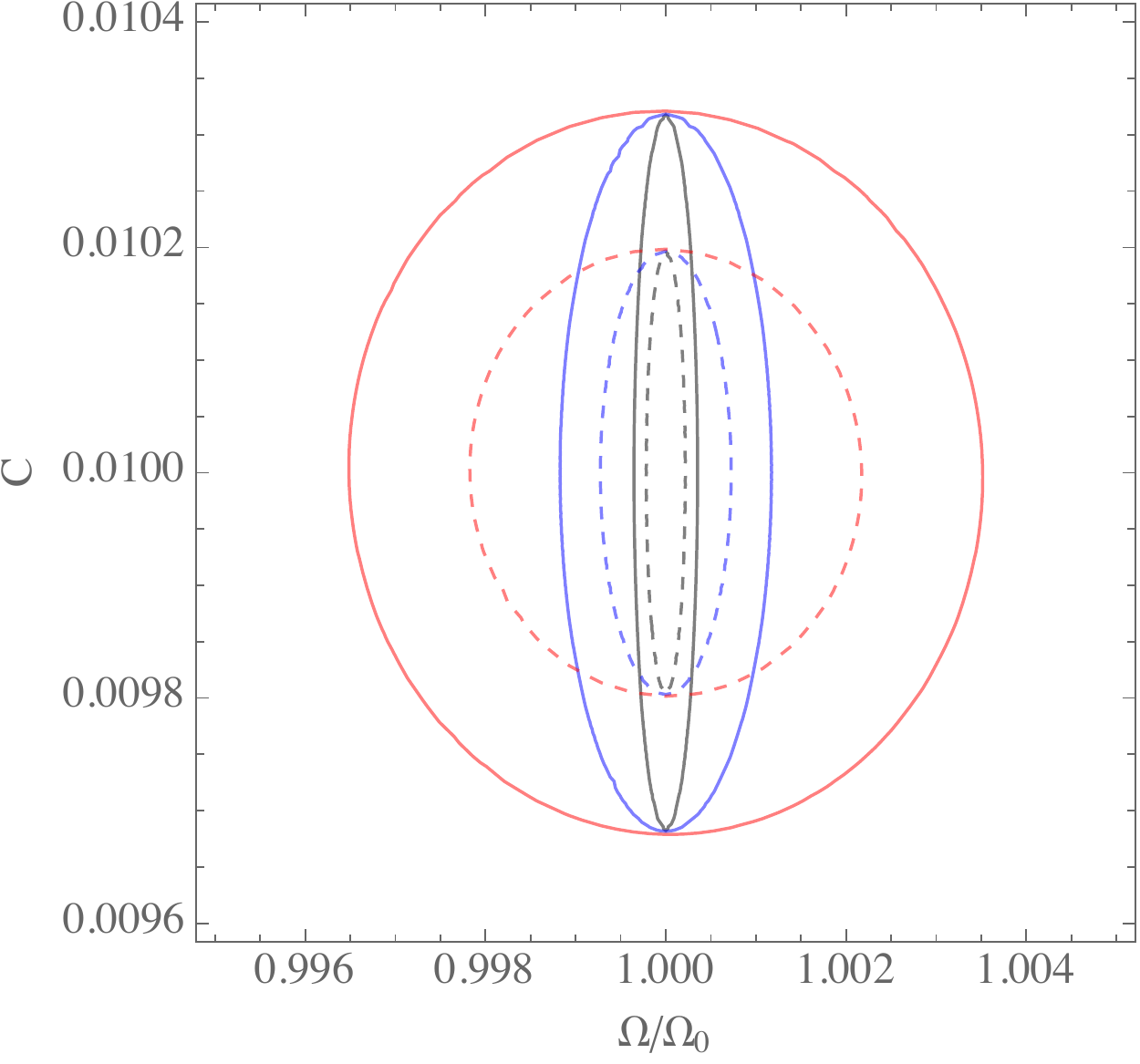}
   \includegraphics[width=3in,valign=t]{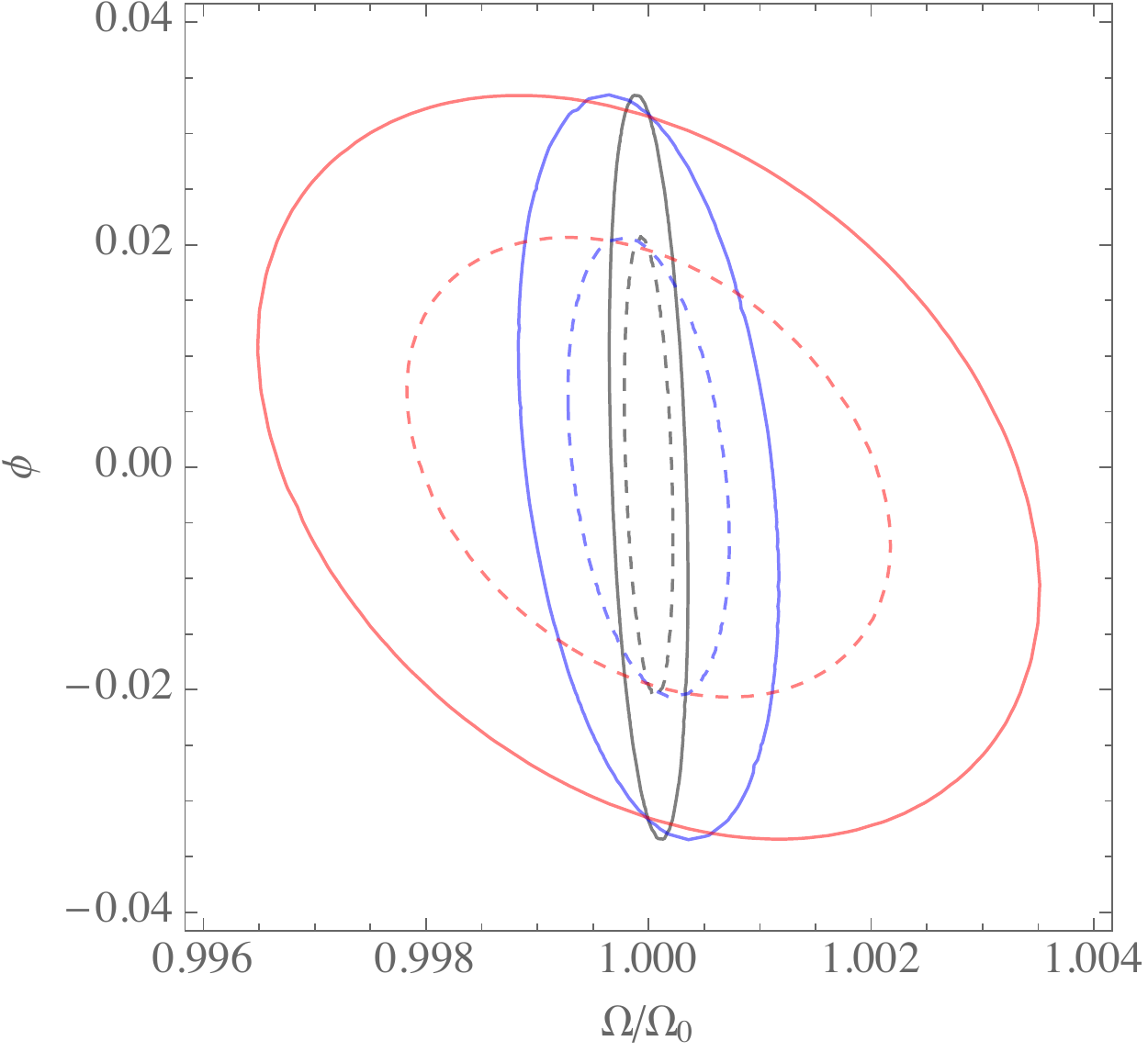}
   \caption{ {\bf Resonance feature:} One and two sigma contours for the amplitude $C$ and the frequency $\Omega$ (left) and  for $\phi$ and $\Omega$ (right). Colors represent $\Omega = 10$ (red), $\Omega = 30$ (blue) and $\Omega = 100$ (black). We find no correlation between the amplitude and other parameters, which is clear from the absence of frequency dependence of the constraint on $C$. As the frequency decreases, $\Omega$ and $\phi$ become more correlated. }
   \label{fig:ContoursLog}
\end{figure}

Unlike the sharp feature, resonance features, which have a logarithmically spaced periodicity, hardly correlate with late time cosmological parameters for low frequencies, because the BAO are linear. In the left panel of  Fig.~\ref{fig:sigmac_resonant} we show $\sigma_C$ as a function of the baseline and the experimental window function for $\Omega = 100$. In the absence of strong degeneracies, the dependence $\sigma_C$ on the baseline and the window function is independent of frequency; we find that we should be able to probe amplitudes as low as $10^{-7}$ for log-type features, as long as they do not decay.  In the right panel we show the constraints on $\Omega$; although there is a small effect from degeneracies between $\Omega$ and $\phi$, the resulting bounds are practically independent of frequency. 


Correlations are explored in the contours of Fig.~\ref{fig:ContoursLog}. We find some correlation between the phase and the frequency. We do not find correlations between the primordial parameters and $H_0$. Such a correlation is found in the CMB, and can be explained by the effect of a shift in phase due to projection from the last scattering surface (see Ref.~\cite{Meerburg2014b}). We expect that this correlation is recovered in a full sky analysis when large scale projection effects are taken into account. This would also change the dependence of $\sigma_{\Omega}$ on $\Omega$, which now is almost independent.

\subsection{Primordial standard clock signals}

In this subsection, we consider the standard clock signals. As mentioned, although the ideal strategy would be to directly apply full standard clock signals to the analyses, due to technical difficulties, the full clock signal is not available analytically in general.
So a practical approach is to single out the clock signal part of the full clock signal and analyze it first. We first study two cases of the clock signals: the inflationary clock signal (Sec.~\ref{Sec:clock_inflation}) and the Ekpyrotic clock signal (Sec.~\ref{Sec:Clock_contraction}). In Sec.~\ref{Sec:Clock_full}, we consider a special example of the full clock signal within the inflationary paradigm.

For the clock signal we need to consider two additional parameters $k_r$ [i.e.~the location of the onset (expanding scenario) or end (contracting scenario) feature] and $p$, increasing the total number of parameters to 6. The addition of extra parameters will weaken the constraints on parameter space. Another important difference between the standard clock signal features and the features we considered previously is that the clock signals are no longer extended to all scales.\footnote{In fact a realistic sharp feature signal does not extend to all scales either, what we are considering is the infinite-sharpness limit of the sharp feature case, for related comments see Sec.~\ref{Sec:Sharp}.}
The clock signals can be considered semi-extended, like that of the inflation or matter contraction scenario. For example, in the inflation scenario, the signal decays towards smaller scales.
The clock signals can also be very localized, like that of the slow contraction (e.g.~Ekpyrotic) or slow expansion scenario. In these cases, if the full feature is resolved within a range of scales, the forecasted constraints cannot be further improved by increasing resolution (both in the radial and angular direction).


\subsubsection{Inflation scenario}
\label{Sec:clock_inflation}
\begin{figure}[htbp] 
   \centering
  \includegraphics[width=1.95in,valign=t]{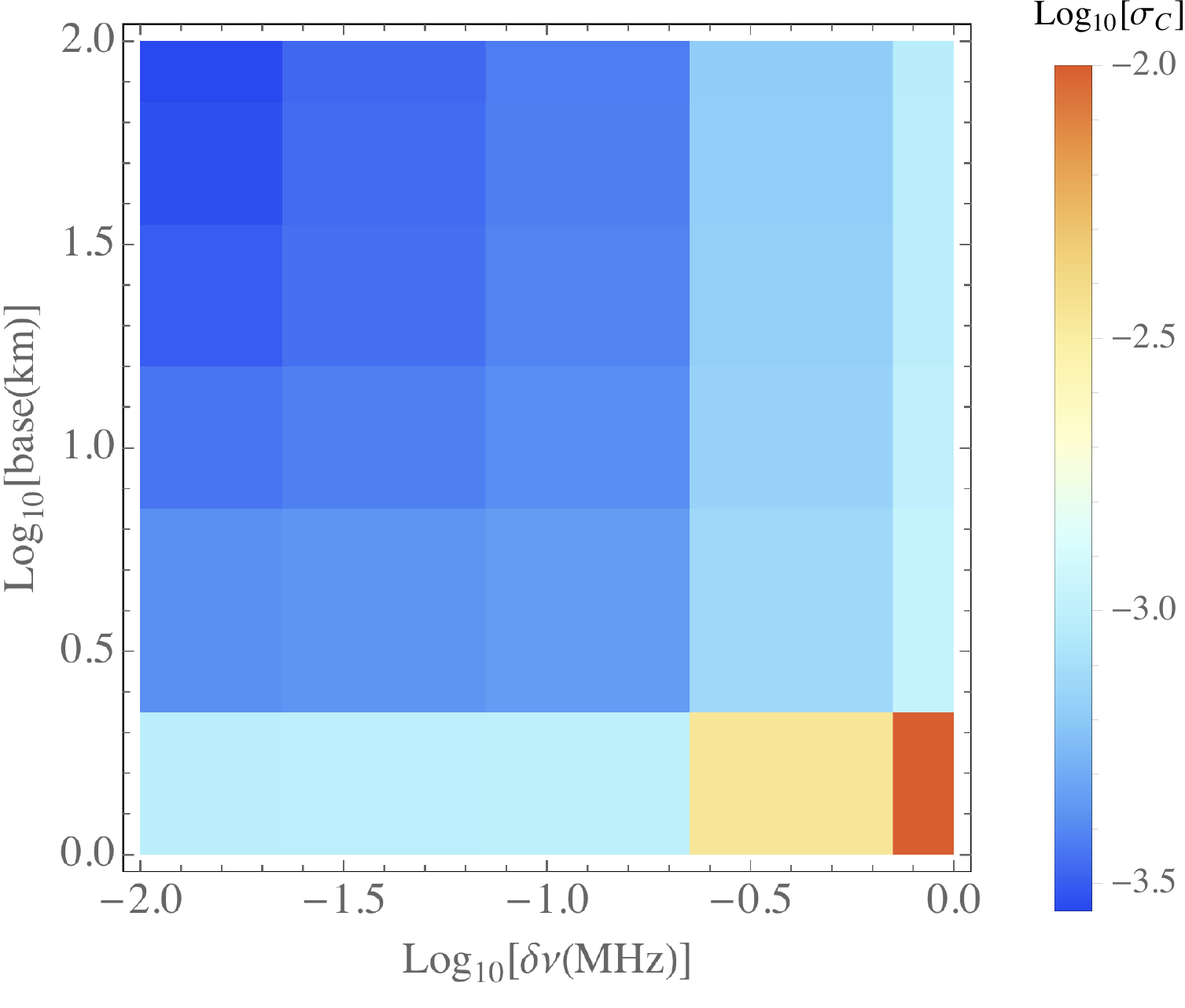}
  \includegraphics[width=1.95in,valign=t]{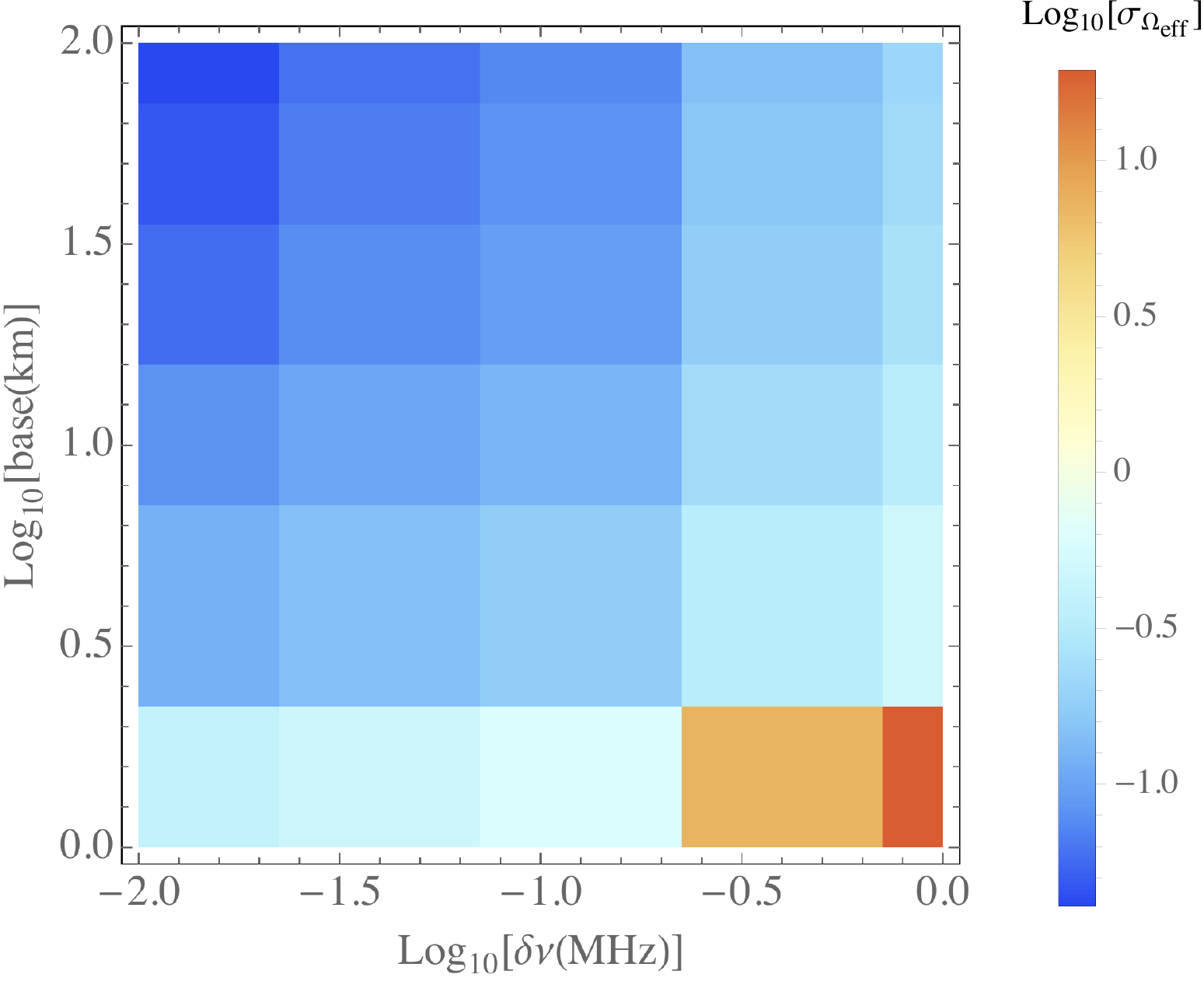}
    \includegraphics[width=1.95in,valign=t]{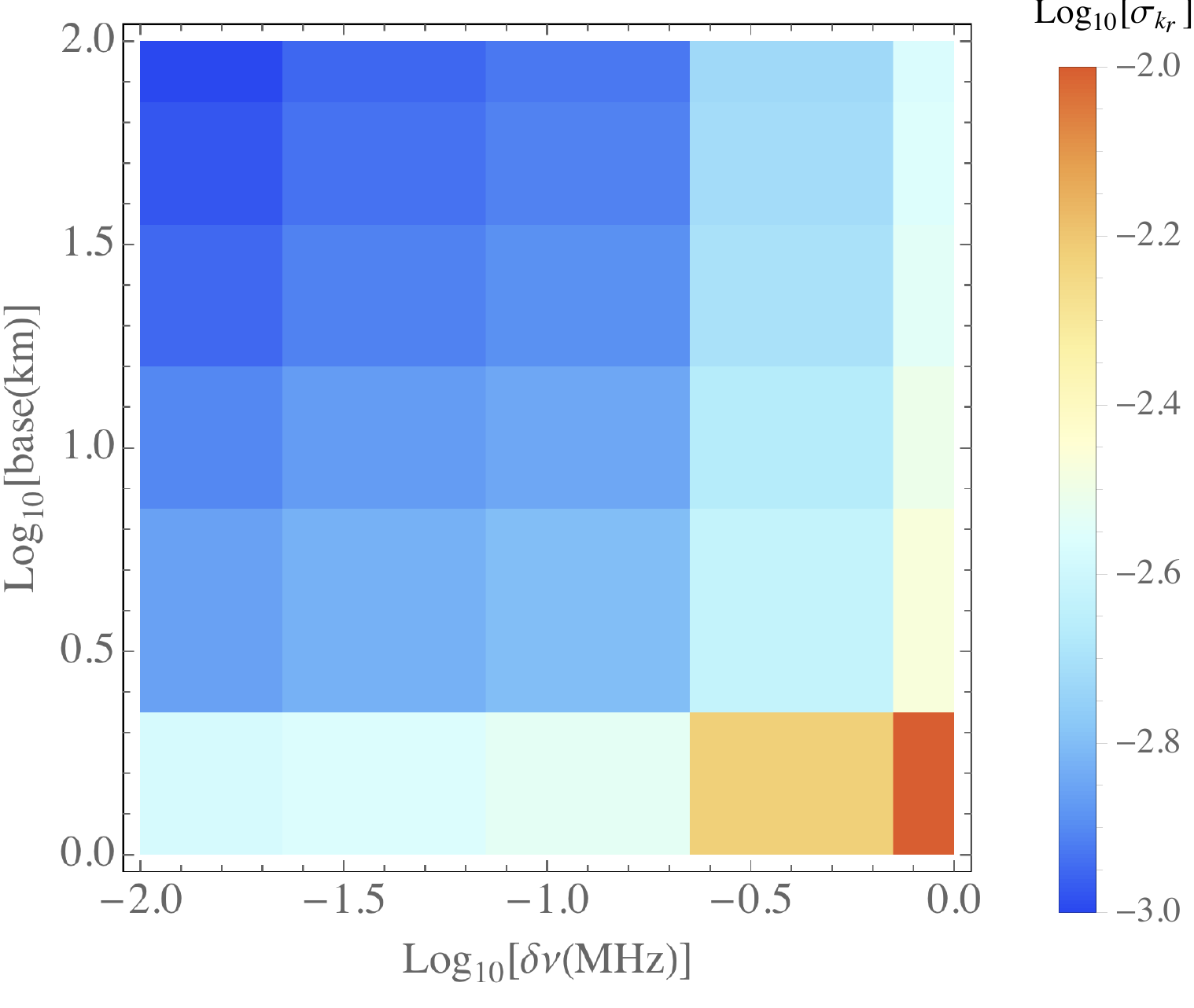}
  \includegraphics[width=1.95in,valign=t]{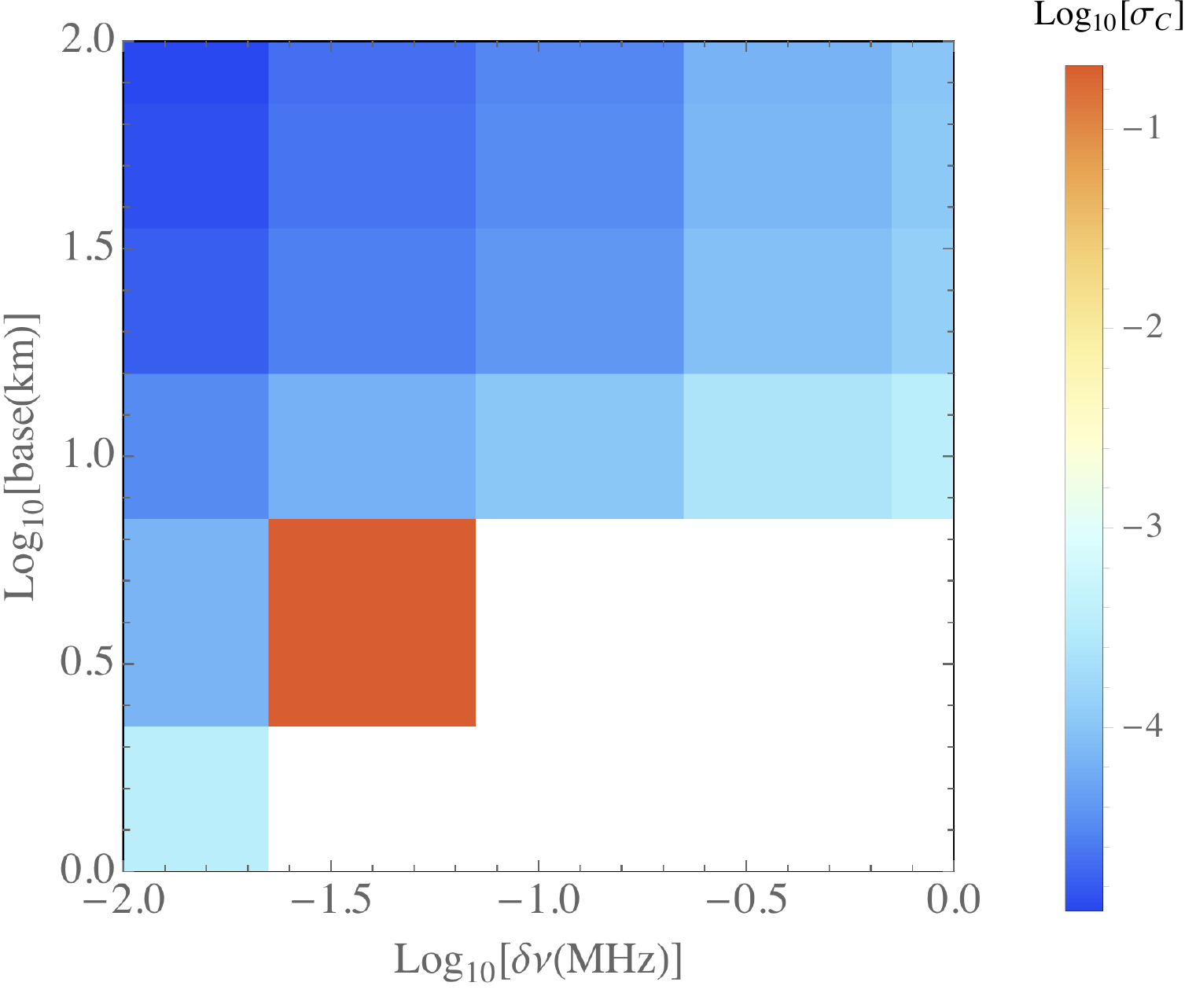}
    \includegraphics[width=1.95in,valign=t]{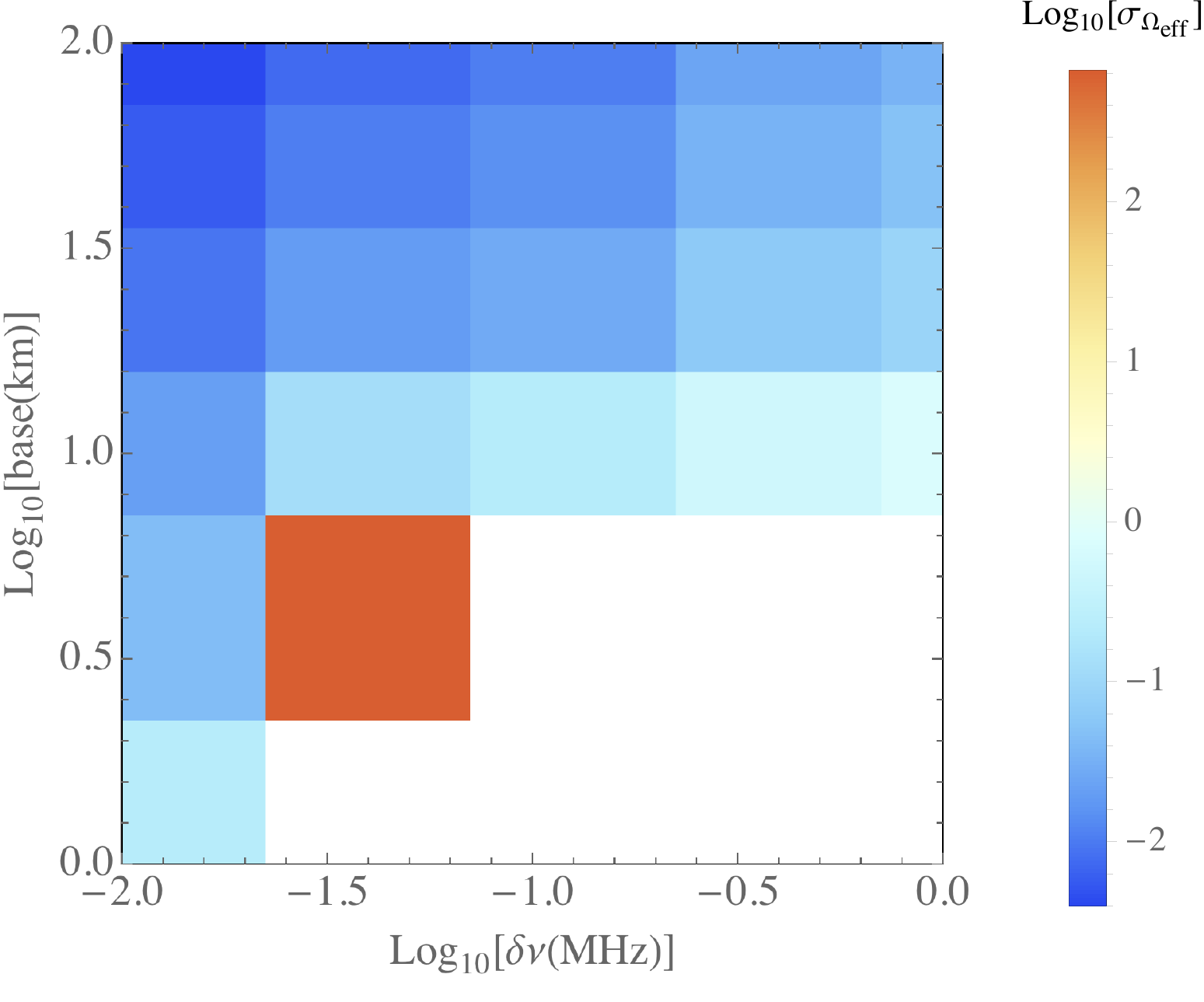}
  \includegraphics[width=1.95in,valign=t]{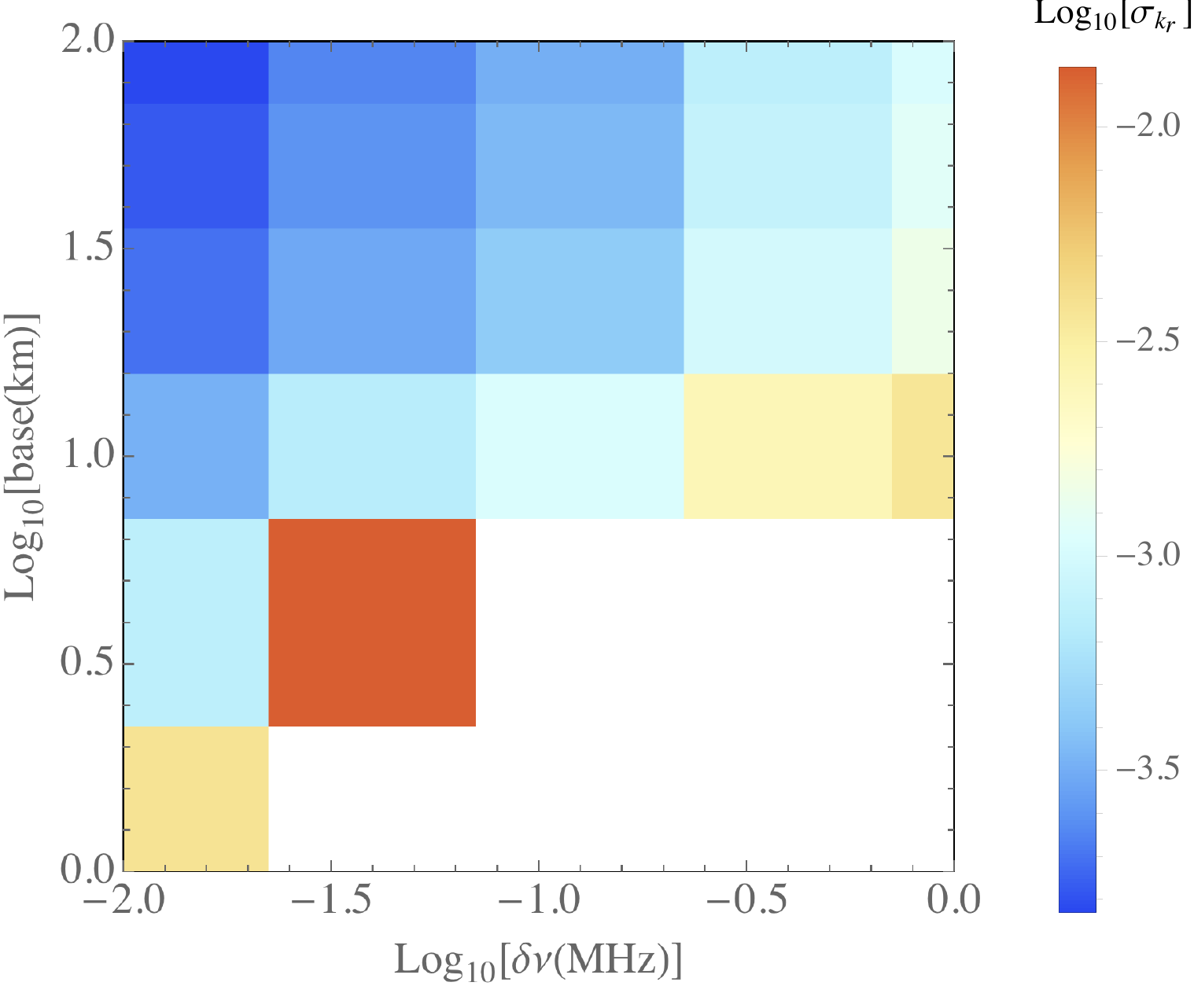}
   \caption{{\bf Clock signal (inflation):} The marginalized absolute errors $\sigma_C$, $\sigma_{\Omega_{\rm eff}}$ and $\sigma_{k_r}$ versus the angular and radial resolution of an experiment for $k_r = 0.1$ Mpc$^{-1}$ at $\Omega_{\rm eff} = 10$ (top) and $k_r = 1$ Mpc$^{-1}$ at $\Omega_{\rm eff} = 30$ (bottom) . The semi-extended nature of the feature is apparent through the local improvement of the sensitivity, which, as expected, shifts towards smaller scales as we change $k_r$ to larger values. The white area in the bottom panels are were there is no constraint (since the feature is not resolved at all). }
   \label{fig:sigmac_clock_expanding}
\end{figure}

\begin{figure}[htbp] 
   \centering
    \includegraphics[width=1.95in,valign=t]{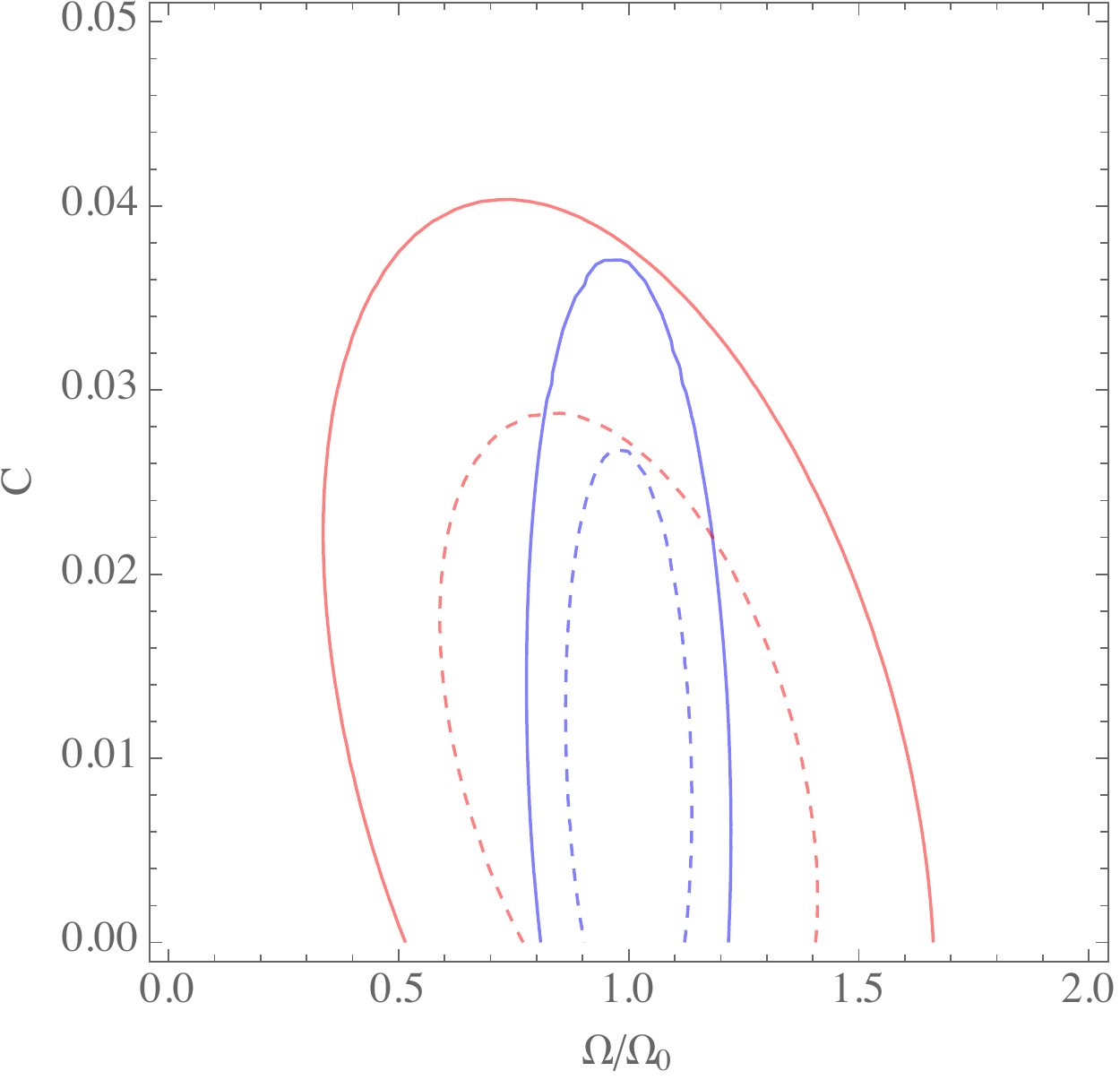}
    \includegraphics[width=1.95in,valign=t]{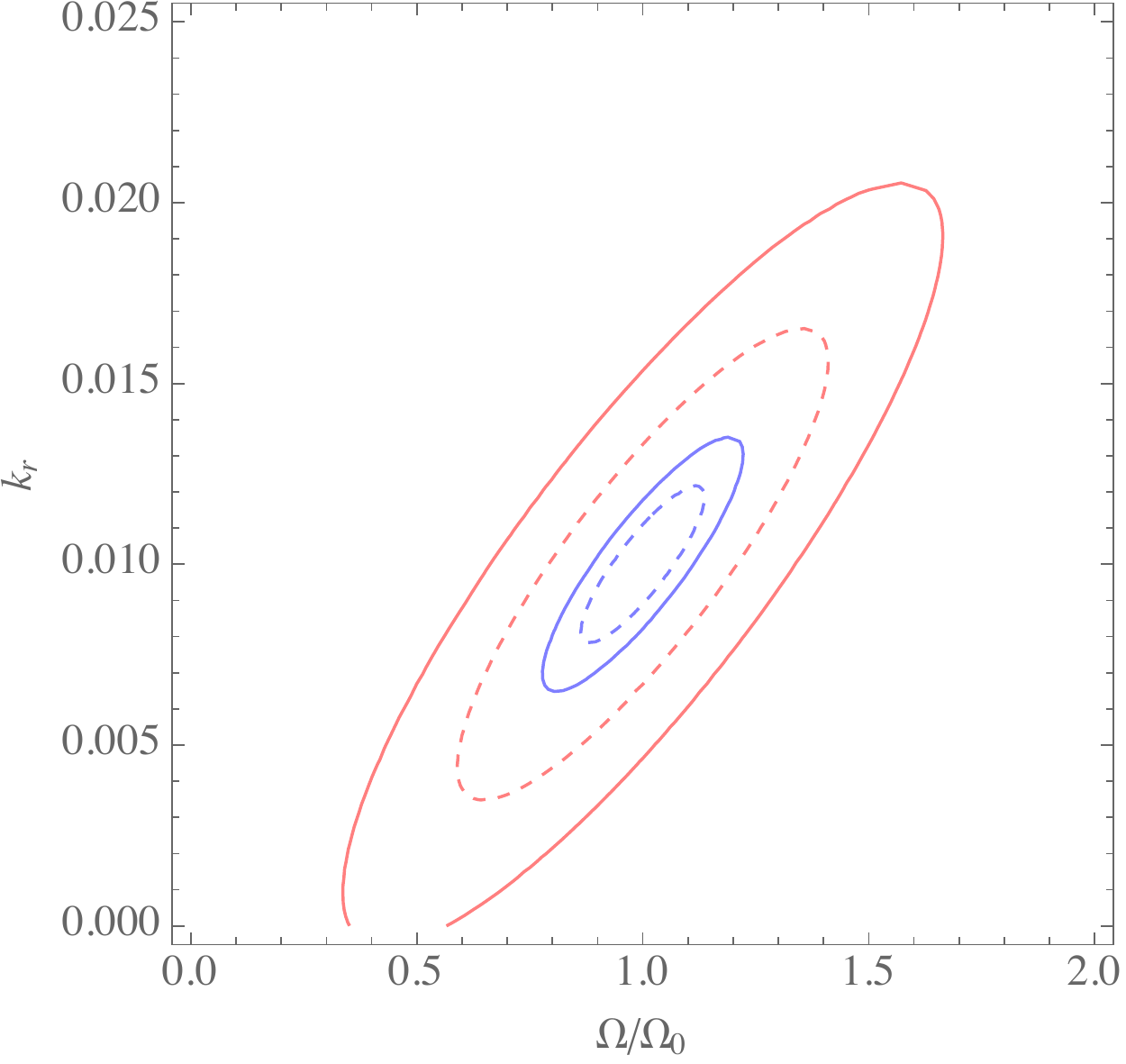}
    \includegraphics[width=1.95in,valign=t]{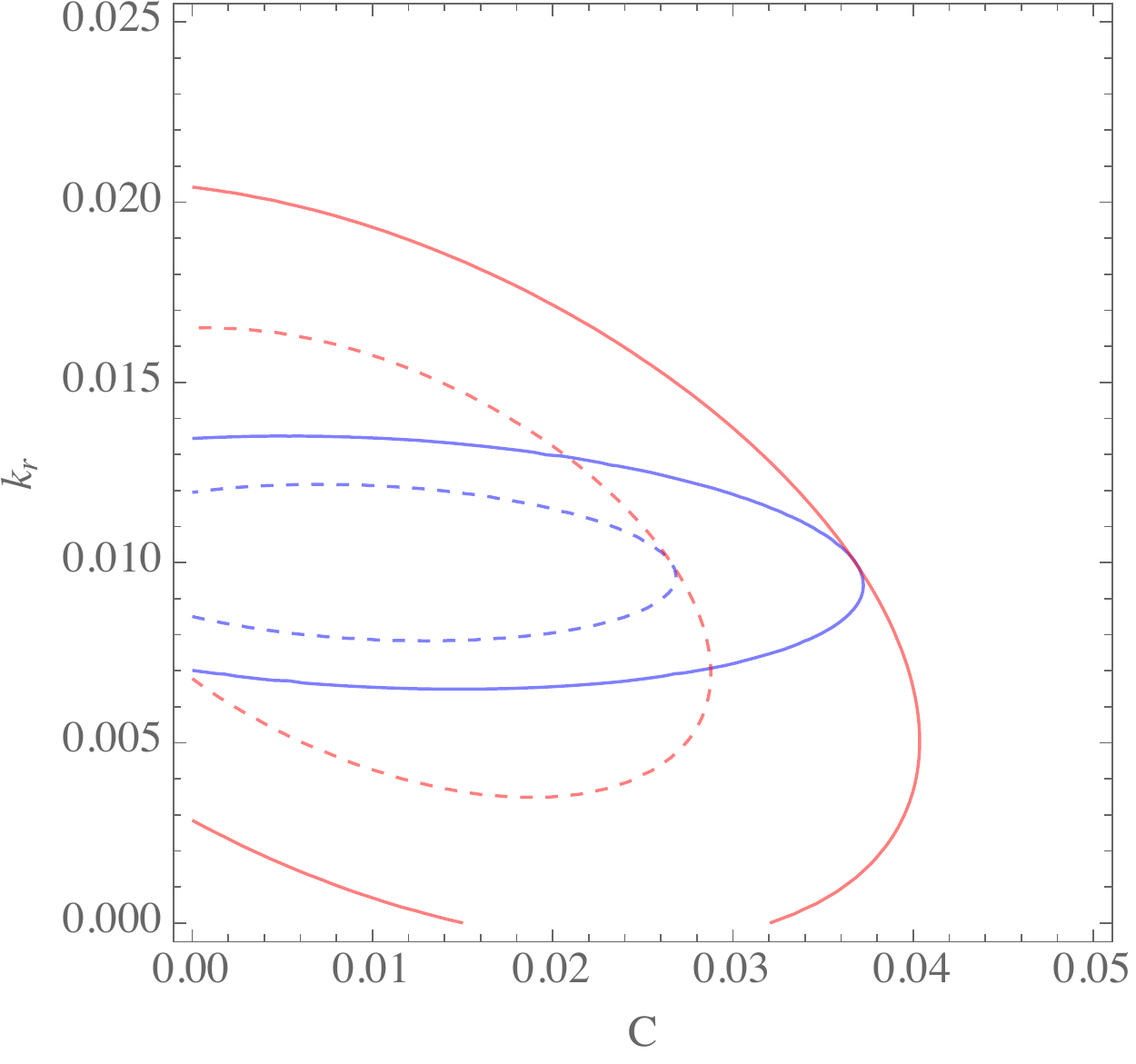}
    \includegraphics[width=1.95in,valign=t]{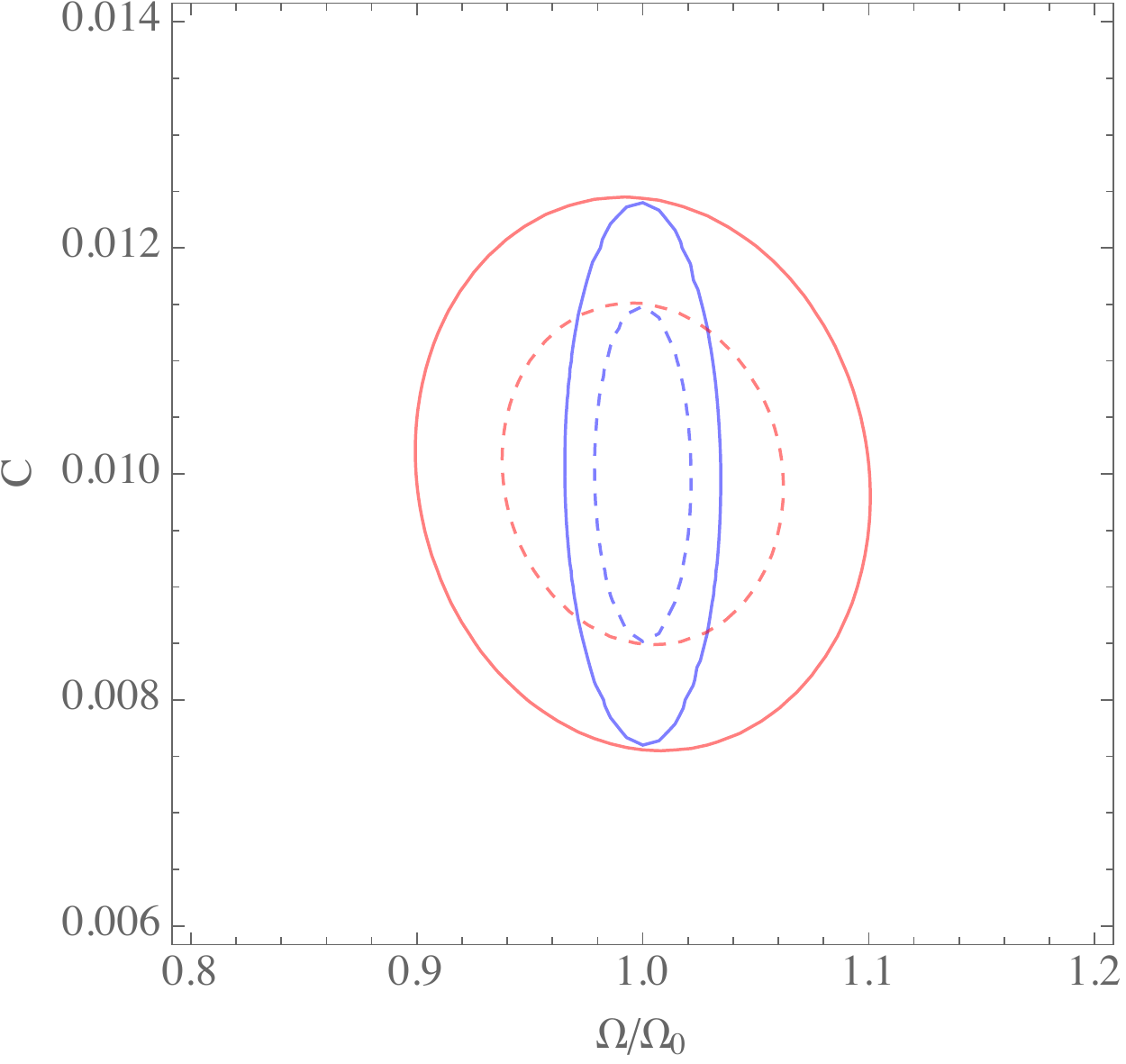}
    \includegraphics[width=1.95in,valign=t]{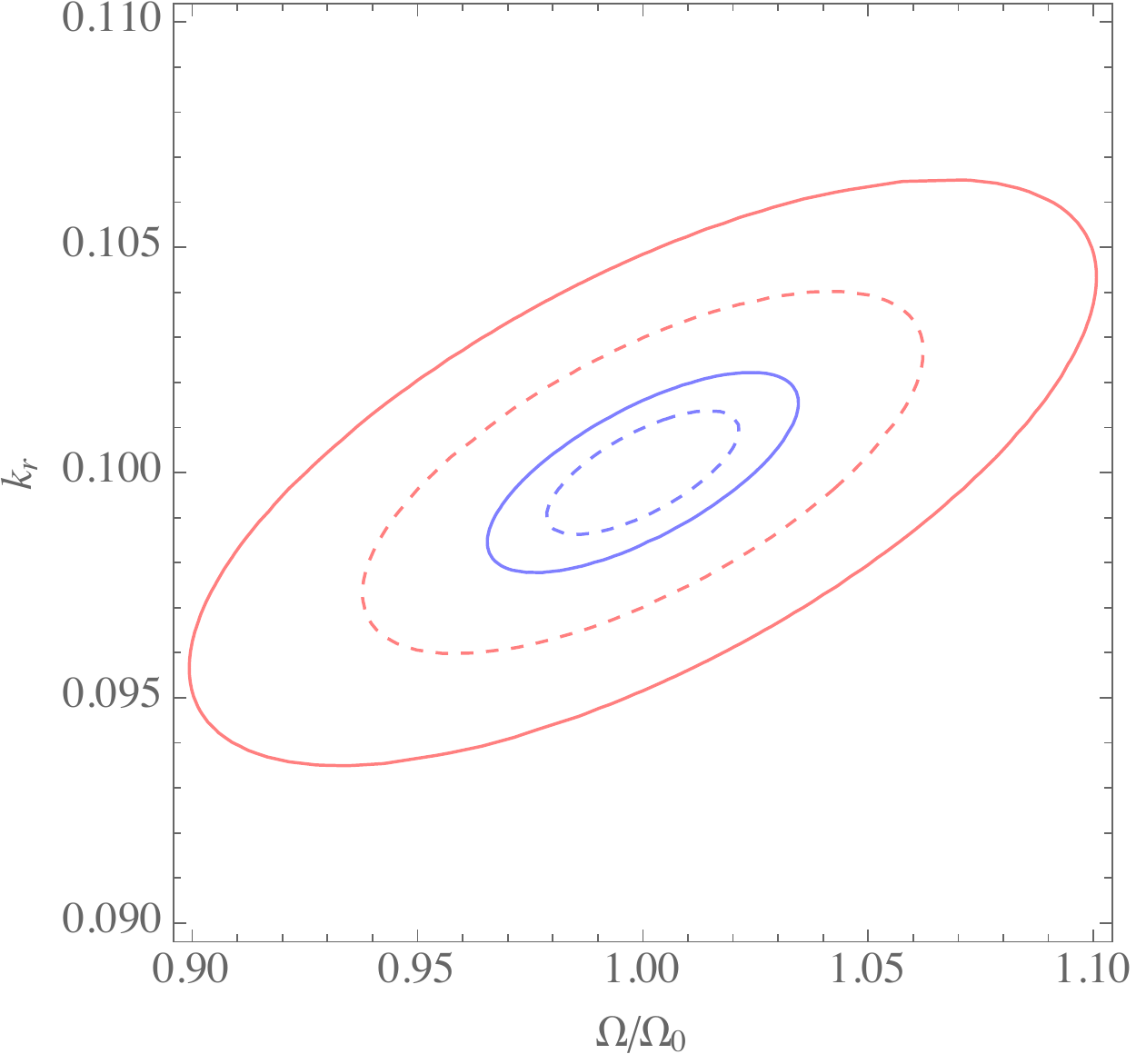}
    \includegraphics[width=1.95in,valign=t]{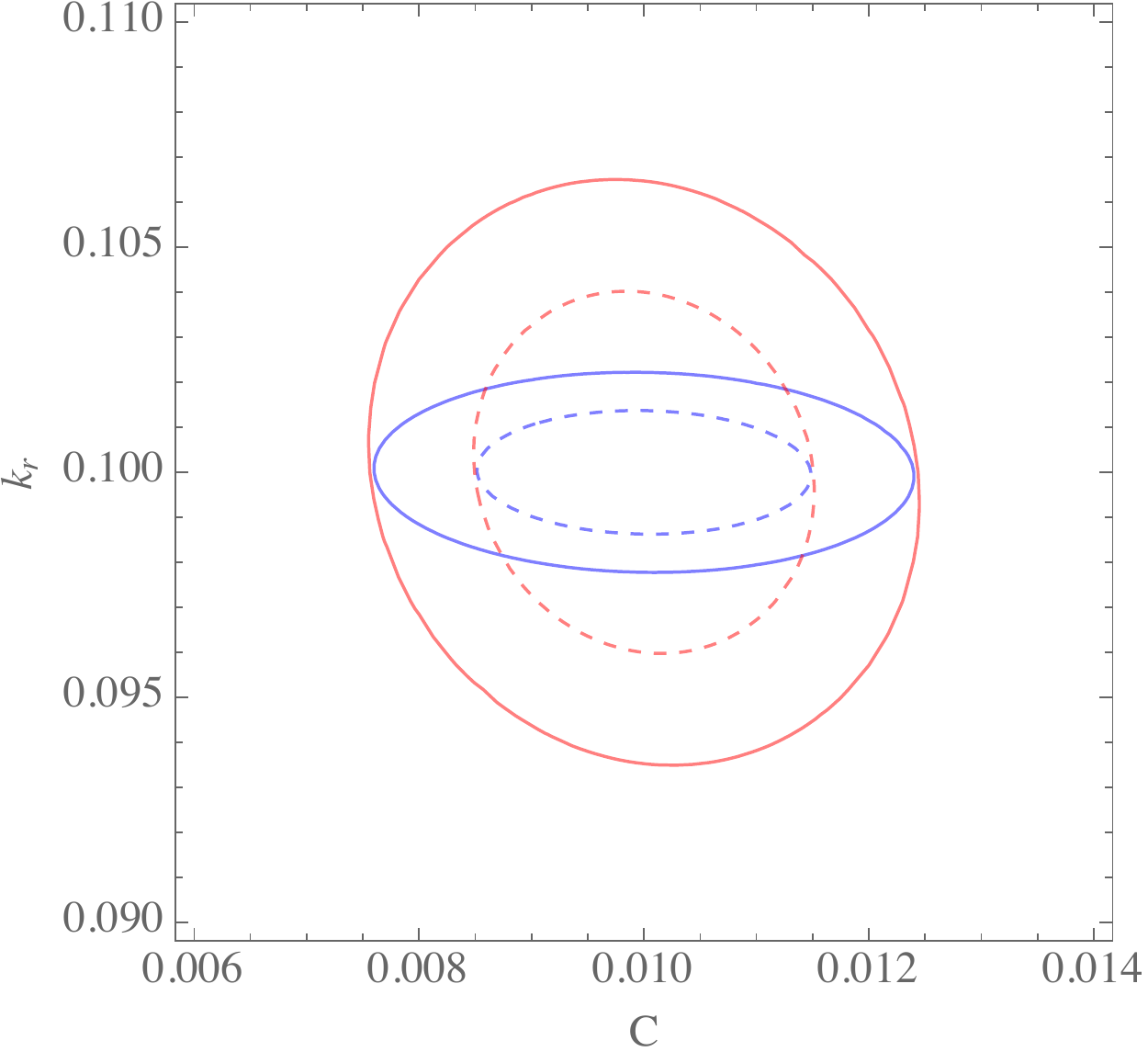}
    \includegraphics[width=1.95in,valign=t]{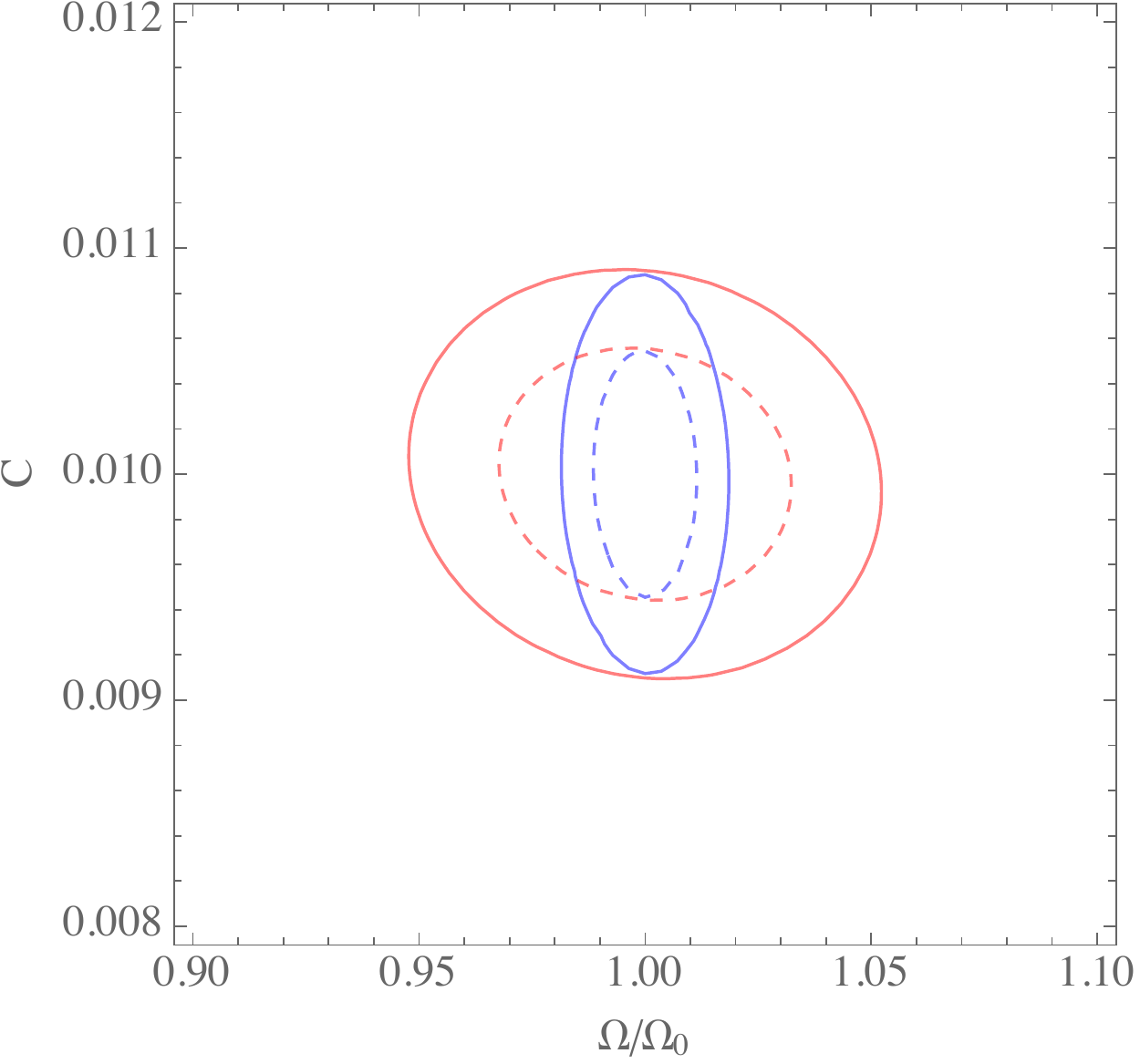}
    \includegraphics[width=1.95in,valign=t]{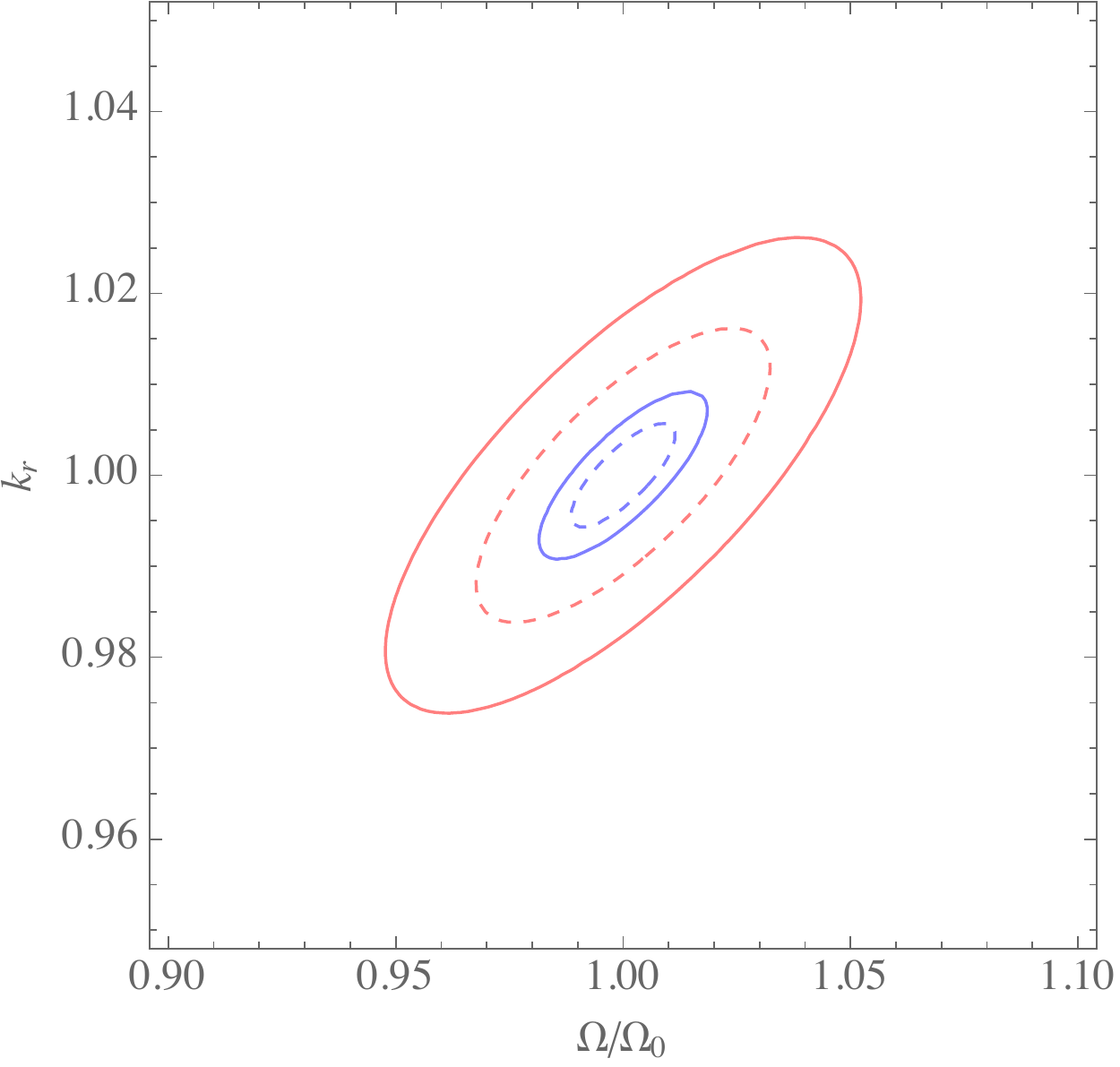}
    \includegraphics[width=1.95in,valign=t]{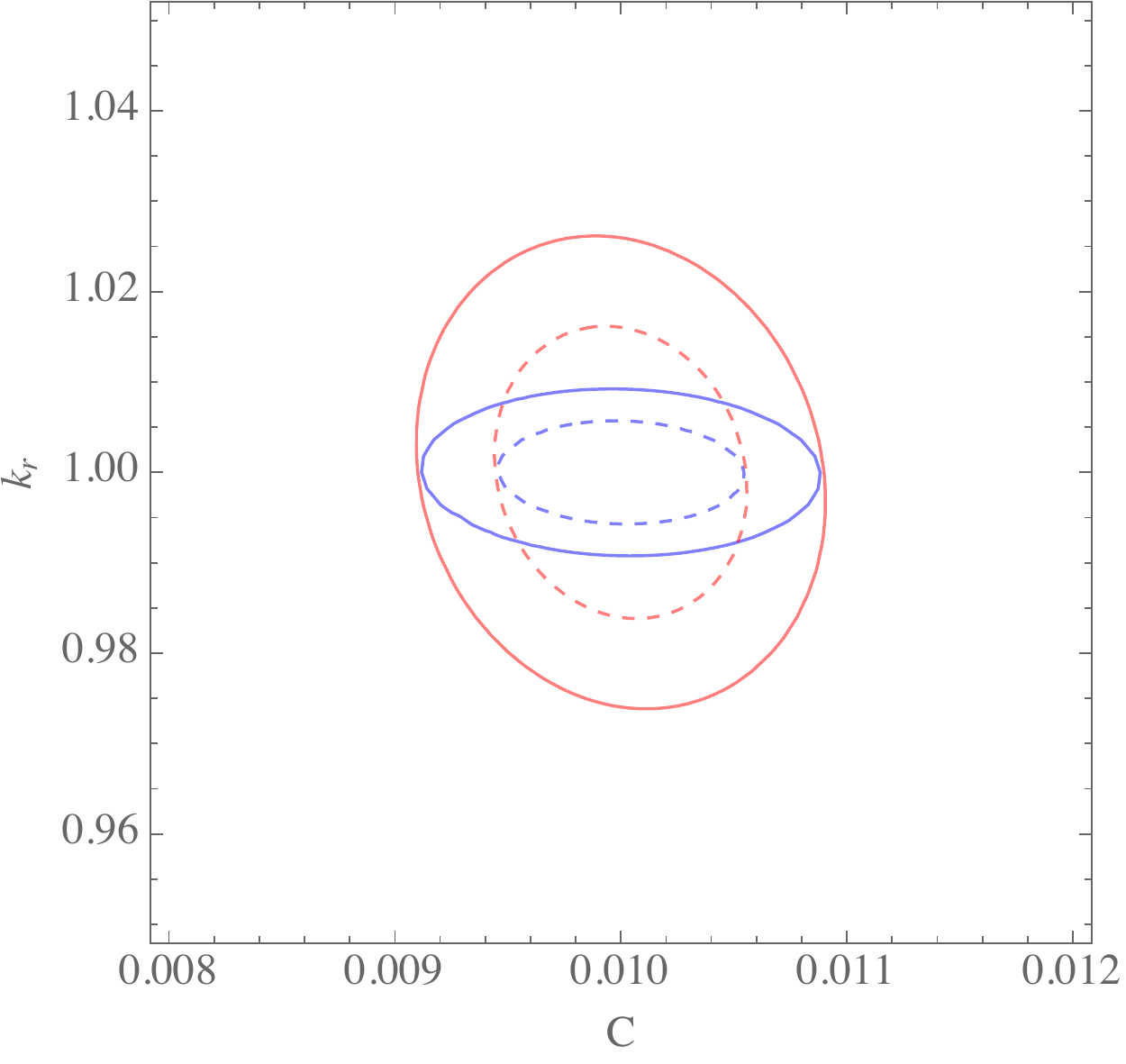}
   \caption{{\bf Clock signal (inflation)} Left: One and two sigma contours showing joined constraints on the amplitude $C$ the frequency $\Omega_{\rm eff}$ and the parameter $k_r$ for the clock signal for $k_r = 0.01$, $k_r=0.1$ and $k_r = 1$ Mpc$^{-1}$ with $\Omega_{\rm eff} = 10$ (red) and $\Omega_{\rm eff} = 30$ (blue) for the inflation scenario. $k_r$ is correlated with $\Omega_{\rm eff}$. A feature on the largest scales requires a large amplitude to be detected with some confidence. As before, increasing the frequency improves parameter constraints. }
   \label{fig:Contoursclock}
\end{figure}

For the inflation scenario, we fall back to the template of Eq.~\eqref{eq:clock_template_large_p}. The reason is that, when $p \gg 1$ it is almost perfectly degenerate with the phase $\phi$ because
\be
p \frac{\Omega_{\rm eff}}{2} \left( \frac{2k}{k_r}  \right)^{\frac{1}{p} } + \phi \xrightarrow{p\gg 1}
p\frac{\Omega_{\rm eff}}{2} + \phi  + \frac{\Omega_{\rm eff}}{2} \ln\frac{2k}{k_r} + \CO(\frac{1}{p}) ~.
\ee
So for this case we should instead use Eq.~\eqref{eq:clock_template_large_p}. We fix $\phi = 0$ for simplicity. We thus are left with only 4 parameters, including $H_0$. This will affect our forecasted constraints on $\Omega_{\rm eff}$ since this parameter has non-negligible correlation with $\phi$. This should be taken into account when considering projected constraints. From our analysis in the previous sections we can assume that this correlation is reduced when the number of oscillations resolved increases; this can happen when we would observe more scales or if the frequency is higher. We also showed that this correlation does not lead to significant loss of constraining power, but it should be taken into account in a future analysis.

We show the marginalized error on the amplitude $C$, $\Omega_{\rm eff}$ and $k_r$ as a function of baseline and window in Fig.~\ref{fig:sigmac_clock_expanding}, for an inflationary universe (e.g.~$p\gg 1$) with $\Omega_{\rm eff} = 10$ for $k_r = 0.1$ Mpc$^{-1}$ (top),  and $\Omega_{\rm eff} = 30$ and $k_r = 1$ Mpc$^{-1}$ (bottom). Due to the localization of the feature, as expected, the largest improvements appear when the feature is just resolved; for the feature at $k_r = 1$ Mpc$^{-1}$ is not resolved at unless $\delta \nu \leq 0.05$ MHz and with a baseline $\geq 1$ km. The feature at $k_r = 0.1$ Mpc$^{-1}$ is just resolved for the largest window and smallest baseline. The projected constraints suggest that inflationary features of this type can be constrained down to $C \simeq 10^{-4}$ as long as the feature is resolved. Higher frequency features at smaller scales are better constrained, which is due to the form of Eq.~\eqref{eq:clock_template_large_p} where $k_r$ suppresses the amplitude and more oscillations generally allow for better parameter constraints. Interestingly, once the first part of the oscillation is resolved, constraints on $k_r$ are tight, with the largest error $\sigma_{k_r} \sim 10^{-2}$.

We study the degeneracies in Fig.~\ref{fig:Contoursclock}, where we show contours for the amplitude $C$ versus $\Omega_{\rm eff}$, $\Omega_{\rm eff}$ vs $k_r$ and $k_r$ vs $C$ for $\Omega_{\rm eff} = 10$ and $30$ and $k_r = 0.01$, 0.1, and 1 Mpc$^{-1}$. For $k_r \leq 0.01$ Mpc$^{-1}$ we cannot rule out features at $C = 0.01$. As we dial up the frequency, parameters as usual get better constrained. Moving the feature towards smaller scales helps in two ways; the effective peak of the feature increases as $k_r^{3/2}$ increases and it is easier to observe the onset of the feature (e.g. for the subvolumes we find at $z = 50$ for example $k_{\rm min} = 0.002$ Mpc$^{-2}$, i.e. it will be hard to constraint features with $k_r < 0.001$  Mpc$^{-1}$ ). 
Planck data contains a potential signal around $k_r=0.1$ Mpc$^{-1}$ with amplitude $C\sim0.05$ which should be straightforward to check, even with the minimal setup we considered here ($\delta \nu = 0.01$ MHz and a baseline of 1 km).

%
%

To summarize, features on very large scales ($k_r \leq 0.01$) require a larger amplitude to be detected, while features below $0.001$ are probably not detectable (they would require large amplitude and are not observed over the full domain) . It is still possible to put constraints on the parameters as can be seen from Fig. \ref{fig:Contoursclock}. Features with larger $k_r$ will be easier to detect, and are currently not constrained by CMB observations simply because the CMB is damped beyond $k = 0.1$ Mpc$^{-1}$. The important advantage then of 21 cm measurements is to explore the parameter range $k_r > 0.1$ Mpc$^{-1}$. We do not find evidence for correlation of this feature with other cosmological parameters, which we attribute to the localized and logarithmic nature of the feature.


 \subsubsection{A contraction scenario}
 \label{Sec:Clock_contraction}

 \begin{figure}[t] 
   \centering
   \includegraphics[width=1.9in,valign=t]{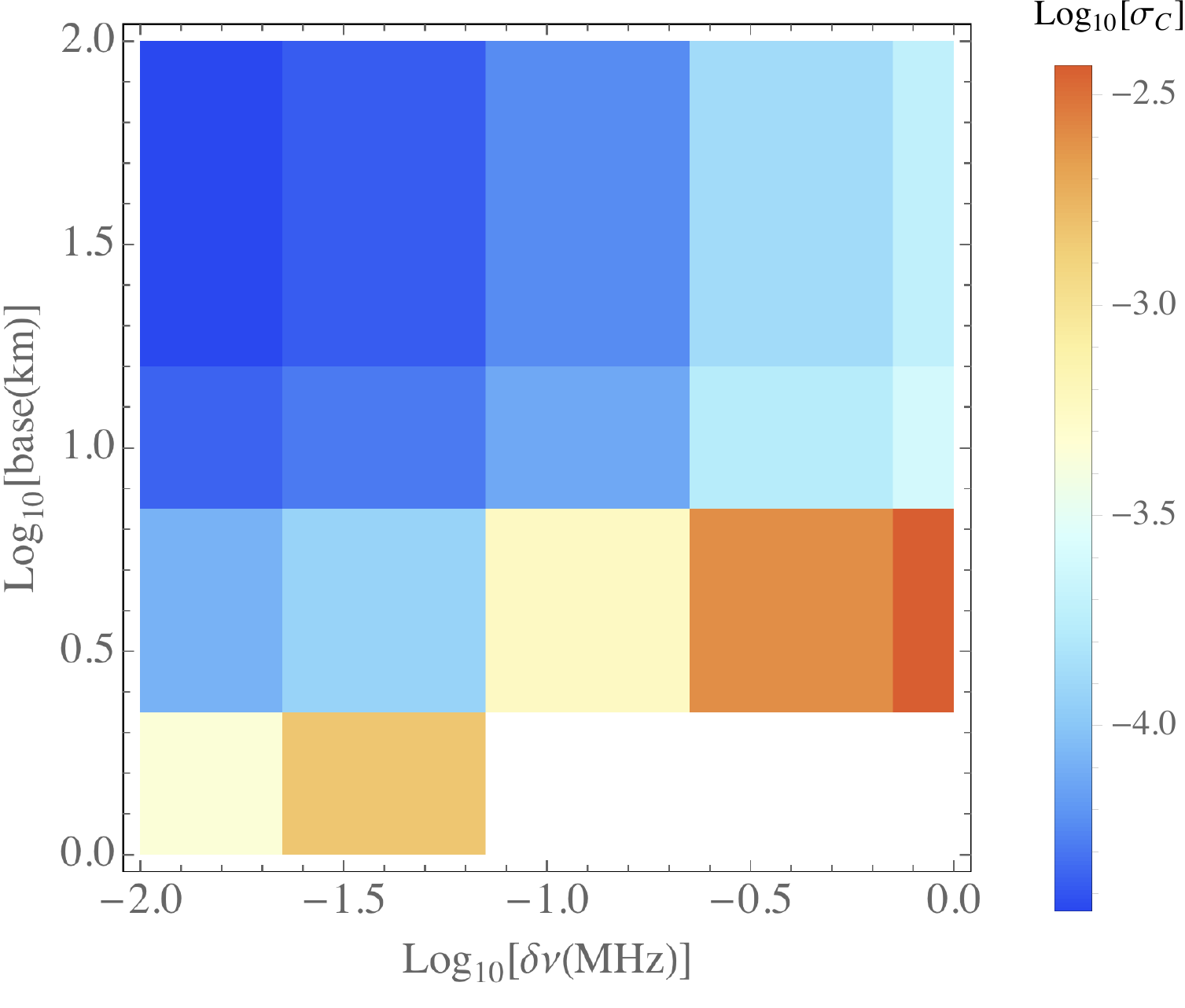}
    \includegraphics[width=1.9in,valign=t]{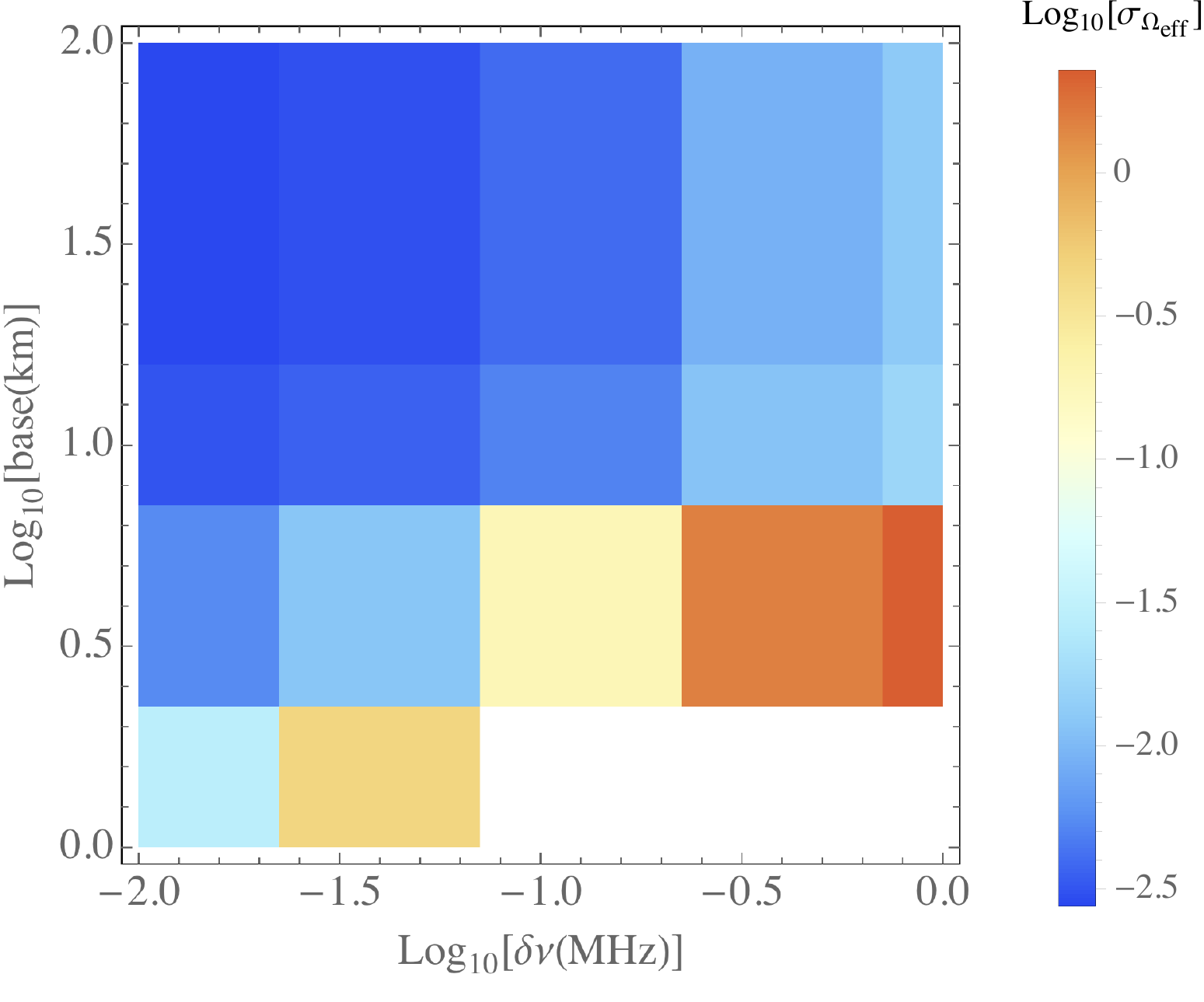}
     \includegraphics[width=1.9in,valign=t]{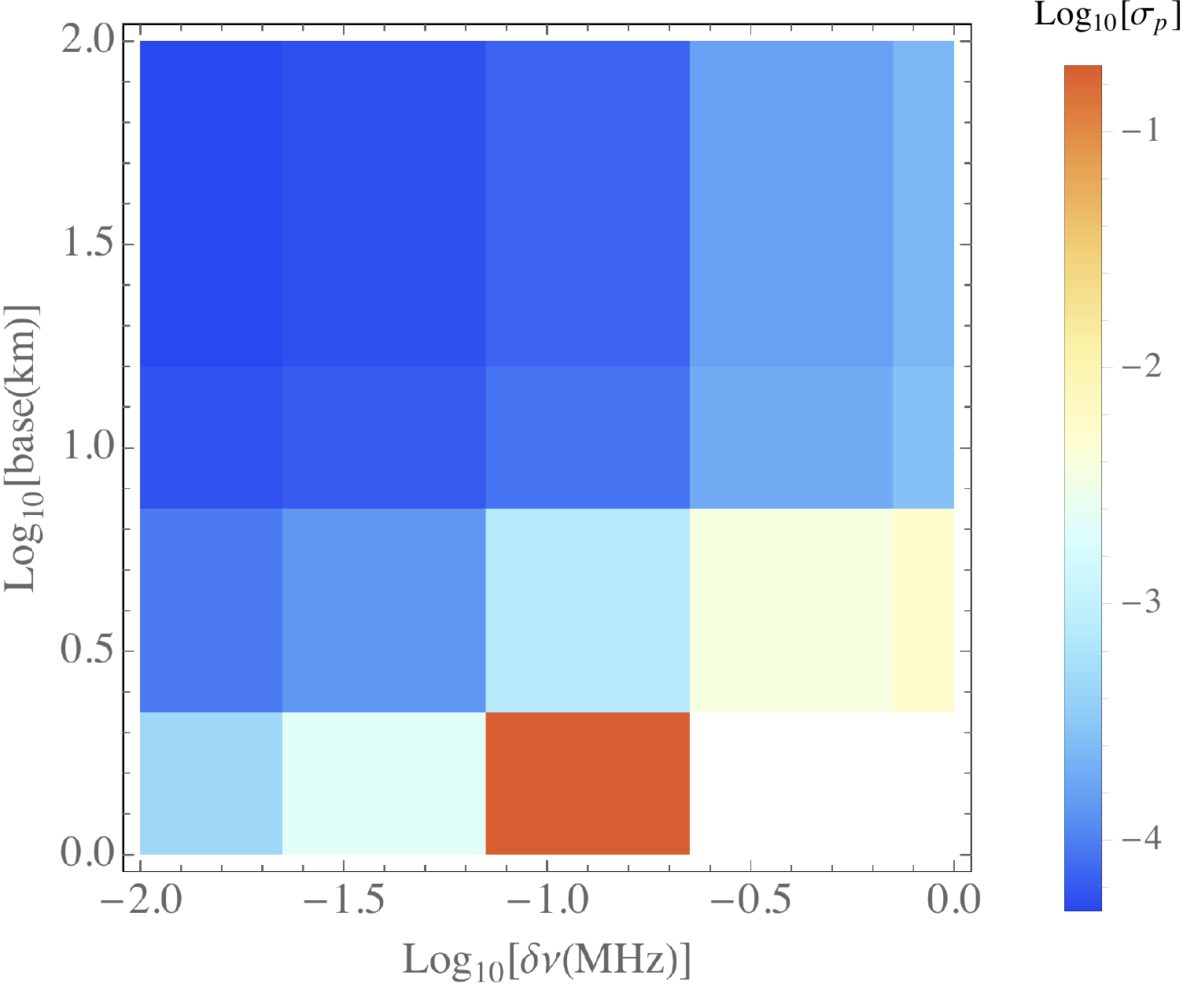}
   \caption{{\bf Clock signal (Ekpyrotic contraction):} The marginalized absolute errors for $\sigma_C$, $\sigma_{\Omega_{\rm eff}}$ and $\sigma_p$ as a function of the experimental configuration for $k_r = 1$ Mpc$^{-1}$,  $\Omega_{\rm eff} = 30$ and $p = 1/5$. The local nature of the feature is apparent through the local improvement of the error; for experiments with $\delta \nu > 0.1$ MHz and a baseline of 1 km, no constraints can be set. When we lower $k_r$ we shift the amplitude and onset of the feature without changing the number of oscillations. Ekpyrotic features can be constrained to $10^{-5}$ levels for optimistic experimental configurations. }
   \label{fig:sigmac_clock_contracting}
\end{figure}

 \begin{figure}[htbp] 
   \centering
   \includegraphics[width=3in,valign=t]{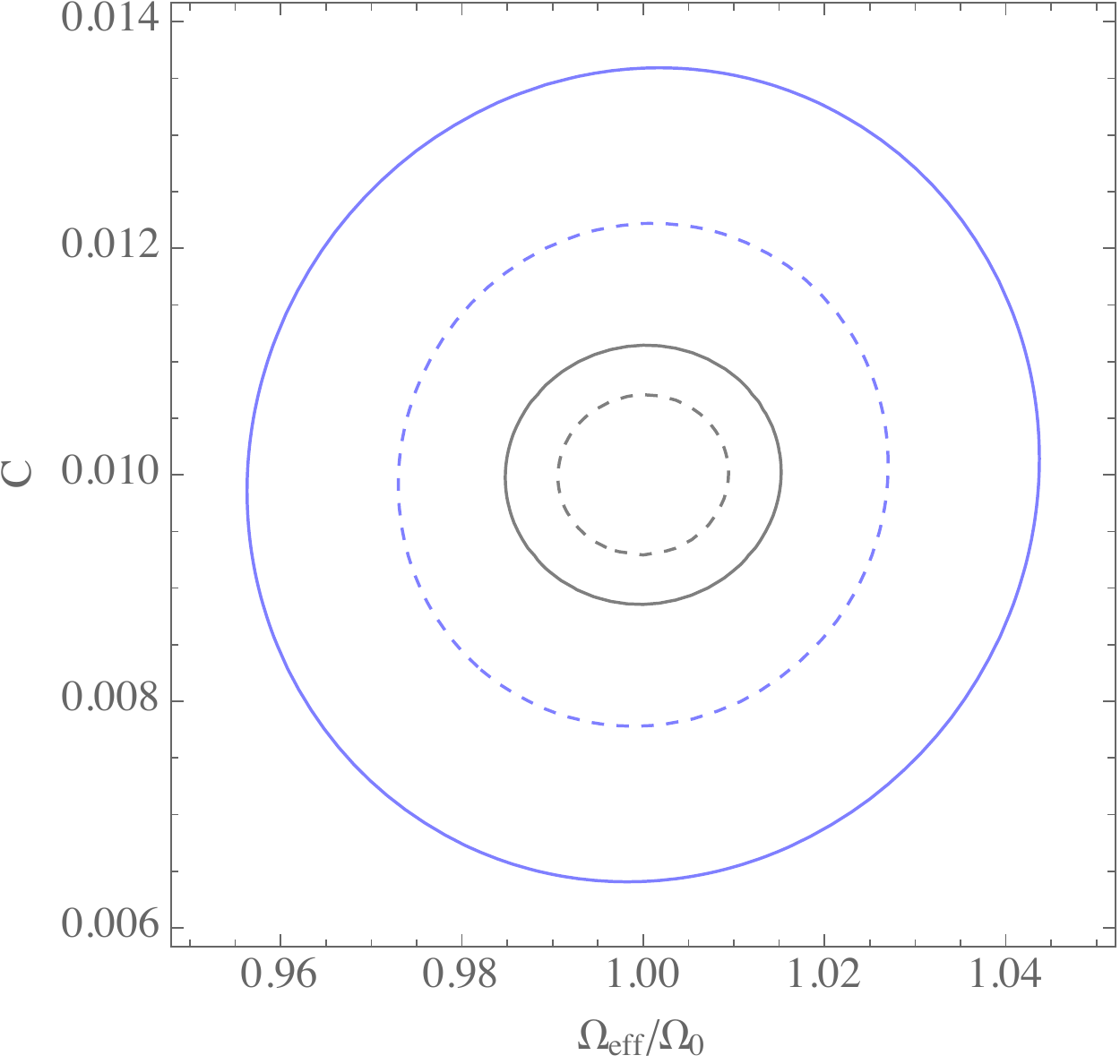}
   \includegraphics[width=3in,valign=t]{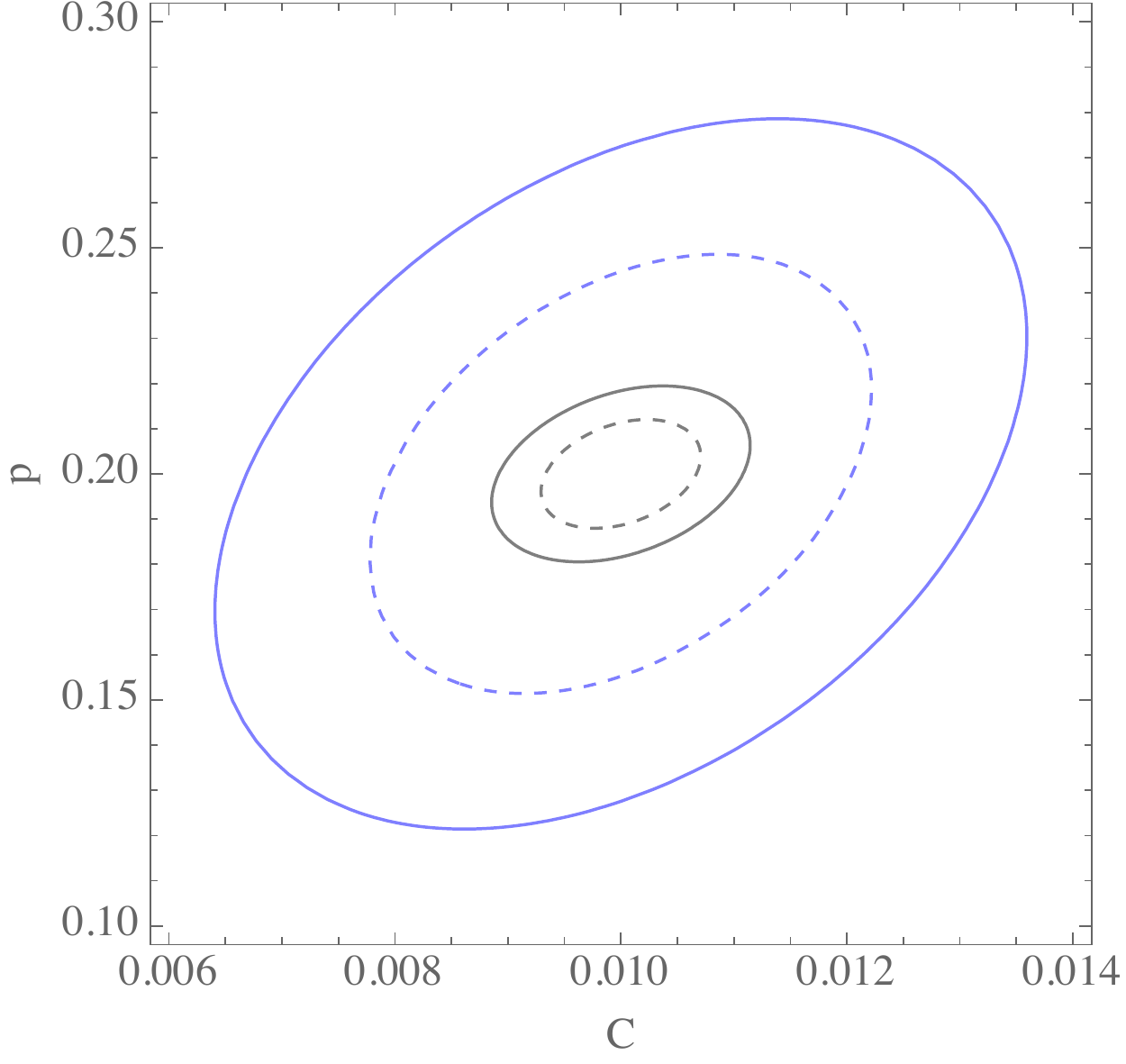}
   \includegraphics[width=3in,valign=t]{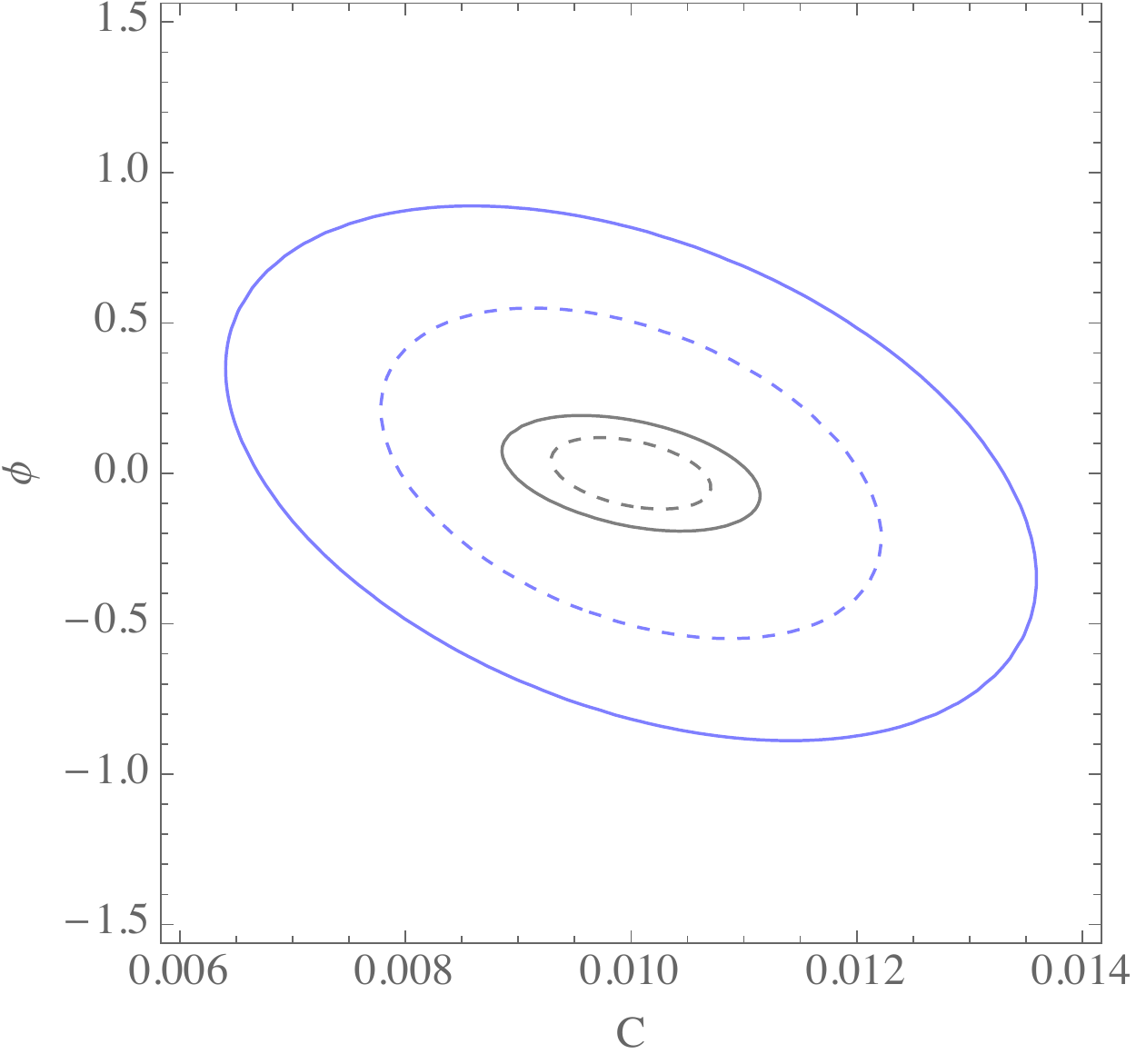}
   \includegraphics[width=2.9in,valign=t]{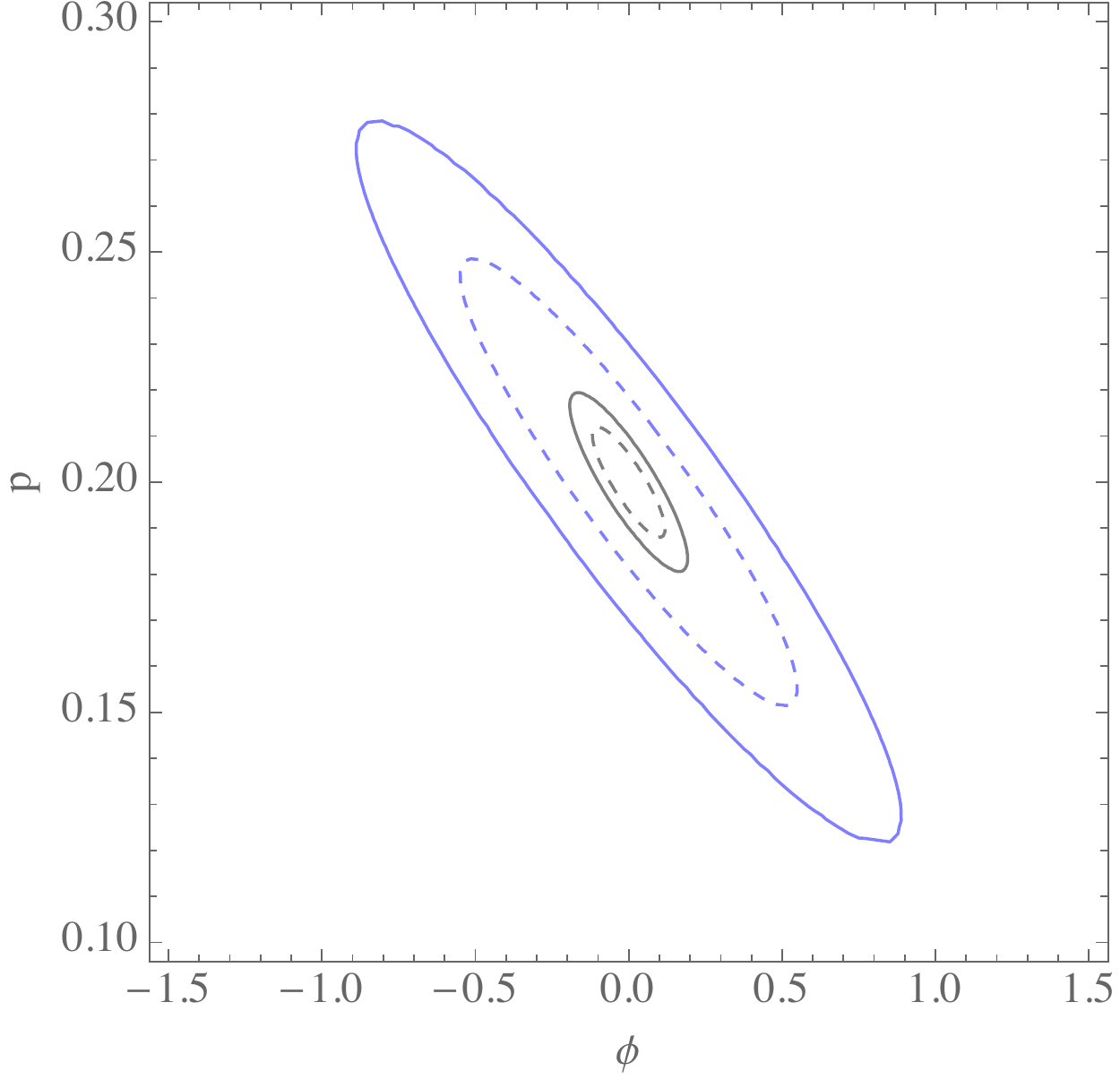}
   \caption{{\bf Clock signal (Ekpyrotic contraction):} Contour plots showing the various joined constraints between parameters for a contracting ekpyrotic universe with a feature at a frequency $\Omega_{\rm eff} = 30$ and $p = 1/5$ with  $k_r=0.1$ Mpc$^{-1}$ (blue) and $k_r=1$ Mpc$^{-1}$ (black). No useful constraint on above parameters can be put for larger values of $k_r$ with an instrument with a baseline of 1 km and a window of $\delta \nu = 0.01$ MHz. There are no correlations with $h_0$.  }
   \label{fig:ContoursclockContracting}
\end{figure}


We consider the ekpyrotic contraction scenario, with $p = 1/5$ and  $\Omega_{\rm eff} = 30$ for $k_r = 0.1$ and $k_r=1$ Mpc$^{-1}$. We find severe issues with inverting the Fisher matrix if we include $k_r$. We therefore fix the value of $k_r$; most likely there exists a strong correlation with frequency due to the presence of both the scale $k_r$ and $\Omega_{\rm eff}$ in the step. We assume that this issue arises due to the semi-analytical nature of our analysis, where we compute the derivatives analytically when possible. We thus vary 5 parameters, including $h_0$.


We show the absolute errors for $\sigma_C$, $\sigma_{\Omega_{\rm eff}}$ and $\sigma_p$ as a function of the experimental configuration for $k_r = 1$ Mpc$^{-1}$ in Fig.~\ref{fig:sigmac_clock_contracting}.
As we lower $k_r$, it becomes increasingly difficult to constrain small primordial amplitudes of the feature.
Because the clock signal for this slowly contracting scenario has very few oscillations, the parameter $p$, which is the fingerprint of the scenario type, has degeneracies with both $\phi$ and $C$ shown in Fig.~\ref{fig:ContoursclockContracting}.


\subsubsection{A full clock signal example}
\label{Sec:Clock_full}

\begin{figure}[htbp] 
   \centering
  \includegraphics[width=1.95in,valign=t]{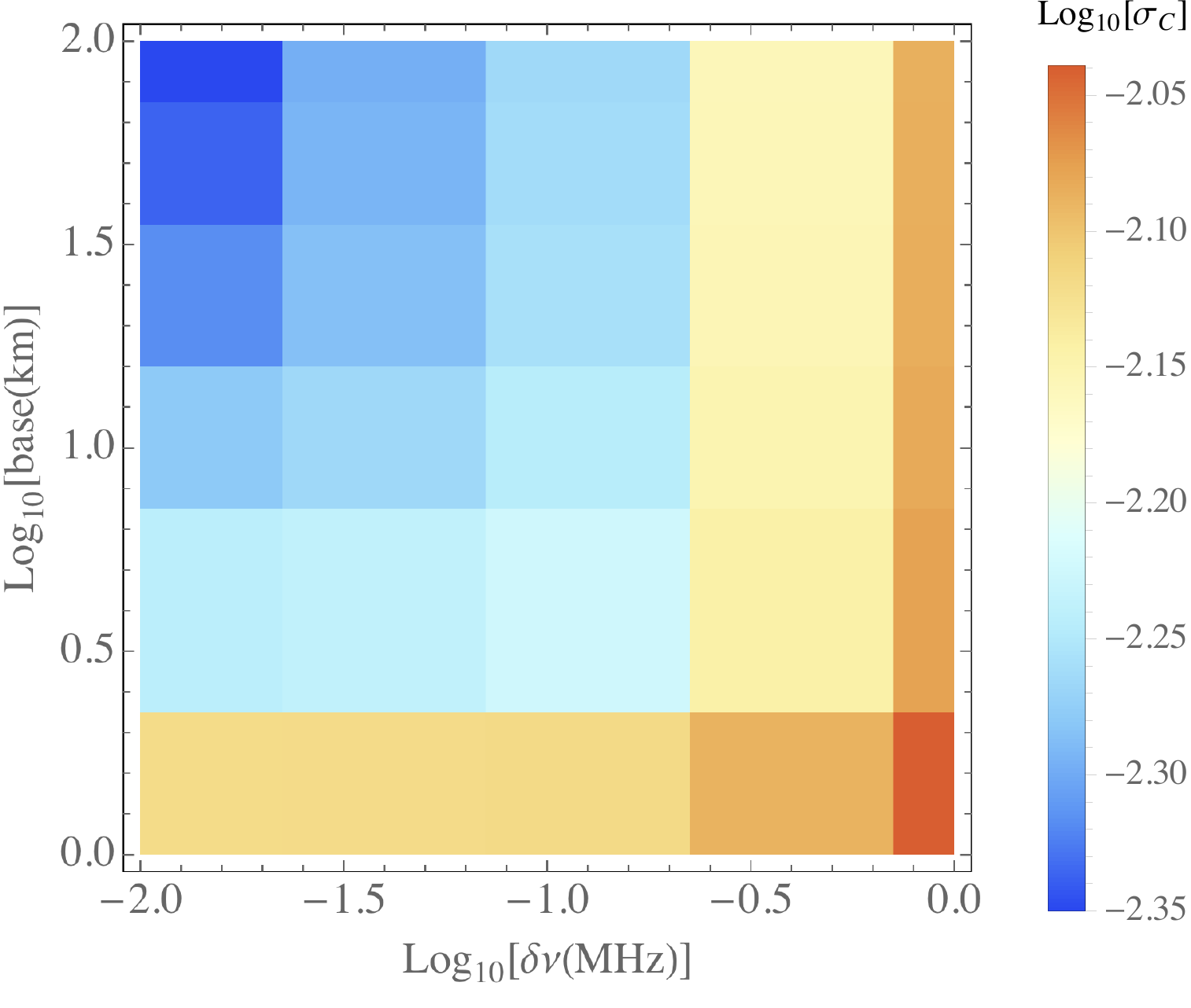}
  \includegraphics[width=1.95in,valign=t]{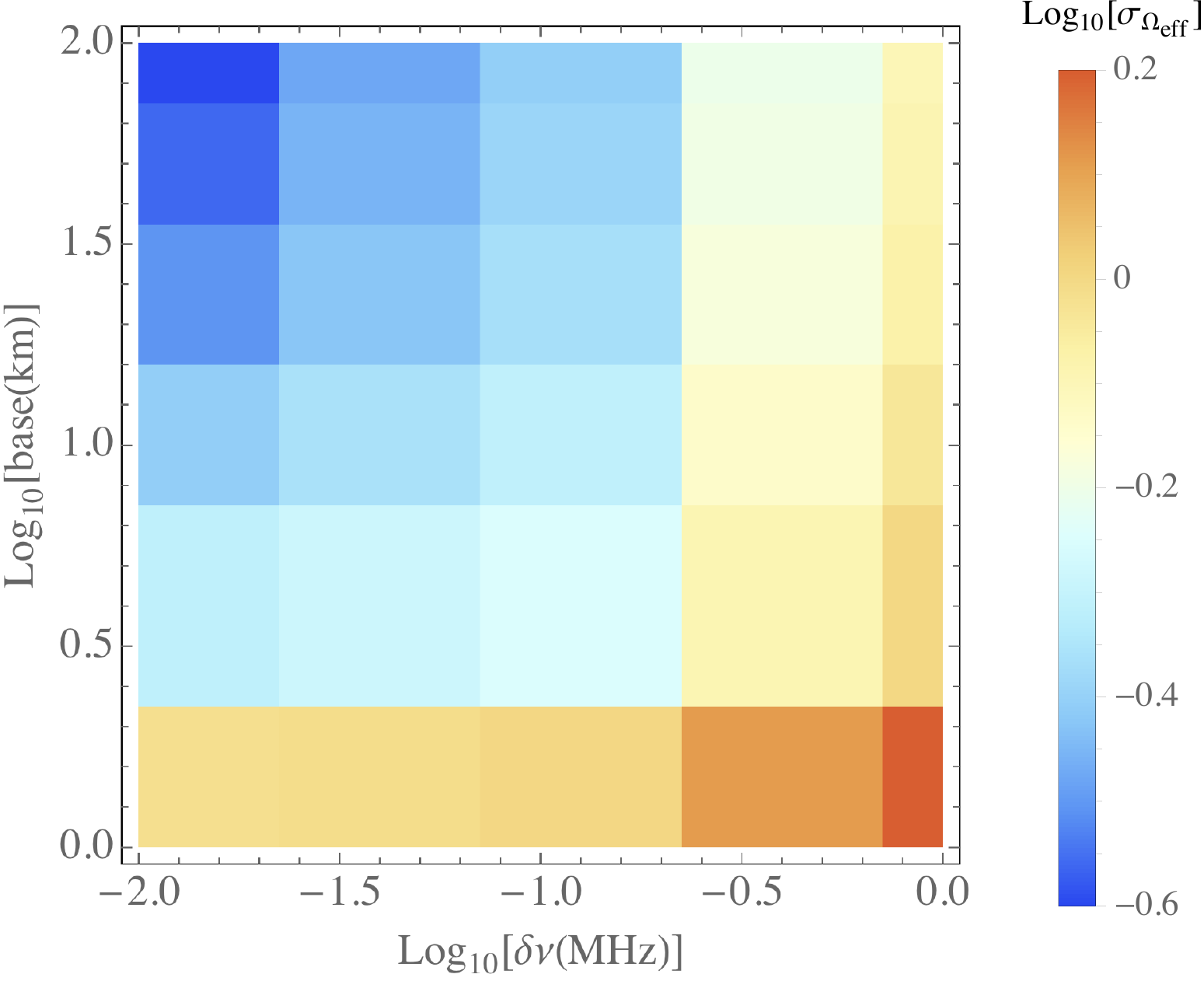}
    \includegraphics[width=1.95in,valign=t]{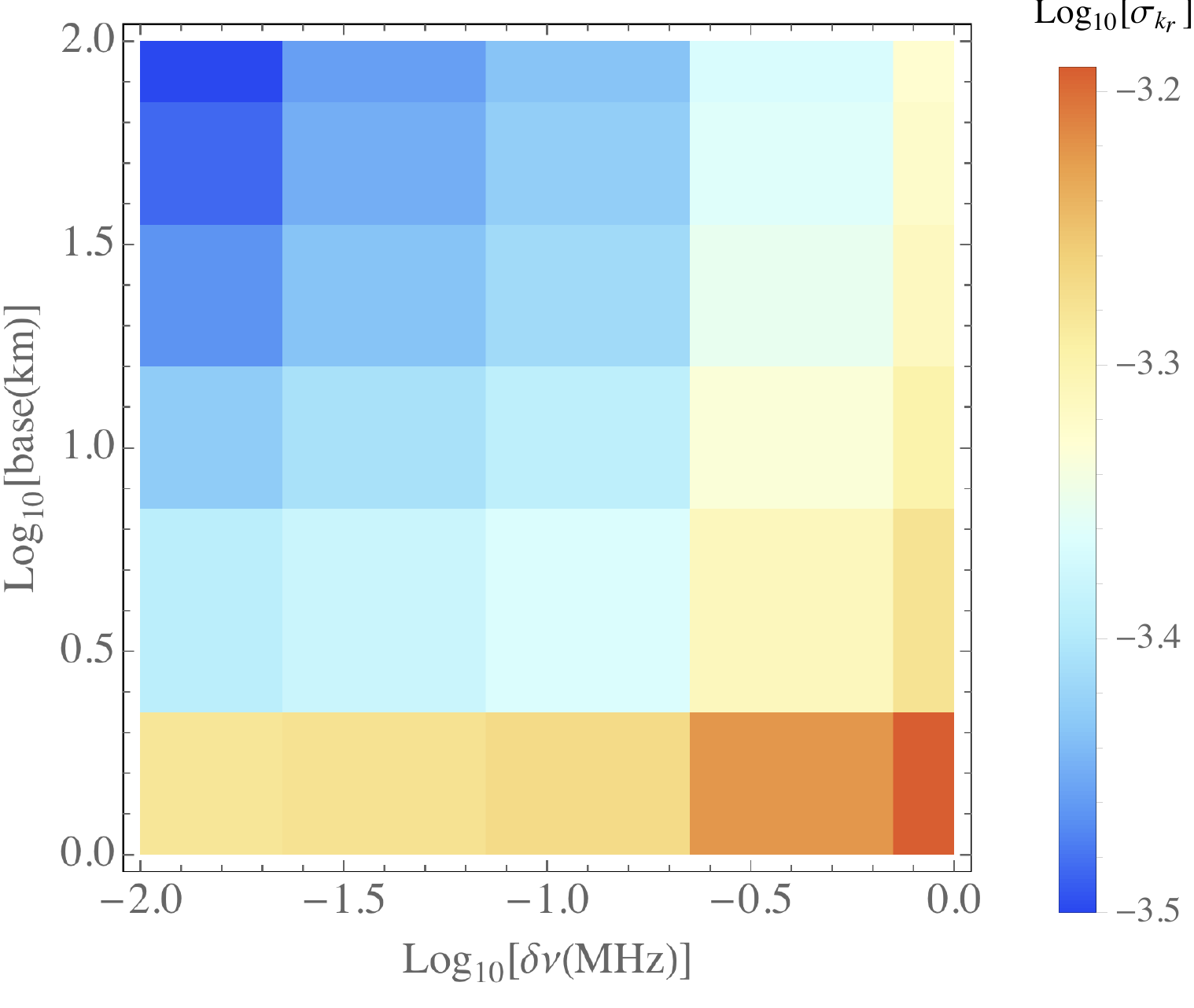}
      \includegraphics[width=1.95in,valign=t]{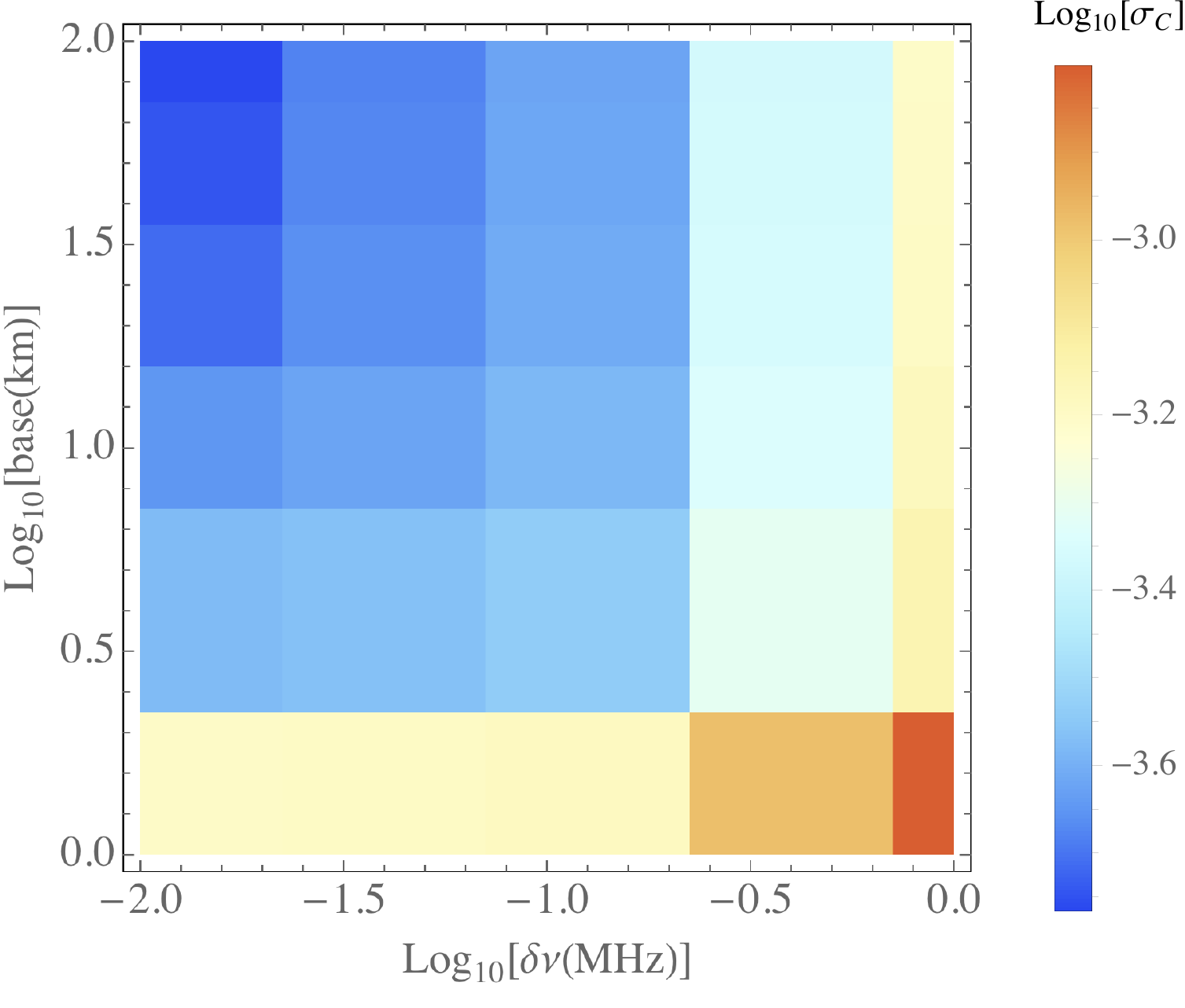}
  \includegraphics[width=1.95in,valign=t]{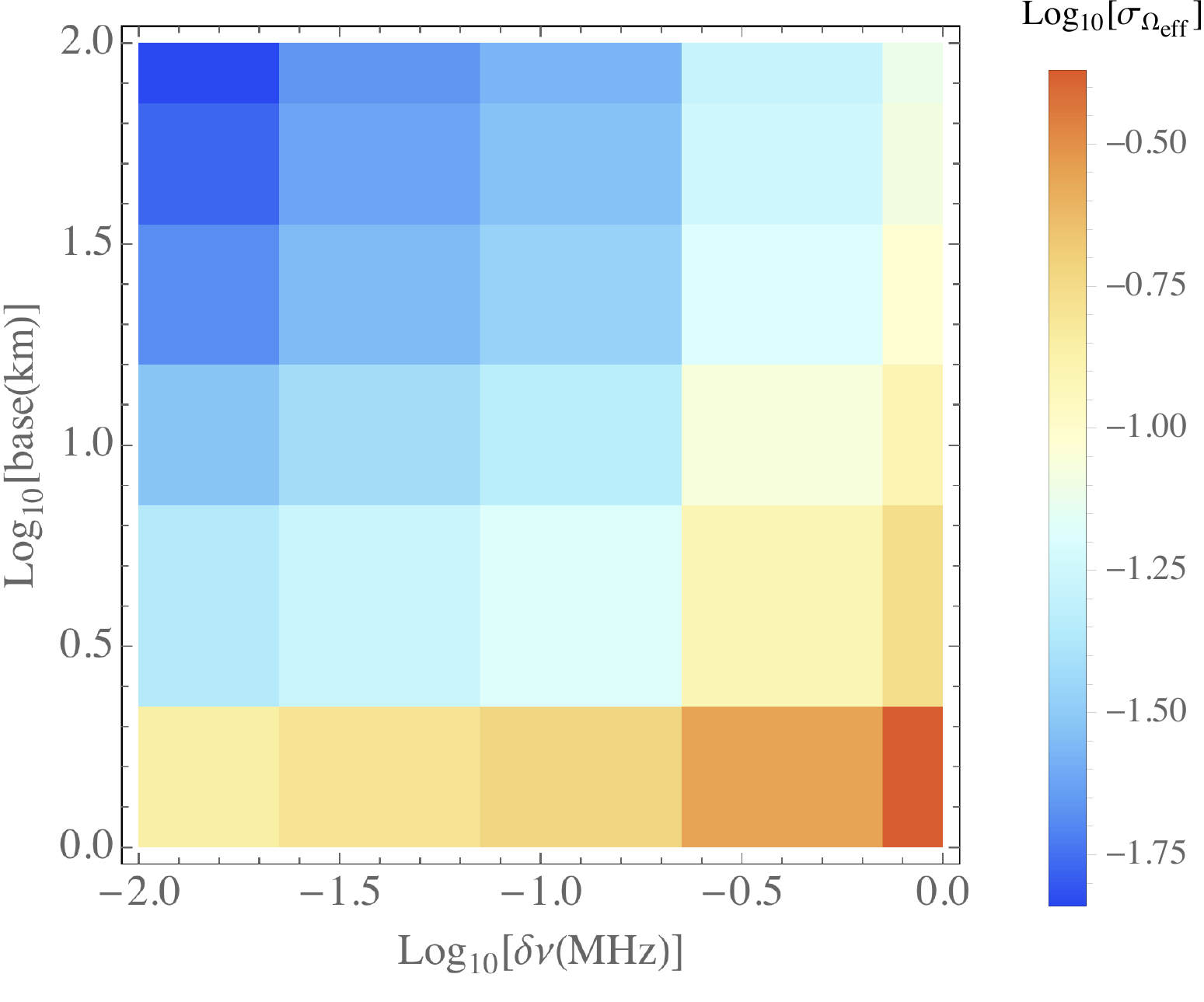}
    \includegraphics[width=1.95in,valign=t]{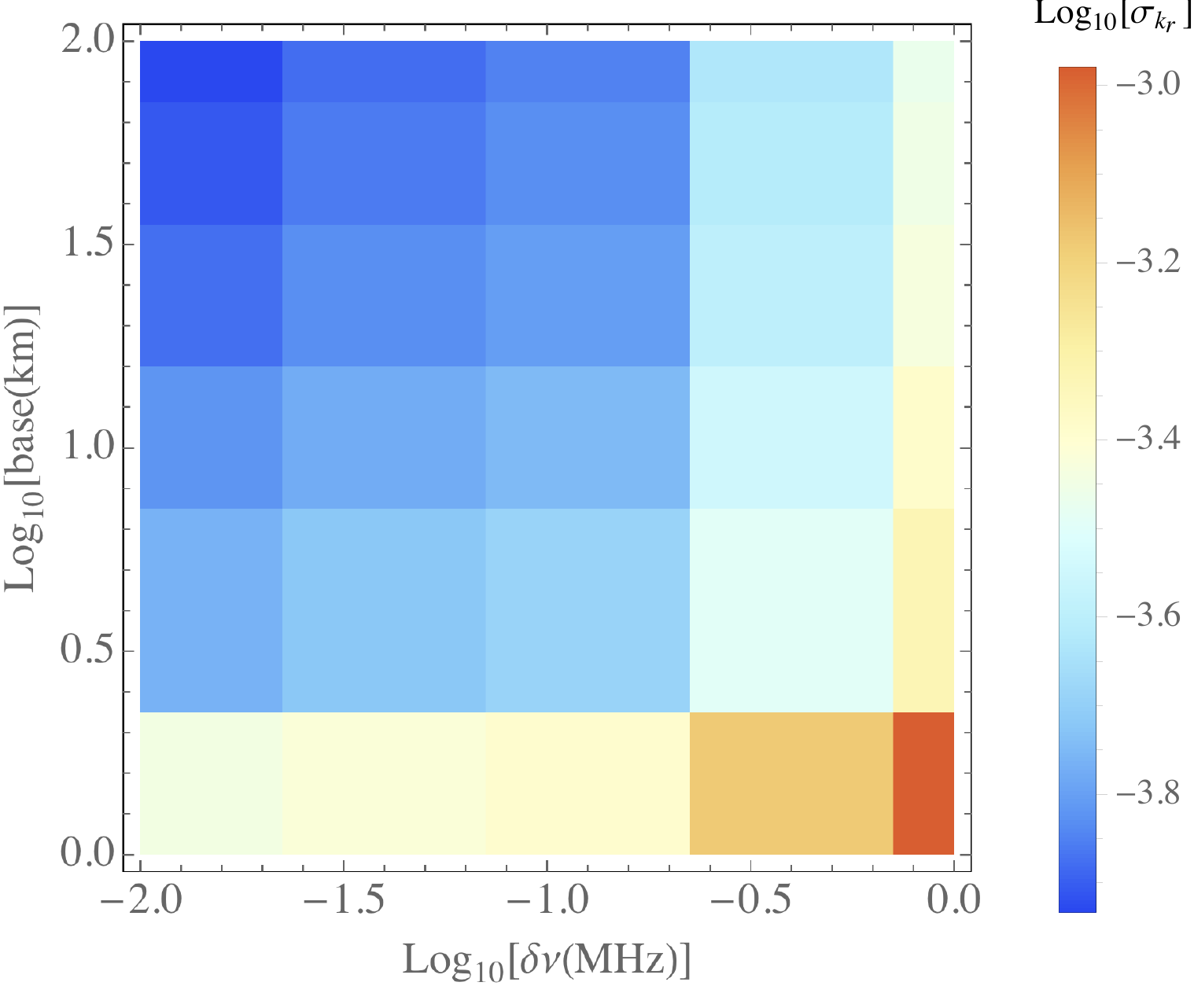}
   \caption{{\bf A full clock signal for inflation:} The marginalized absolute errors $\sigma_C$, $\sigma_{\Omega}$ and $\sigma_{k_r}$ versus the angular and radial resolution of an experiment for $k_r = 0.01$ Mpc$^{-1}$ (top) and $k_r = 0.1$ Mpc$^{-1}$ at $\Omega_{\rm} = 30$ (bottom). A feature with $k_r = 0.01$ Mpc$^{-1}$ with an amplitude $C = 0.01$ is barely detectable at 1 sigma (with $\sigma_C \simeq 10^{-2}$). For features on larger scales should be detectable with high significance. As expected, we find that a feature on smaller scales (e.g. $k_r \geq 1$ Mpc$^{-1}$)  requires a narrow window (i.e. $\delta \nu \leq 0.1$) to be detected. Interestingly, the frequency and the location of the peak can be detected with high significance even on large scales, while the amplitude constraints range from $10^{-2}$ for $k_r = 0.01$ Mpc$^{-1}$ and $10^{-5}$ for $k_r = 1$ Mpc$^{-1}$ (not shown).  }
    \label{fig:sigmac_full}
\end{figure}

%

\begin{figure}[tbp] 
   \centering
     \includegraphics[width=2.02in,valign=t]{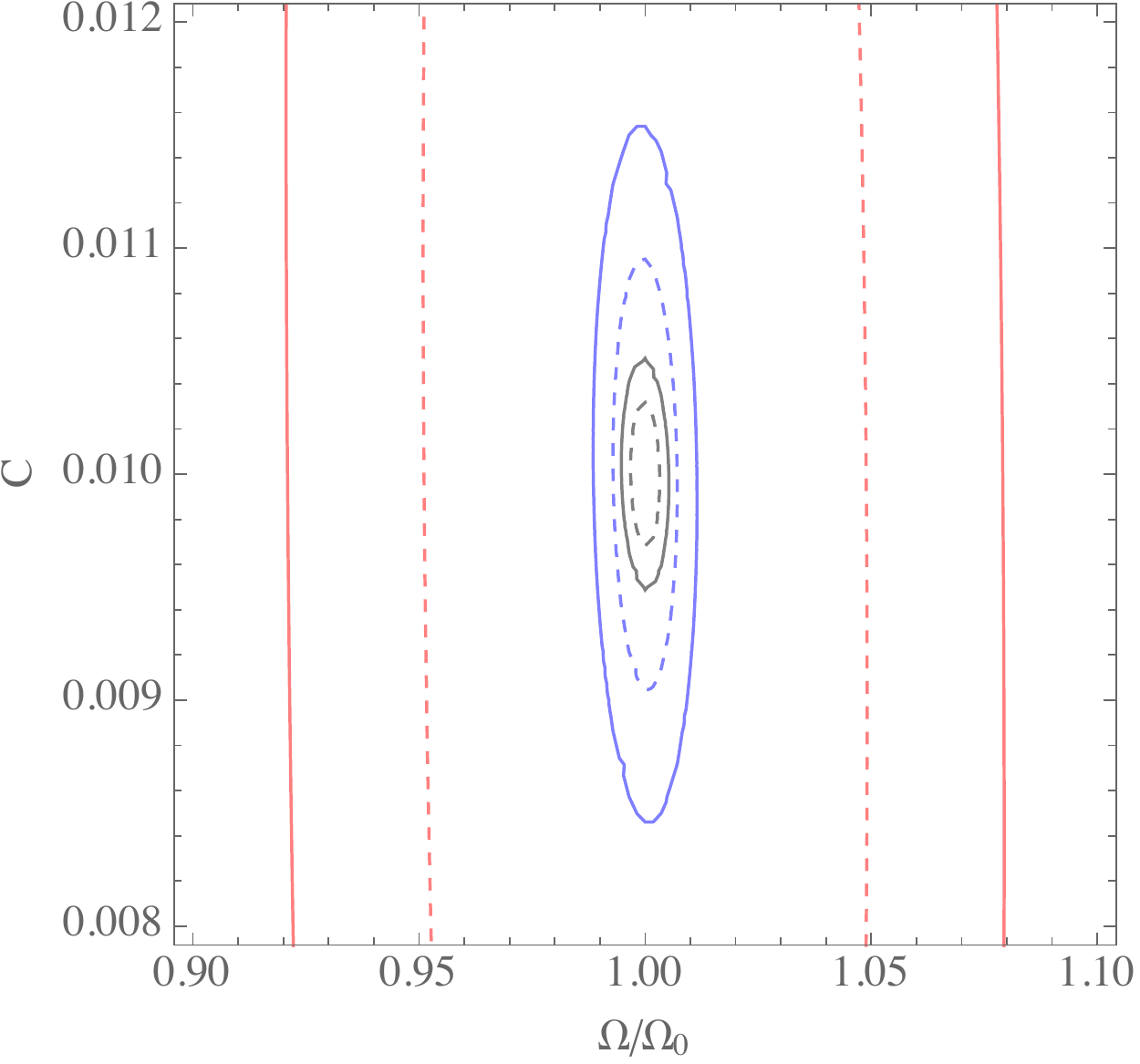}
     \includegraphics[width=1.95in,valign=t]{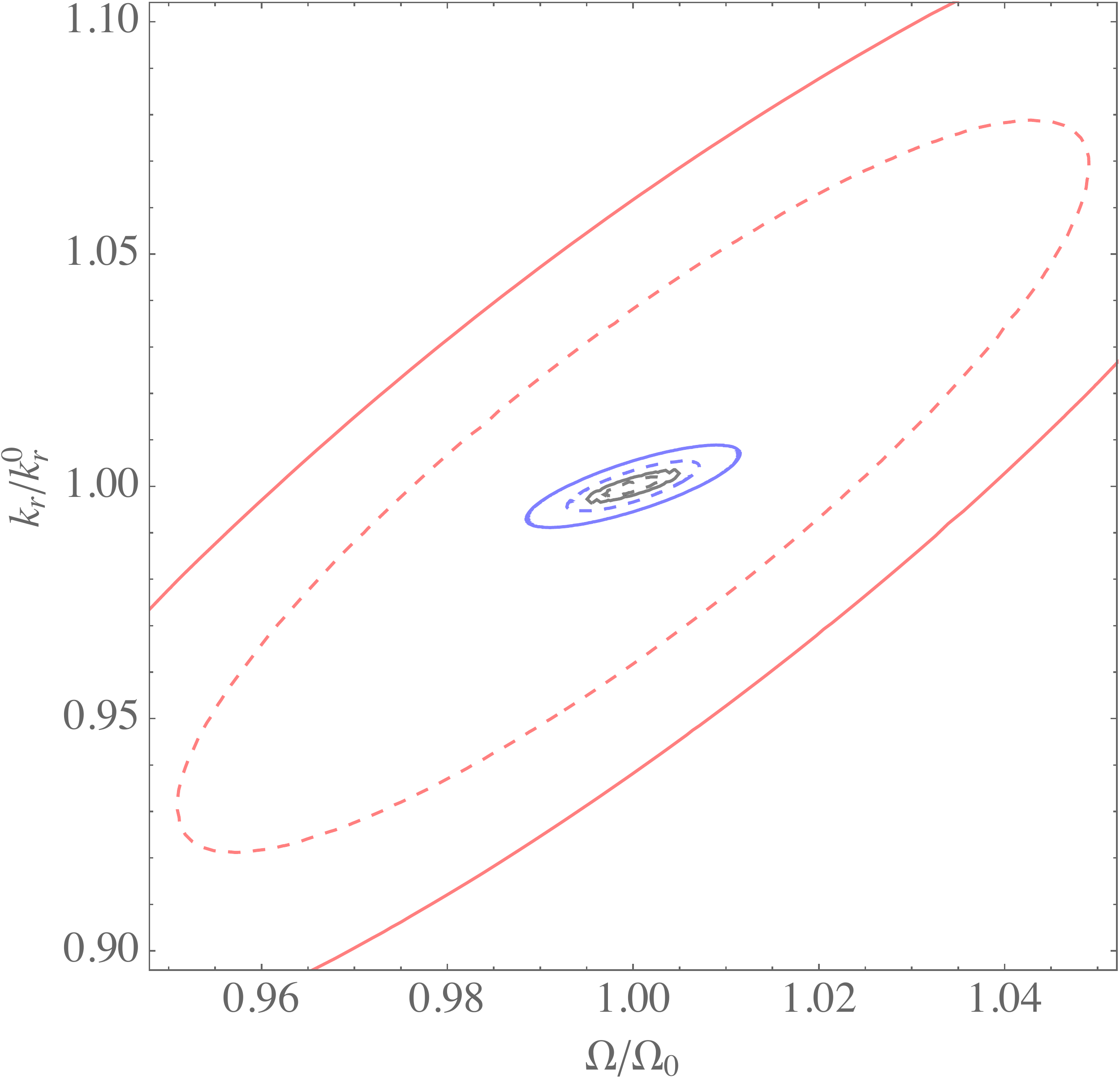}
      \includegraphics[width=2.02in,valign=t]{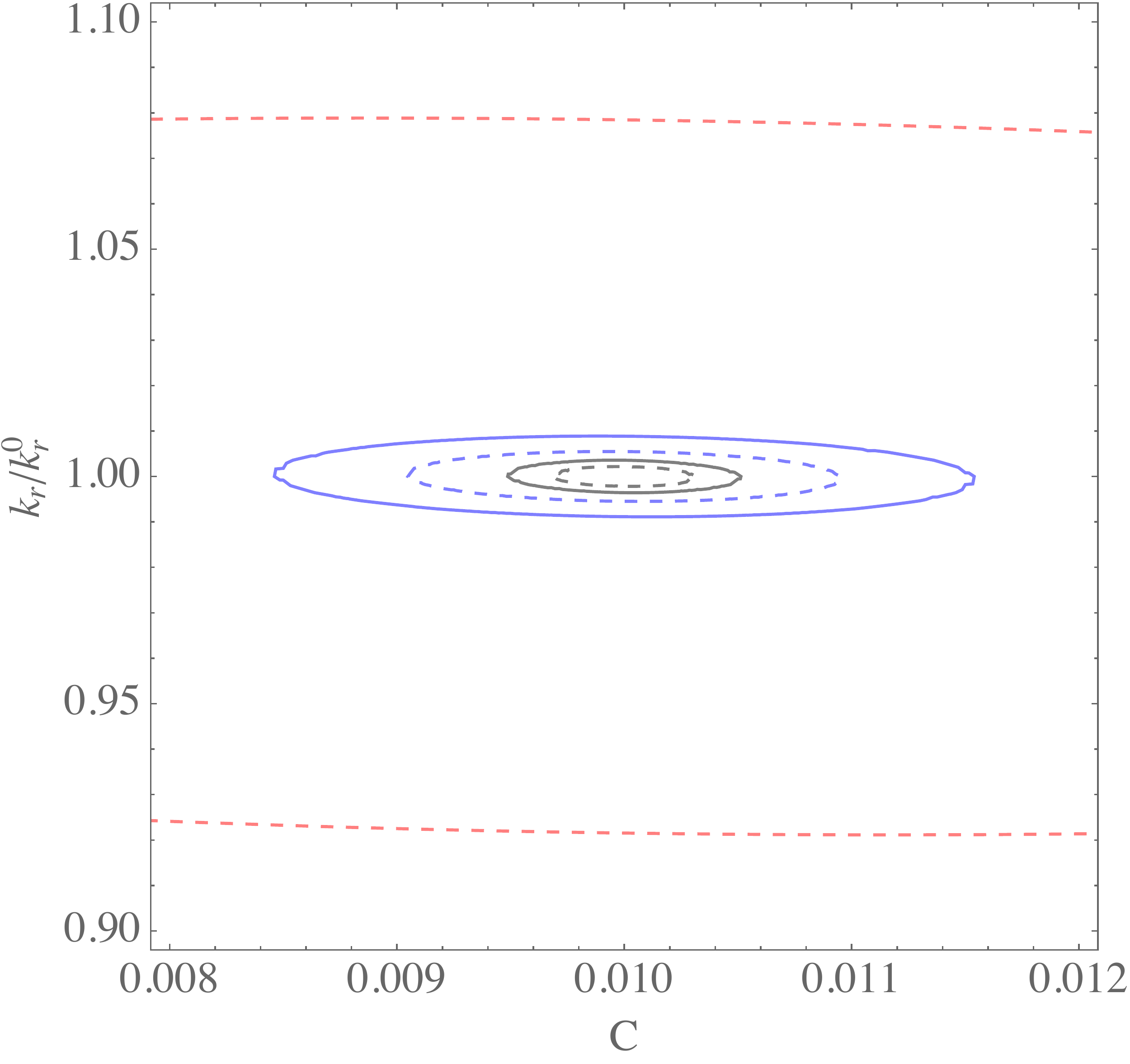}

   \caption{{\bf A full clock signal for inflation: }Showing joined one and two sigma contours for $C$, $\Omega$ and $k_r$, with $\Omega = 30$ and $k_r$ = 0.01 (red), 0.1 (blue) and 1 (black) Mpc$^{-1}$. For a large scale feature, we can not exclude amplitudes $C < 0.01$ at more than 1 sigma. Small scale features are very well constrained. There is a correlation between $k_r$ and $C$. Although the numbers are similar to the clock signal, the fitted form above yields slightly better constraints due to having more scales contribute (while the form of Eq.~\eqref{clock_template} is artificially cutoff at low $k$). }
   \label{fig:ContoursFullClock}
\end{figure}

For the full clock signal example (\ref{Template_SC_full}) we reduce the total number of primordial fitting parameters to 2, but we will include $\Omega$ in the Fisher matrix (i.e. $k_r$, $C$, $h_0$ and $\Omega$).  We consider 2 different values of $k_r$, $k_r = 0.01$ and $k_r=0.1$ Mpc$^{-1}$ around $\Omega = 30$. We again compute the error on the amplitude $C$, $\Omega$ and $k_r$ as function of experimental configurations in Fig.~\ref{fig:sigmac_full}. We also show the two dimensional confidence contours for $k_r =0.1$ and $k_r=1$ Mpc$^{-1}$ of $C$ versus the relative error in $k_r$ in Fig.~\ref{fig:ContoursFullClock}. Smaller values of $k_r$ are harder to measure, with $k_r = 0.01$ Mpc$^{-1}$ just detectible at $1\sigma$ for an amplitude $C=0.01$ and increasing the baseline or narrow the radial resolution does not significantly improve the constraints. Interestingly, the location of the associated feature as well as its frequency can be determined.

The `fitted' shape (\ref{Template_SC_full}) is continuous and includes both the sharp feature and the clock signal. Increasing $k_r$ shifts the wave form to larger $k$ with increased $k$-range. As in the clock-signal-only case in Sec.~\ref{Sec:clock_inflation}, more modes are available in shorter scales and the constraint can be significantly improved if such a feature is present at larger $k_r$.

We show contours of joined probability in Fig.~\ref{fig:ContoursFullClock} for $k_r = 0.01$, $0.1$ and 1 Mpc$^{-1}$ at $\Omega = 30$. Like the clock signal template, there is a correlation between the frequency $\Omega$ and $k_r$. The contours confirm that features on smaller scales are better detectible and that although the amplitude can not always be excluded at $10^{-2}$ (for $k_r < 0.01$), the frequency $\Omega$ and $k_r$ can. We do not find correlations with $h_0$.

In conclusion, the smaller physical scale the feature appears, the easier it becomes to measure using 21 cm observations, with a threshold set by cosmic variance $k_r \geq 0.01$ Mpc$^{-1}$ for amplitudes of order 0.01.
Interestingly, the current best-fit model with Planck data has $k_r\approx 0.1$ Mpc$^{-1}$ and $C\approx 0.03$ \cite{Chen:2014cwa}, so in principle it could be well tested with 21 cm tomography.

Also note that, with the same value of $k_r$, we expect that the full signal template be better constrained than the clock-signal-only template because the signal in the region $k<k_r$ (i.e.~the sharp feature signal part) is no longer cut off and hence there exist less degeneracy with the parameter $k_r$.

\section{Conclusions} \label{Sec:conclusions}

In this paper we have forecasted the constraining power of 21 cm tomography in the search for primordial features. Primordial features are one of the most important extensions beyond the Standard Model of cosmology. As we reviewed, depending on their nature, detecting these features could reveal details of the inflation models, discriminate inflation from alternative-to-inflation scenarios, and potentially discover new massive particles.
Therefore a detection would imply a major discovery in the field. Current constraints come from the CMB. Although tentative candidates have been singled out, there is no convincing evidence that these are not just sourced by fluctuations in the noise or the result of cosmic variance. In the future, polarization measurements could provide additional evidence, or, further constrain parameter space. Beyond polarization, our next best hope is large scale structure \cite{FeaturesLargeScaleStructure,LSSnonGaussianity2014,FteauresLargeScaleStructure2,Chen:2016vvw}. In this paper we took this one step further, by considering very futuristic constraints derived from mapping out the 21 cm fluctuations against the CMB background during the Dark Ages. Our goal was to point out any limitations of such an experiment; in the most optimistic scenario, what kind of features would be observable?

We also investigated correlations between both primordial parameters and other cosmological parameters. We recover earlier found correlations for the the sharp feature mode in CMB analysis and forecasts. However, once the first oscillation of the feature is seen in the power spectrum, most degeneracies vanish, and in principle all parameters can be constrained independently (as long as the amplitude of the feature is sufficient to beat cosmic variance). In the flat-sky  $k$ space analysis we do not recover correlations found earlier in the resonant model, due to the absence of high oscillatory tranfer functions. Performing a similar analysis on the full sky and in multipole space does lead to such correlations.

Here we summarize the most important findings. We use $C$ to denote the ratio of the amplitude of the primordial features to that of the featureless primordial amplitude. Note that, although in this paper we take the fiducial value $C=0.01$, the one-sigma error for $C$ is independent of this value as long as it is small. The errors for other parameters such as the frequencies may change with the fiducial value. Our conclusions are based the error of the amplitude $C$, which is the most important parameter.

\begin{itemize}
\item Features that do not decay benefit most from having additional modes, with a radial resolution as high as $\delta \nu = 0.01$~MHz and an instrument with a baseline as long as 100 km. In particular:
\begin{itemize}
\item Sharp feature signals that extend all the way to small scales (and do not decay) will have a 1-sigma error bar for $C$ of order $10^{-5}$ to $10^{-6}$.
    The details depend on the correlations with other cosmological parameters. For example, for the sharp feature, at low frequencies there are correlations with $H_0$ which inflate the error bars. Increasing the frequency or extending the analysis to higher resolution breaks these degeneracies and the constraint on the amplitude becomes frequency independent. Other parameters associated with the feature are constrained better if more oscillations are resolved.
    Note the current CMB temperature data in Planck 2015 constrains the $C$ of sharp features to be below a few percent. For a couple of statistically insignificant best-fit feature models, locally, the one-sigma error on $C$ is $\sim 0.01$. So potentially, 21 cm tomography can improve the constraints by 3 or 4 orders of magnitude. Also note that the template we use here apply to features that are infinitely sharp. Realistically the signals are semi-extended and decay towards shorter scales depending on the sharpness of the feature. They are expected to be less constrained. 
The details of this resolution dependence can be read off from Fig.~\ref{fig:sigmac_feature}.

\item Resonance feature signals that extend all the way to small scales will have a 1-sigma error bars on $C$ of order $10^{-5}$ to $10^{-6}$. Due to the absence of strong correlations with cosmological parameters, the constraints on the amplitude is practically independent of the frequency. In the flat sky there is no rapidly oscillating transfer function which leads to partial cancellation of the signal; in a full-sky analysis we expect higher frequencies to be more difficult to measure.
Currently, Planck 2015 constrains $C$ to be below a few percent and the one-sigma error bar of $C$ for the best-fit model is $\sim 0.01$, so again we can hope to have 3 or 4 orders of magnitude of improvement from using  21 cm tomography.
The resolution dependence of the error bar can be read off from Fig.~\ref{fig:sigmac_resonant}.
\end{itemize}

\item Features that are semi-extended or localized in scales are harder to constrain because they only have support over a limited range of scales. Features that have more free parameters are harder to constrain because the additional degrees of freedom inflate the error bars on all primordial parameters. In particular:
\begin{itemize}
\item Features that are localized at very large scales, $k_r\lesssim 0.01 \;{\rm Mpc}^{-1}$, do not benefit significantly from 21 cm tomography and projected constraints from the CMB are comparable. A full 3D reconstruction, recovering the largest scales, could potentially improve this. However, from an observational point of view, those modes are usually masked since they contain most of the foreground power. In summary, features on large scales will generally be hard to constrain with 21 cm observations.

\item Around the scales $k_r \sim 0.1 \;{\rm Mpc}^{-1}$, in our very optimistic analysis we find 1-sigma error bar for $C$ of the inflationary clock signal (semi-extended) and ekpyrotic clock signal (localized) to be of order $\sim 10^{-3}$ and features with $C< 0.01$ would not be detected on these scales. The current inflationary clock signal candidate found in CMB \cite{Chen:2014cwa} has $C\sim 0.05$, so such a candidate should be testable. In fact, because these scales are also probed by galaxy surveys, 21 cm tomography is not quite advantageous for signals localized on these scales -- analyses show that the future low-$z$ galaxy surveys covering the same volume can reach a precision that is of the same order of magnitude and only worse by a factor of few \cite{Chen:2016vvw}.

\item Beyond the scales $k_r>1\;{\rm Mpc}^{-1}$, the CMB is damped and contaminated by astrophysical effects while computations of structure formation at low redshifts are completely non-perturbative; high redshift 21 cm tomography opens up a unique window into primordial features that are semi-extended or localized. Due to large amounts of modes, the projected 1-sigma error bars on the amplitude of features $C$ of both the inflationary and ekpyrotic clock signals are of order $2-4\times 10^{-4}$ or even better if one considers signals on even smaller scales. Such features should be detectable down to amplitudes $C \sim 0.0001$ or lower. This sets 21 cm apart from CMB and low $z$ LSS experiments which are not able to probe these scales.
\end{itemize}

\item Comparing the constraint for the clock-signal-only template and a full-clock-signal template, the error of the latter is reduced. This is mostly due to the artificial reduction of the number of parameters in the fitting procedure, and partly due to the loss of information in the analytical clock signal template used above, in which the sharp feature signal has to be cut off. Therefore, it is important to construct full standard clock templates that include both the sharp feature and clock signal.

\item Degeneracies between primordial parameters are caused by 1) having too few modes to resolve a full oscillation (only for localized features) 2) by the details of the template.

\item Degeneracies between primordial parameters and late time parameters are sourced by mimicking BAO-like features (linear features) or by projection effects (log features); the latter is not recovered in our flat-sky analysis.  Local features have minimal degeneracies with late time parameters.

\item{The analysis here is optimistic, but not optimal as further information can be harvested from cross-correlating redshift slices to recover modes from projection. One can further investigate the optimal window function to maximize the science output or consider different windows when searching for different signals, given the projected noise. }

\end{itemize}

In this paper we decided not to include noise or foregrounds. Our main goal was to show the limitations of 21 cm tomography in the search for features, in the ideal case. It is well known that foregrounds peak many order of magnitude above the actual signal \cite{loeb2013first}. However, current forecasts including noise and foregrounds would be highly speculative, since actual measurements of the 21 cm signal have not yet been made for redshifts $z \geq 30$. That being said, the power of 21 cm observations lies in the access to the smallest scales, were it is expected that foregrounds are suppressed. Our analysis concludes that localized features on large scales are very hard to constrain with 21 cm observations and using low-$z$ LSS surveys might prove a better alternative for constraining such features \cite{Chen:2016vvw}.

Besides noise and foregrounds, the extend to which the forecasted constraints can be realized relies on whether one is ever able to build an experiment that measures the 21 cm field in absorption in this frequency range. From earth it is theoretically possible to probe up to $z \simeq 45$ \cite{Carilli_2007} which would give access to a subvolume of the presented analysis here. The advantage is that from earth one can build a larger baseline and therefore, in principle, get access to smaller scales (with baselines $> 100$ km). Of course, this would require tremendous sensitivity (i.e. filling of the array), but it is worthwhile pointing out that such an experiment is more realistic than building an interferometer on the moon or in space. At the same time however, current models of the gas evolution suggest that at these lower redshifts the gas is diluted sufficiently to prevent direct coupling of the spin temperature to the gas \cite{Pritchard:2011xb}, setting $T_s$ close to $T_{\rm cmb}$ driving the observed brightness close to zero. In a realistic experiment including foregrounds and instrumental noise, this would lower the signal to noise of an experiment covering the redshift range observable from earth relative to the higher redshifts. We will leave a more detailed calculation for a future study.

We also considered the 21 cm signal as a stand-alone experiment, not taking into account existing cosmological constraints. Low $z$ LSS as well as CMB measurements provide independent constraints on all $\Lambda$CDM parameters, which set stringent priors on those parameters. This is partially taken into account by fixing all cosmological parameters except $H_0$. However, primordial parameters would also benefit from using multiple tracers.
Future LSS and CMB measurements can also be used to constrain primordial features, and are complimentary to 21 cm tomography \cite{Chen:2016vvw}.
Ultimately, the constraining power of all observations should be combined to put the strongest bounds on features.

The power spectrum is one possible observable that contains features that are sourced by new physics. Higher order correlation functions, such as the bispectrum, are predicted to contain features \cite{Chen:2006xjb,NonBDBispectrum2009,NonBDBispectrum2010,NonBDBispectrum2010b,nonBDbispectrum2015,Chen:2008wn,Adshead:2011jq,UnwindingInflation2013,Fergusson:2014hya,Fergusson:2014tza} that can be related to those in the power spectra. Currently, there is no significant evidence for non-Gaussianities, let alone one that contains features (although the ones with features have the highest significance, see e.g. \cite{PlanckNGs2015}). The power of 21 cm tomography in constraining non-Gaussianities with features is currently being investigated. Note that it is also possible to have features in the bispectrum, which are absent in the power spectrum \cite{LargeBispectrumGreen2012}.

The future of primordial features is promising given the forecast we have presented here. Because there is no theoretical lower-limit on the amplitudes of these feature models, it is important to know the experimental sensitivity that could be achieved in principle by this type of experiment. In this paper we have quantified these limits in the most optimistic scenario. We can only hope that nature was kind enough to provide us with a signal that falls within the observational limitations.

\medskip
\section*{Acknowledgments}
We thank Cora Dvorkin, Anastasia Fialkov, Zhiqi Huang, Avi Loeb, Mohammad Hossein Namjoo, Licia Verde, Benjamin Wandelt and the anonymous referee for helpful discussions and comments.
XC is supported in part by the NSF grant PHY-1417421.
PDM would like to thank the hospitality of Dutch-ITP, at the University of Amsterdam, where most of the this work was completed.

%

\bibliographystyle{JHEP}
\bibliography{references}

\end{document}